\documentclass[lineno]{jfm}

\usepackage{caption}
\usepackage{subcaption}

\usepackage{graphicx}
\usepackage{newtxtext}
\usepackage{newtxmath}
\usepackage{natbib}
\usepackage{hyperref}
\hypersetup{
    colorlinks = true,
    urlcolor   = blue,
    citecolor  = black,
}

\newcommand{\RomanNumeralCaps}[1]
\linenumbers

\title{Large Eddy Simulations and modal reconstruction of laminar transonic buffet}

\author{Pradeep Moise\aff{1}\corresp{\email{pradeep890@gmail.com}},
  Markus Zauner\aff{1,2}
 \and Neil Sandham\aff{1}}

\affiliation{\aff{1}University of Southampton, Southampton, Hampshire, SO17 1BJ, UK
\aff{2}ONERA - The French Aerospace Lab, Meudon, Île-de-France, 92190, France}

\begin{document}
\maketitle

\begin{abstract}
Transonic buffet refers to the self-sustained periodic motion of shock waves observed in transonic flows over wings and limits the flight envelope of aircraft. Based on the boundary layer characteristics at the shock foot, buffet has been classified as laminar or turbulent and the mechanisms underlying the two have been proposed to be different (Dandois \textit{et al.}, 2018, J. Fluid Mech., vol. 18,  pp. 156--178). The effect of various flow parameters (freestream Mach and Reynolds numbers and sweep and incidence angles) on laminar transonic buffet on an infinite wing (Dassault Aviation's supercritical V2C aerofoil) is reported here by performing Large-Eddy Simulations (LES) for a wide range of parameters. A spectral proper orthogonal decomposition identified the presence of a low-frequency mode associated with buffet and high-frequency wake modes related to vortex shedding. A flow reconstruction based only on the former shows periodic boundary-layer separation and reattachment accompanying shock wave motion. A modal reconstruction based only on the wake mode suggests that the separation bubble breathing phenomenon reported by Dandois \textit{et al.} is due to this mode. Together, these results indicate that the physical mechanisms governing laminar and turbulent buffet are the same. Buffet was also simulated at zero incidence. Shock waves appear on both aerofoil surfaces and oscillate out of phase with each other indicating the occurrence of a Type I buffet (Giannelis \textit{et al.}, 2018, Aerosp. Sci. Technol., vol. 18, pp. 89--101) on a supercritical aerofoil. These results suggest that the mechanisms underlying different buffet types are the same. % word count: 250

%Additionally, the upstream shock motion was always accompanied by a pressure buildup near the trailing edge. %Methods to predict and mitigate buffet are crucial for next-generation laminar flow wing sections. % Buffet frequency increased with $M$ while it remained approximately a constant when $\alpha$ or $\Rey$ is increased. Multiple shock wave structures occur at low $\Rey$, but their numbers reduce with increasing $\Rey$. %  at a Strouhal number, $St \approx 0.1$ % with $1 \leq St \leq 4$
%  (Giannelis \textit{et al.}, 2018, Aerosp. Sci. Technol., vol. 18, pp. 89--101) The reference case is selected as an unswept wing at an incidence angle of $\alpha = 4^\circ$ with freestream Mach and Reynolds numbers as $M = 0.7$ and $\Rey = 5\times 10^5$ and each parameter is individually varied about this and their effect documented. % Motivated by the possibility of three-dimensional buffet cells, simulations were performed at different sweep angles. For all, no buffet cells were observed and the flow features were similar. % , with an approximately 90-degree phase lag present between shock wave position and separation length for all cases
%2do: 249
\end{abstract}


%%%%%%%%%%%%%%%%%%%%%%%%%%%%%%%%%%%%%%%%
\section{Introduction}
\label{secIntro}
%%%%%%%%%%%%%%%%%%%%%%%%%%%%%%%%%%%%%%%%

Interaction between shock waves and the boundary layer (BL) developing on rigid wings can lead to self-sustained periodic flow oscillations referred to as transonic buffet \citep{Helmut1974}. Transonic buffet, which we'll henceforth refer to as buffet for brevity, can cause strong variations in lift coefficients thus limiting the flight envelope of commercial aircraft and the manoeuvrability of combat aircraft \citep{Roos1980,Lee2001,Giannelis2017}. Most studies on buffet focus on `turbulent buffet' where a turbulent BL develops well upstream of the shock wave (\textit{e.g.}, due to fully turbulent inflow conditions or forced transition \citep{Jacquin2009} or even under free transition conditions at high $\Rey$ \citep{Lee1989}). Although the fundamental mechanisms that drive this buffet type remain unclear \citep{Giannelis2017}, recent studies have shown that this phenomenon arises through a supercritical Hopf bifurcation associated with an unstable global mode \citep{Crouch2007, Crouch2009, Sartor2015, Crouch2019, Timme2020}. In contrast to turbulent buffet, relatively few studies have examined `laminar buffet', for which the BL remains laminar from the leading edge (LE) until approximately the shock foot location. The mechanisms governing laminar buffet has been proposed to be distinct from those of turbulent buffet \citep{Dandois2018}, while the effects of various flow parameters on it remain largely unexplored. Motivated by these, we examine the influence of flow parameters on laminar buffet by studying the flow past an infinite wing-section based on a laminar supercritical aerofoil using Large-Eddy Simulations (LES) while varying the incidence and sweep angles ($\alpha$ and $\Lambda$), and the freestream Mach and Reynolds numbers ($M$ and $\Rey$). The three main objectives of this study are: (1) to provide a numerical database for laminar buffet over a large parametric range based on scale-resolved simulations for future studies to compare with, (2) to find underlying aspects that are common to buffet over the entire range in which it is observed, and use these to (3) assess buffet models and (4) make comparisons with turbulent buffet.

Buffet on supercritical aerofoils has been examined extensively in various experimental studies \citep{Tijdeman1976, Lee1989, Jacquin2009, Hartmann2012, Brion2020}. Similarly, computational studies have successfully simulated buffet using a wide range of numerical approaches from integral boundary layer equations \citep{Tijdeman1980}, unsteady Reynolds-Averaged Navier-Stokes equations (URANS) \citep{Xiao2006a}, detached eddy simulations (DES) \citep{Deck2005}, wall-modelled LES \citep{Fukushima2018}, wall-resolved LES \citep{Dandois2018} and direct numerical simulations (DNS) \citep{Zauner2019}. Based on these studies, several models have been proposed to explain buffet on supercritical aerofoils, although the physical mechanisms that drive it remain unclear, as elaborated below.

Based on previous studies on shock wave responses to a trailing edge (TE) flap \citep{Tijdeman1976,TijdemanReport}, a model for buffet based on a feedback loop was proposed in \citet{Lee1990}, as follows. Pressure waves generated by the shock wave motion convect from the shock foot to the TE along the BL in a time, $t_\mathrm{down}$. These pressure waves interact with the TE, generating ``Kutta waves" \citep{TijdemanReport}. The Kutta waves propagate upstream outside the BL and reach the shock wave near its foot in a time, $t_\mathrm{up}$. At this point, they interact with the shock foot and induce shock wave motion, completing the loop. The buffet time period is then predicted as $\tau_\mathrm{Lee} = t_\mathrm{down}+t_\mathrm{up}$. There is some evidence supporting this model. For example, the presence of propagating pressure waves in the flow field is well-documented \citep{Jacquin2009,Hartmann2013,Zauner2020}, while the predicted frequency has been shown to match with observed buffet frequency in some studies \citep{Deck2005, Xiao2006a}. However, other studies have not found the predictions to match \citep{Jacquin2009,Garnier2010} and modifications that alter the distance travelled by the Kutta waves have been proposed \citep{Jacquin2009,Hartmann2013}. Furthermore, Lee's model does not explain how the Kutta waves generated at the TE lead to shock motion. This question was partly addressed in \citet{Hartmann2013}, where it was suggested that an increase or decrease in intensity of acoustic waves generated at the TE causes upstream or downstream shock wave motion. This intensity was predicted to increase/decrease due to strong/weak vortices generated by shock-induced BL separation interacting with the TE when the shock wave is at its most downstream/upstream position. 

Other studies have proposed that buffet might occur due to the interplay of shock wave strength and BL separation \citep{McDevitt1976, TijdemanReport, McDevitt1985, Gibb1988, Raghunathan1998}. If the shock wave is perturbed to move upstream, it would strengthen, causing stronger shock-induced separation. The decambering effects caused by shock-induced separation would lead to the shock moving further upstream, which in turn would increase the shock strength due to increasing effective Mach number upstream to the shock. \citet{Iovnovich2012}  further refined this model by proposing that the shock strengthening and weakening could be governed by the wedge, dynamic and curvature effects with the influence of each varying based on the phase of the cycle. 

As noted above, strong evidence supporting the notion that buffet occurs as a global instability was first provided in \citet{Crouch2007} using a global linear stability analysis. However, a physical mechanism explaining what \textit{drives} the global instability is not evident from this result, implying that other physical models that can be interpreted to rely on instabilities \citep{McDevitt1985, Gibb1988, Raghunathan1998, Iovnovich2012} might complement it. Importantly, all of the models discussed above are based on turbulent buffet, while their applicability to laminar buffet has not been tested. 

Motivated by requirements of reducing civilian aircraft emissions, buffet has also been investigated on \textit{laminar} supercritical aerofoils \citep{Dor1989, Brion2020, Dandois2018, Memmolo2018, Zauner2019, Zauner2020, Plante2020}. In an experimental study of the laminar supercritical OALT25 profile, \citet{Brion2020} observed that unlike turbulent buffet which occurs (when BL tripping is employed) at a low Strouhal number (based on chord and freestream velocity) of $St \approx 0.05$, laminar buffet (no trip) is dominated by pressure fluctuations at $St \approx 1.2$. LES at the same flow conditions ($\Rey = 3\times 10^6$) were carried out in \citet{Dandois2018}. Based on the LES results, the authors concluded that, unlike turbulent buffet, the former is driven by a separation bubble breathing phenomenon. However, the authors reported only temporal variations in the position of the shock foot and not the entire shock wave. Laminar buffet at a lower Reynolds number ($\Rey = 5 \times 10^5$) was examined for the V2C profile in our previous studies \citep{Zauner2019, Zauner2020a, Zauner2020}. Unlike \citet{Dandois2018}, a shock system comprising of multiple shock wave structures was observed, with the entire shock system exhibiting periodic oscillations. This has been recently confirmed by an ongoing experimental campaign at ONERA \citep{ZaunerUKFluids2021}.

From the above discussion it is evident that further exploration is required in assessing the various models proposed for both laminar and turbulent buffet. In this regard, comprehensive parameter studies for multiple aerofoils are useful for assessing the validity of various models and also for quantifying the sensitivity of buffet to the profile shape. Here, we continue our studies on laminar buffet \citep{Zauner2019,Zauner2020,Zauner2020a} focusing on Dassault aviation's supercritical laminar V2C profile which complements those on the OALT25 profile carried out by others \citep{Brion2020,Dandois2018}. Furthermore, we have opted to employ wall-resolved LES of laminar buffet on the V2C profile, which has several advantages over other methods. Various challenges faced in experimental studies when examining buffet are avoided, including confinement and side-wall effects \citep{Davidson2016}, surface quality/manufacturing tolerances, fluid-structure interaction, measurement uncertainties, wind-tunnel noise and free-stream turbulence levels \citep{Giannelis2017}. Furthermore, although major insights into buffet features have come from URANS studies \citep{Crouch2009, Iovnovich2012, Sartor2015, Xiao2006a, Iovnovich2015}, these too have several drawbacks related to sensitivity of results to turbulence closure models \citep{Grossi2014,Giannelis2017}, accuracy of modelling free transition, and the capturing of vortex shedding \citep{Sartor2015,Grossi2014,Poplingher2019}. Indeed, due to the strong sensitivity of predicted buffet features to the turbulence closure model adopted, \citet{Giannelis2017} concluded that ``the simulation of shock buffet through URANS becomes more so an art than a science", which motivates the present LES study. 

In addition to the commonly investigated parameters of $M$, $\alpha$ and $\Rey$, we also briefly examine the effect of sweep here. While buffet remains essentially two-dimensional (2D) for unswept infinite wings, \citet{Iovnovich2015} reported three-dimensional (3D) ``buffet cells" that occur with the introduction of sweep. Subsequent studies have shown this feature to occur over finite wings \citep{Dandois2016}, arise as a global instability \citep{Crouch2019, Paladini2019, Timme2020} and is related to stall cells superimposed on two-dimensional buffet \citep{Plante2020}. 

The rest of the article is organised as follows. The methodology used for the simulations and modal decomposition are discussed in  \S\ref{secMethod}. A description of the flow states that occur as the different parameters are varied is provided in  \S\ref{secResults}. Subsequently, the coherent features of these flows are scrutinised in \S\ref{secSPOD} by performing a spectral proper orthogonal decomposition and reconstructing the flow field based on relevant modes obtained from the same. The implications of these results are discussed in \S\ref{secDisc}, while  \S\ref{secConc} concludes the study. 

%%%%%%%%%%%%%%%%%%%%%%%%%%%%%%%%%%%%%%%%
\section{Methodology}
\label{secMethod}
%%%%%%%%%%%%%%%%%%%%%%%%%%%%%%%%%%%%%%%%

%%%%%%%%%%%%%%%%%%%%%%%%%%%%%%%%%%%%%%%%
\subsection{Numerical simulations}
\label{subsecMethodLES}
%%%%%%%%%%%%%%%%%%%%%%%%%%%%%%%%%%%%%%%%
The numerical simulations were carried out using the in-house code, SBLI \citep{Yao2009}, which is a scalable compressible flow solver with multi-block and shock-capturing capabilities and has been used previously to study buffet \citep{Zauner2019,Zauner2020a,Zauner2020}. SBLI solves for the compressible Navier-Stokes equations which govern the flow evolution in %given by (using Einstein's summation convention)
%$$
%\frac{\partial \rho}{\partial t} + \frac{\partial \rho u_j}{\partial x_j} = 0,
%$$
%$$
%\frac{\partial \rho u_i}{\partial t} + \frac{\partial \rho u_i u_j}{\partial x_j} = -\frac{\partial p}{\partial x_i} + \frac{1}{\Rey}\frac{\partial\tau_{ij}}{\partial x_j},
%$$
%$$
%\frac{E_T}{\partial t} + \frac{\partial E_T u_j}{\partial x_j} = -\frac{\partial pu_j}{\partial x_j} + %\frac{\partial q_j}{\partial x_j} +  \frac{1}{\Rey}\frac{\partial u_i\tau_{ij}}{\partial x_j},
%$$
%representing the conservation of mass, momentum and energy. Here, the volume specific total energy, $E_T = \rho e_{int} + u_ju_j$ and the volume specific internal energy $e_{int} = T/(\gamma(\gamma-1)M_l^2)$ The scales used to arrive at this dimensionless form are the aerofoil chord ($c^*$) and freestream density ($\rho^*_\infty$), axial velocity component ($U_\infty^*$) and temperature ($T_\infty^*$). 
a dimensionless form \citep[see][pp. 3-4]{Zauner2020}). The aerofoil chord, the freestream density, streamwise velocity (unswept) and temperature are used as reference scales implying that their corresponding dimensionless equivalents are given by $c = \rho_\infty = U_\infty = T_\infty = 1$, respectively. Fourth-order finite difference schemes (central at interior and the Carpenter scheme \citep{Carpenter1999} at boundaries) are employed for spatial discretisation, while a low-storage third-order Runge-Kutta scheme is used for time discretisation. A constant dimensionless time-step of $3.2\times10^{-5}$ is used. The fluid is assumed to be a perfect gas with a specific heat ratio, $\gamma = 1.4$ which satisfies Fourier's law of heat conduction (Prandtl number, $\Pran = 0.72$). It is also assumed to be Newtonian, with its viscosity variation with temperature governed by the Sutherland's law (Sutherland coefficient, $C_\mathrm{Suth} = {110.4}/{268.67} \approx 0.41$). 

To perform the LES, we adopt the spectral-error based implicit LES approach which has been validated against DNS for flows where buffet is observed \citep{Zauner2020}. This approach utilises the error estimator proposed in \citet{Jacobs2018} for identifying regions of insufficient grid resolution in DNS. Based on this estimator, a low-pass filter is used on all conserved variables to locally correct spectral deviations whenever and wherever they occur. Sixth-order compact finite difference schemes for filtering applications \citep{Lele1992} are used for this purpose. A blending function is used to reduce the impact of filtering at higher wavenumbers. This is given by $q_{\mathrm{updated}} = q_{\mathrm{unfilt.}} - a_{\mathrm{lim}} (q_{\mathrm{unfilt.}}- q_{\mathrm{filt.}})$, where the updated flow field is computed as an affine combination of the unfiltered and filtered flow fields \citep{BOGEY2004194} with the constant, $a_{\mathrm{lim}} = 0.4$. The parametric values used here are the same as those used in \citet{Zauner2020} (see their table 1). To capture features of shock waves, we use a total variation diminishing scheme, details of which can be found in \citet{Zauner2019}.

\begin{figure} 
\centerline{\includegraphics[trim={1cm 0cm 2cm 0cm},clip,width=0.495\textwidth]{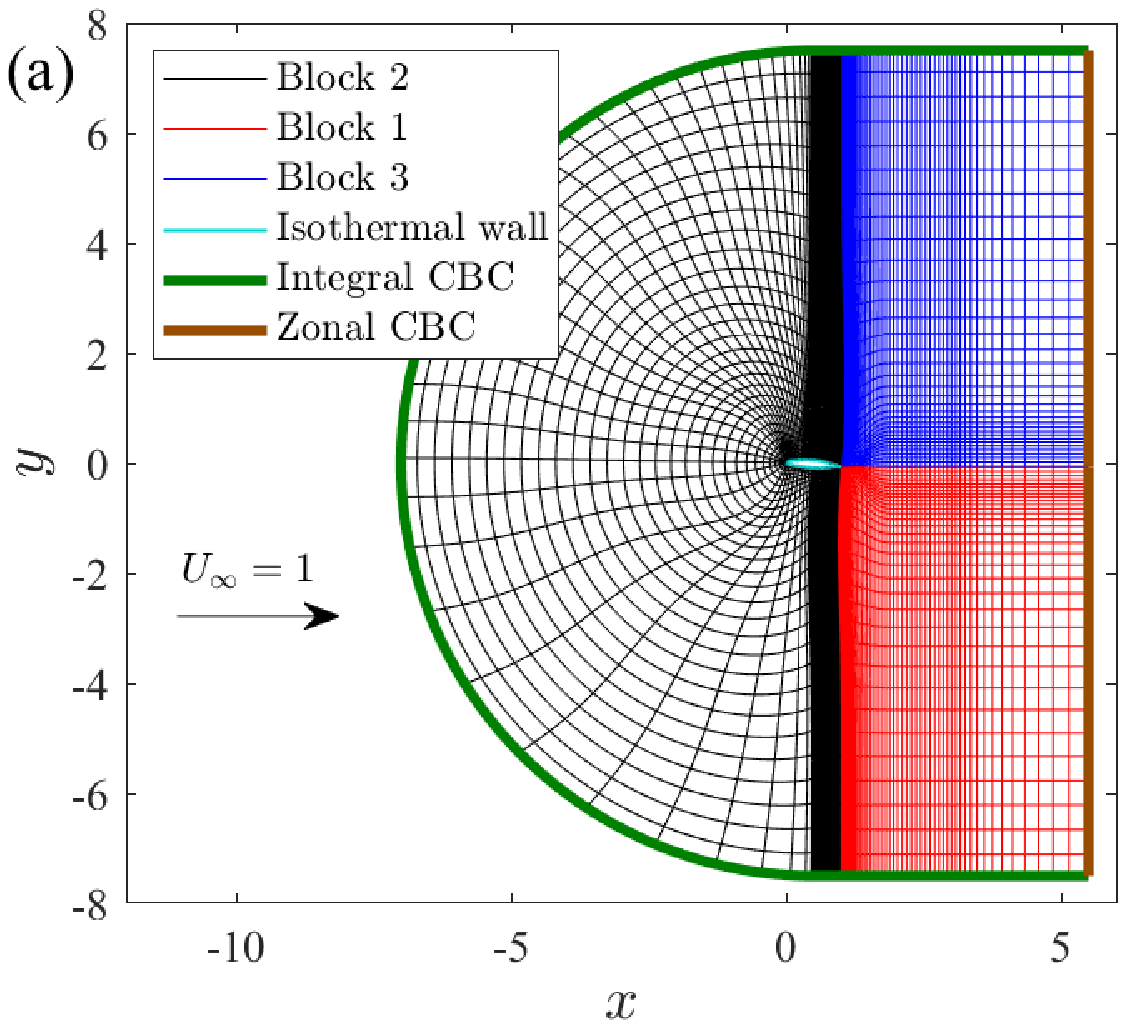}
\includegraphics[trim={1cm 0cm 2cm .5cm},clip,width=0.495\textwidth]{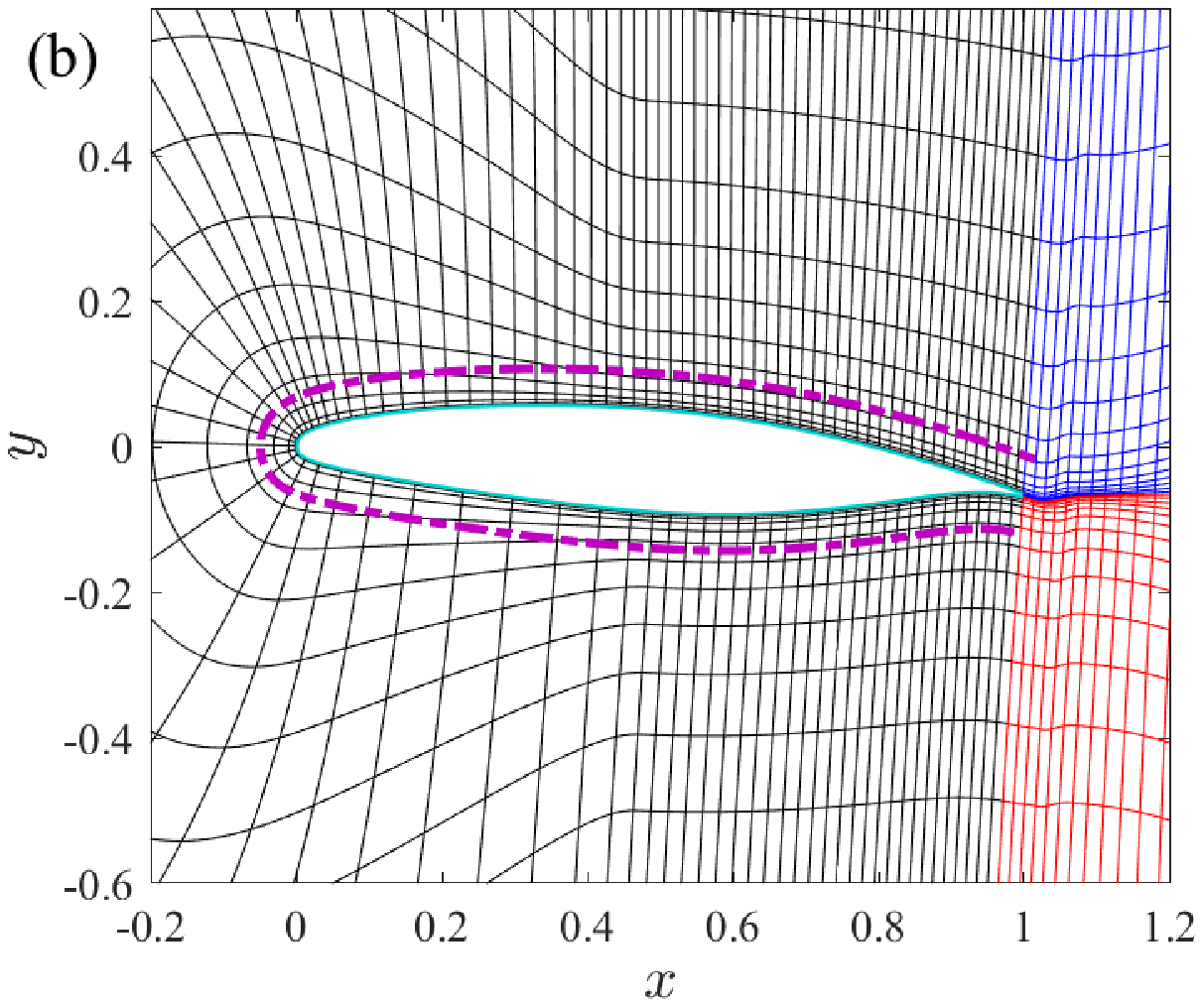}}
\caption{Multi-block grid shown by plotting every $15^\mathrm{th}$ grid point in $\xi$ and $\eta$ directions for the case of $\alpha = 4^\circ$: (a) entire domain and (b) vicinity of aerofoil.}  
    \label{fig:gridAll}
\end{figure}

Dassault Aviation's supercritical, laminar V2C profile with a blunt trailing edge, as used in the TFAST project \citep{TFAST}, is employed in the simulations. To accurately model the blunt trailing edge, maintain grid smoothness, and ensure flexibility in the distribution of grid points, C-H multiblock structured grids were generated using an in-house, open-source code \citep{Zauner2018} for different incidence angles. Features of a typical case of $\alpha = 4^\circ$ (``reference grid'') are highlighted in figure \ref{fig:gridAll}\textit{a}. The aerofoil (cyan) is treated as an isothermal wall with the temperature set equal to that of the freestream, while integral and zonal characteristic boundary conditions (CBC) \citep{sandhu1994boundary, Sandberg2006} are used on the freestream (green) and outflow boundaries (brown), respectively. A localised filter is adopted to handle the singular points at the corners of the blunt trailing edge by employing the strategy proposed in \citet{Jones2006}. For the unswept cases, $x$ and $z$ represent the dimensionless streamwise and spanwise Cartesian coordinates based on the freestream, respectively, while $y$ represents the coordinate orthogonal to $x$ and $z$ and $x'$, the chordwise coordinate. The curvilinear circumferential and radial coordinates are $\xi$ and $\eta$, respectively. The reference grid's features in the aerofoil's vicinity are highlighted in figure \ref{fig:gridAll}\textit{b}. A curve at a constant wall-normal distance of $0.05c$ from the aerofoil's suction surface, is also shown (dashed curve). The latter is used for monitoring shock wave motion and will be referred to as C5 henceforth. The grid clustering is relatively denser close to the aerofoil surface, in its wake, and in the region where the shock wave is expected.

We examine the effect of various flow parameters ($M$, $\alpha$, $\Rey$ and $\Lambda$) on buffet by starting from a baseline ``reference" case with $M = 0.7$, $\alpha = 4^\circ$, $\Rey = 5\times 10^5$ and $\Lambda = 0^\circ$ and varying the value of only one of these parameters while keeping the other parameters the same as that of the reference. In addition to these, a case at $M = 0.8$ and $\alpha = 0^\circ$, with other parameters at reference value, was also simulated to examine buffet features at zero incidence. This is referred to as the A0M8 case (see  \S\ref{subSecZeroAoA}). A complete list of all cases studied and their parametric values is given in table \ref{tableCases}. Note that when $\Lambda = 0^\circ$, the flow configuration is of an unswept wing and the $x-$coordinate is based on the streamwise direction. Thus, the freestream $x-$velocity component is unity while other components are zero. For these unswept cases, the grid was extruded in the third dimension to achieve a span of $L_z = 0.05$ (``narrow domain") with the spanwise boundaries assumed periodic (to model an infinite unswept wing). A uniform grid-spacing of $10^{-3}$ is used. Note that the grid resolution close to the leading edge is relatively coarser than at other regions of the aerofoil (figure \ref{fig:gridAll}\textit{b}). This was found to introduce minor grid-level oscillations in the flow field only for the highest incidence angle considered ($\alpha = 6^\circ$, see table \ref{tableCases}). This occurred  in parts of the buffet cycle when the shock wave moved in close proximity to the leading edge due to the high amplitude of buffet at this $\alpha$. However, this was only during a small part of the buffet cycle and no significant effect on the buffet flow features could be discerned. Nevertheless, higher angles of attack ($\alpha > 6^\circ$) were not examined for this reason.

The effect of sweep angle was examined by varying the freestream spanwise velocity component as $u_z = \tan\Lambda$, while $u_x$ is maintained as unity. Note that $x$ is no longer associated with the streamwise direction when $\Lambda\neq 0^\circ$, but is normal to the spanwise direction and oriented at an angle $\Lambda$ with the freestream velocity vector. This choice of $u_z$ allows to examine the validity of the Independence Principle (also known as sweep-independence principle), which goes back to Prandtl \citep[see][]{Selby1983}. This principle can be derived for flows which are spanwise invariant under assumptions of incompressible laminar flow, by showing that the Navier Stokes equations reduce to a set of equations in which the velocity components, $u_x$ and $u_y$, pressure, and the spanwise vorticity component are uncoupled from the other velocity and vorticity components \citep[see Eqs. 4-7][p. 488]{Hetsch2009}). Similar expressions have also been derived for compressible flows \citep{Jones1946,Struminsky1951} and form the basis for sweepback in supersonic aircraft. This implies that irrespective of the sweep angle, the features in the plane normal to the spanwise direction are the same as that of the unswept case. While strictly valid only for laminar flows, this principle has also been shown to hold for a significant range of sweep angles even if the flow is turbulent \citep{Selby1983}. Note that here the velocity component normal to the aerofoil's leading edge does not change with $\Lambda$, since the incidence angle is maintained constant ($\alpha = 4^\circ$) for these cases. To check for the occurrence of 3D buffet cells which are generally predicted to have spanwise wavelengths of the order of the aerofoil chord \citep{Iovnovich2015,Crouch2019,Paladini2019}, we used $L_z = 1$ with the same uniform spanwise grid spacing. This grid is referred to as a ``wide domain" and is used only when examining effects of $\Lambda$. 

\begin{table}
\begin{center}
\def~{\hphantom{0}}
\begin{tabular}{lcccccc}
Name & $M$ & $\alpha$ & $\Rey$ & $\Lambda$ & $L_z$ & Parameter values\\[6pt]
Reference & 0.7 & $4^\circ$ & $5\times 10^5$ & $0^\circ$ & 0.05 & --\\
Effect of $M$ & -- & $4^\circ$ & $5\times 10^5$ & $0^\circ$ & 0.05 & 0.5, 0.6, 0.65, 0.68, 0.69, \textbf{0.7}, \textbf{0.735}, \textbf{0.75}, \textbf{0.775}, \textbf{0.8}, \textbf{0.85}, 0.9\\
Effect of $\alpha$ & 0.7 & -- & $5\times 10^5$ & $0^\circ$ & 0.05 & $\boldsymbol{3^\circ}$, $\boldsymbol{4^\circ}$, $\boldsymbol{5^\circ}$, $\boldsymbol{6^\circ}$\\
Effect of $\Rey$ & 0.7 & $4^\circ$ & -- & $0^\circ$ & 0.05 & $2\times 10^5$, $\boldsymbol{5\times 10^5}$, $\boldsymbol{1\times 10^6}$, $\boldsymbol{1.5\times 10^6}$ \\
% Effect of $\Lambda$ (narrow domain) & 0.7 & $4^\circ$ & $5\times 10^5$ & -- & 0.05 & $\boldsymbol{0^\circ}$, $\boldsymbol{20^\circ}$, $\boldsymbol{40^\circ}$\\
Effect of $\Lambda$ & 0.7 & $4^\circ$ & $5\times 10^5$ & -- & 1 & $\boldsymbol{0^\circ}$, $\boldsymbol{20^\circ}$, $\boldsymbol{40^\circ}$\\
A0M8 & 0.8 & $0^\circ$ & $5\times 10^5$ & $0^\circ$ & 0.05 & --\\
\end{tabular}
\caption{Parameter values for various cases simulated (cases with buffet in boldface).}
\label{tableCases}
\end{center}
\end{table}

For all cases except those where $\alpha$ is varied, the simulations were initialised with freestream conditions used at all grid points. For the study on varying $\alpha$, because it was originally initiated as part of a different study, the initial conditions were chosen as the fully evolved solutions of the reference case. Note that the choice of initial conditions have been shown to not affect flow features that occur past the transients for supercritical aerofoils \citep{Xiao2006a}. Beyond the transients, the simulations were run until at least ten buffet cycles were completed (or for 20 time units when buffet is absent) for the narrow-domain cases so as to improve statistical convergence of mean quantities and the accuracy of modal decomposition. Only two or three cycles were run for the wide-domain cases examining sweep effects due to associated numerical expense.

\subsection{Validity of simulation results}
\label{subSecValid}

The LES approach used here is the same as that in \citet{Zauner2020}, and has been validated in that study for the reference case by comparing with a DNS. Key flow features, including the frequency and amplitude of buffet agree well between the two. As noted in  \S\ref{secIntro}, an ongoing experimental campaign at ONERA on laminar buffet at Reynolds numbers similar to those studied here has confirmed several results observed in this study including the presence of multiple shock wave structures, shock system oscillation and a buffet frequency of $St \approx 0.1$ \citep{ZaunerUKFluids2021}. Further support for the present simulations, including the choice of spanwise width and grid resolution are provided below.

\subsubsection{Spanwise width}
The validation of the LES with DNS in \citet{Zauner2020} was performed in a narrow domain of $L_z = 0.05$. To examine the effect of spanwise width, an LES for a wide domain  of $L_z = 1$ was also performed. It was observed that some differences exist between the two, including increases in the buffet amplitude and regularity of the oscillations, but the major flow features, including the development of multiple shock wave structures, their propagation features and buffet frequency remain unaffected. Since the focus of this study is on examining qualitative aspects of buffet, we have chosen $L_z = 0.05$. Results from other previous studies on turbulent buffet also support this choice. Experiments reported in \citet{Jacquin2009} have shown that although weak 3D features associated with flow separation are present in the flow, buffet on unswept wings is characterised by a 2D mode. Wall-modelled LES in a domain of span 0.065$c$ was shown in \citet{Fukushima2018} to accurately capture buffet features observed in experiments. Similarly, in \citet{Grossi2014}, 3D structures with the spanwise wavelength between $0.029-0.04c$ were observed in DES. 

\subsubsection{Grid resolution}

In the present simulations, the main feature of the flow is a laminar BL that develops from the leading edge up to the shock foot. To examine if this is accurately captured, a new grid with 25\% more points was generated, with the grid spacing in $\xi$ and $\eta$ directions in the laminar BL region ($0 \leq x \leq 0.5$) approximately halved. The simulation was carried out at the highest Reynolds number studied here ($\Rey = 1.5\times10^6$). The results, including instantaneous lift coefficient evolution and mean surface coefficients, match well for the two grids (not shown, provided as supplementary material), confirming the adequacy of the original grid in capturing the laminar BL. 

In the regions where the mean flow is attached and turbulent, for most cases studied, the grid spacing in wall units at the aerofoil surface computed based on the mean wall shear stress, $\Delta \xi^+$, $\Delta \eta^+$ and $\Delta z^+$, were found to be approximately 10, 1 and 10, respectively. This indicates that the grid resolution is more than adequate for LES \citep{Garnier2010, Dandois2018}. For a few cases close to buffet onset ($\alpha = 3^\circ$ and $M \leq 0.69$), the $\Delta \eta^+$ value is higher ($\approx 2$), but only in a small region downstream to the shock (approximately $0.85 \leq x \leq 0.9$) and well downstream to the mean shock wave position ($x \approx 0.6$, see figure \ref{fig:M65N69DensGrad}\textit{b}). Additionally, for $\Rey > 5\times 10^5$, we found $\Delta \xi^+,\:\Delta z^+ \approx 30$ and $\Delta \eta^+ \approx 5$ in some regions of the flow. While the $\Delta \xi^+$ and $\Delta z^+$ are still adequate (for comparison, \citet{Garnier2010} used $\Delta \xi^+ \approx 50$ and $\Delta z^+ \approx 20$ for LES), the $\Delta \eta^+$ values are higher than recommended. We do not expect this to play a significant role in affecting buffet dynamics since we have a laminar BL upstream to the shock wave, and downstream, a separation bubble which reattaches only in parts of the buffet cycle to form a turbulent BL which persists only a short distance before reaching the trailing edge. Care has been taken to ensure that the regions associated with separated flow are well-resolved, based on visual checks of instantaneous flow fields and using the spectral-error indicator to check filter activity. Also, as noted above, a grid study at $\Rey = 1.5 \times 10^6$, did not show any significant changes in buffet features. Thus, the adequacy of the grid is supported by a combination of wall-unit checks, grid sensitivity studies and careful monitoring of simulations using the spectral error detector.

\subsection{Modal decomposition and reconstruction}
\label{subSecSPODMethod}

We employ spectral proper orthogonal decomposition (SPOD) to examine coherent structures observed in the flow field \citep{Lumley1970, Glauser1987, Picard2000, Towne2018}. This approach has several advantages over other modal decomposition techniques. Compared to the classic proper orthogonal decomposition, the modes extracted using SPOD are temporally orthogonal and monochromatic. Also, as compared to the dynamic mode decomposition (DMD), which also obtains modes with such properties, it is shown in \citet{Towne2018} that the SPOD modes are optimally-averaged DMD modes. Thus, SPOD provides better estimates of coherent features by reducing the `noise' that can accompany DMD modes. This is particularly useful in the modal reconstruction that we have implemented here. Furthermore, the problem of spurious modes that are observed for DMD \citep{Ohmichi2018,Zauner2020a} are avoided here. 

\subsubsection{Formulation}
Given an ensemble of spatio-temporal realisations that represent a stochastic process, $\{\boldsymbol{q}_\zeta(\boldsymbol{x},t)\}$, with $\zeta$ denoting the realisation's index, a set of orthogonal basis functions, $\{\boldsymbol{\phi}_i(\boldsymbol{x},t)\}$, that ideally represents the coherent aspects of the process is determined by maximising a utility functional \citep{Lumley1970,Towne2018},
\begin{equation}
J(\boldsymbol{\phi}) = E_\zeta\left[\frac{|\langle\boldsymbol{q}_\zeta,\boldsymbol{\phi}\rangle_{(\boldsymbol{x},t)}|^2}{\langle\boldsymbol{\phi},\boldsymbol{\phi}\rangle_{(\boldsymbol{x},t)}}\right]. \label{eqnSPODCostFunc}
\end{equation}
Here the $E_\zeta[]$ is the expectation operator, while $\langle,\rangle_{(\boldsymbol{x},t)}$ denotes the appropriate inner product over space, $\boldsymbol{x}$, and time, $t$. This allows for choosing each basis function such that the projection of $\boldsymbol{q_\zeta}$ on $\boldsymbol{\phi}_i$, (\textit{i.e.}, $\langle\boldsymbol{q}_\zeta,\boldsymbol{\phi}_i\rangle_{(\boldsymbol{x},t)}$), is maximised over $\zeta$ in a least-square sense. It is possible to show (\textit{e.g.}, by assuming that $\boldsymbol{\phi}_i$ are normal and reformulating using Lagrange multipliers) that the extrema of this functional satisfy the eigenvalue problem
\begin{equation}
    \langle \boldsymbol{\mathsfbi{C}}(\boldsymbol x,\boldsymbol{x'},t,t'),\boldsymbol{\phi}_i(\boldsymbol{x'},t')\rangle_{(\boldsymbol{x'},t')} = \lambda_i\boldsymbol{\phi}_i(\boldsymbol x,t), \label{eqnEVPC}
\end{equation}
where $\boldsymbol{\mathsfbi{C}}$ is the two-point spatio-temporal correlation tensor based on any two points $\boldsymbol x$ and $\boldsymbol{x'}$ in the spatial domain and $t$ and $t'$ in time, while $\boldsymbol{\phi}_i$ and $\lambda_i$ are the $i-$th eigenfunction and eigenvalue, respectively. Note that, as is common for eigenvalue problems, the index $i$ on the right hand side does not denote a summation. Here, $\lambda_i = J(\boldsymbol\phi_i)$, implying that the maximum of $J$ is given by the maximum eigenvalue. The expansion of any realisation, $\boldsymbol{q}_\zeta$, using the basis $\boldsymbol{\phi}$ is given by
\begin{equation}
 \boldsymbol{q}_\zeta(\boldsymbol x,t) = \sum_i  \langle\boldsymbol{q}_\zeta,\boldsymbol{\phi}_i\rangle_{(\boldsymbol{x},t)} \boldsymbol{\phi}_i(\boldsymbol x,t) = \sum_i a_{i,\zeta} \boldsymbol{\phi}_i(\boldsymbol x,t),  \label{eqnSPODBasisExpans}
\end{equation}
where $a_{i,\zeta}$ are the expansion coefficients. It follows from the above equations and the orthonormality of the basis functions ($\langle\boldsymbol{\phi}_j, \boldsymbol{\phi}_k\rangle_{(\boldsymbol{x},t)} = \delta_{jk}$) that 

$$
E_\zeta\left[a_{j,\zeta}a_{k,\zeta}^*\right] = \langle\langle \boldsymbol{\mathsfbi{C}}(\boldsymbol x,\boldsymbol{x'},t,t'),\boldsymbol{\phi}_j(\boldsymbol{x'},t')\rangle_{(\boldsymbol{x'},t')},\boldsymbol{\phi}_k(\boldsymbol{x},t)\rangle_{(\boldsymbol{x},t)} 
$$
\begin{equation}
\Rightarrow E_\zeta\left[a_{j,\zeta}a_{k,\zeta}^*\right] = \lambda_j \delta_{jk}, \label{eqnSPODExpCoeff}
\end{equation}
where $\delta_{jk}$ is the Kronecker Delta function and $()^*$ denotes complex conjugation.

For a zero-mean, statistically stationary flow, the correlation tensor is dependent only on the time interval, $\tau = t'-t$ and thus, a Fourier transform in time can be used to compute the cross-spectral density tensor, $\boldsymbol{\mathsfbi{S}}$. The relations between the two are
\begin{equation}
    \boldsymbol{\mathsfbi{S}}(\boldsymbol x,\boldsymbol{x'},f) = \int_{-\infty}^\infty \boldsymbol{\mathsfbi{C}}(\boldsymbol x,\boldsymbol{x'},\tau) \mathrm{e}^{-2\mathrm{i}\upi f\tau} d\tau; \:\:\: \boldsymbol{\mathsfbi{C}}(\boldsymbol x,\boldsymbol{x'},\tau) = \int_{-\infty}^\infty \boldsymbol{\mathsfbi{S}}(\boldsymbol x,\boldsymbol{x'},f) \mathrm{e}^{2\mathrm{i}\upi f\tau} df, \label{eqnC-S}
\end{equation}
where $f$ is the frequency. Following the proof outlined in \citet{Towne2018} (pp. 859-860), we can relate the eigenvalue problem of $\boldsymbol{\mathsfbi{C}}$ with that of $\boldsymbol{\mathsfbi{S}}$. Assuming the inner product to be a weighted integral in space and time, and substituting for $\boldsymbol{\mathsfbi{C}}$ in equation \ref{eqnC-S} into equation \ref{eqnEVPC}, we get
$$
 \int_{-\infty}^\infty\int_\Omega\int_{-\infty}^\infty \boldsymbol{\mathsfbi{S}}(\boldsymbol x,\boldsymbol{x'},f) \mathrm{e}^{2\mathrm{i}\upi f(t-t')} df \:\boldsymbol{\mathsfbi{W}}(\boldsymbol x') \boldsymbol{\phi}_i(\boldsymbol{x'},t')dx'dt' = \lambda_i\boldsymbol{\phi}_i(\boldsymbol x,t),
$$
where $\boldsymbol{\mathsfbi{W}}$ is the weight matrix associated with the quadrature on the curvilinear grid on the spatial domain, $\Omega$. From the definition of a Fourier transform in time of $\boldsymbol{{\phi}}_i(\boldsymbol{x'},t')$, $\boldsymbol{\hat{\phi}}_i(\boldsymbol{x'},f)$, we get
\begin{equation}
\int_{-\infty}^\infty\int_\Omega \boldsymbol{\mathsfbi{S}}(\boldsymbol x,\boldsymbol{x'},f)   \:\boldsymbol{\mathsfbi{W}}(\boldsymbol x') \boldsymbol{\hat{\phi}}_i(\boldsymbol{x'},f) \mathrm{e}^{2\mathrm{i}\upi ft} dx'df =  \int_{-\infty}^\infty \lambda_i\boldsymbol{\hat\phi}_i(\boldsymbol x,f)\mathrm{e}^{2\mathrm{i}\upi ft}df.    \label{eqSPODPhiHat}
\end{equation}
To solve further, we assume the following:
\begin{equation}
\boldsymbol{\hat\phi}_i(\boldsymbol x,f) = \boldsymbol{\psi}_i(\boldsymbol x,f_0)\delta(f-f_0). \label{eqnSPODModeHat}
\end{equation}
Substituting in equation \ref{eqSPODPhiHat} and eliminating $\exp{(2\mathrm{i}\upi f_0t)}$ on both sides leads to the eigenvalue problem:
\begin{equation}
\int_\Omega \boldsymbol{\mathsfbi{S}}(\boldsymbol x,\boldsymbol{x'},f_0)   \:\boldsymbol{\mathsfbi{W}}(\boldsymbol x') \boldsymbol{{\psi}}_i(\boldsymbol{x'},f_0) dx' =  \lambda_i\boldsymbol{\psi}_i(\boldsymbol x,f_0).
\end{equation}
Here, $\boldsymbol{\psi}_i(\boldsymbol x,f_0)$ and  $\lambda_i$ are the $i-$th eigenfunction and eigenvalue at a given frequency, $f_0$, respectively, with the former referred to as an SPOD mode. 

Flow fields were reconstructed using the SPOD mode at a given $f_0$ by reverting back to the time domain. Using an inverse Fourier transform in time in equation \ref{eqnSPODModeHat}, we have 
\begin{equation}
\boldsymbol{\phi}_i(\boldsymbol{x},t) = \int_{-\infty}^\infty \boldsymbol{\psi}_i(\boldsymbol x,f_0)\delta(f-f_0)\mathrm{e}^{2\mathrm{i}\upi ft} df =  \boldsymbol{\psi}_i(\boldsymbol{x},f_0) \exp{(2\upi\mathrm{i} f_0 t)}. \label{eqnSPODPhiNPsi}
\end{equation}
Additionally, although the expansion coefficients, $a_{i,\zeta}$, vary for different realisations, we are interested in one that best captures the entire ensemble. Based on equation \ref{eqnSPODExpCoeff}, we achieve this by choosing an ideal realisation with an `average' expansion coefficient 
\begin{equation}
a_{i,\zeta_0} = \sqrt{E_\zeta\left[|a_{i,\zeta}|^2\right]} = \sqrt\lambda_i \label{eqnSPODIdealExp}
\end{equation}
Note that SPOD is carried out after subtracting the mean flow field. Thus, the reconstruction based on the required SPOD mode is obtained by summing up the mean with the real part of the truncated basis expansion. From equations \ref{eqnSPODBasisExpans}, \ref{eqnSPODPhiNPsi} and \ref{eqnSPODIdealExp} we get
\begin{equation}
    \boldsymbol{\tilde{q}}(\boldsymbol{x},t) =  \boldsymbol{\overline{q}}(\boldsymbol{x},t) + \mathrm{Re}\left\{\sqrt{\lambda_i}\boldsymbol{\psi}_i(\boldsymbol{x},f_0) \exp{(2\upi\mathrm{i} f_0 t)}\right\}, \label{eqnReconstr}
\end{equation}
where $\boldsymbol{\tilde{q}}$ and $\boldsymbol{\overline{q}}$ represent reconstructed and mean quantities, respectively. This approach is sufficient for the present study, but we note that a more general, low-rank flow reconstruction using SPOD has been implemented for a compressible turbulent jet at a low Mach number in a recently published study \citep{Nekkanti2021}. % 2do: finish reading!

\subsubsection{Implementation}
\label{subsubSecSPODImpl}
Indexing the eigenvalues such that $\lambda_1 > \lambda_2 > \lambda_3 ...$, we have the most-energetic SPOD mode as $\boldsymbol{\psi_1}$ with higher indices representing lower energies. The first two SPOD modes were computed in this study using the memory-efficient streaming algorithm proposed in \citet{SCHMIDT201998}. Negligible energy content was observed for $\boldsymbol{\psi_2}$ for all cases studied and thus only features of the $\boldsymbol{\psi_1}$ are reported. 

The SPOD modes were computed for the flow field variables density, velocity vector, and pressure. Instantaneous `snapshots' based on these variables, extracted from the $z = 0$ plane, were stored at regular time intervals of $\Delta t = 0.08$ (sampling frequency, $F_s = 12.5$). These were then divided into blocks of total time, $T_\zeta$, each block representing a realisation of the stochastic process. Their number is further increased by using a 50\% overlap and this ensemble is used to compute $\boldsymbol{\mathsfbi{S}}$ using the Welch's method. As noted previously, at least ten buffet cycles were simulated for all cases where buffet occurs. To examine the low frequency buffet which occurs at $St \approx 0.1$, it was ensured that at least four cycles occur in a block by choosing $T_\zeta = 44$ (frequency resolution, $\Delta F_\zeta \approx 0.02$). In addition to buffet, we also observed high-frequency coherent structures that we refer to as `wake' modes ($St \sim O(1)$, see figure \ref{fig:RefSPODSpectrum}\textit{a}). To get a better spectral estimate of the energy associated with these modes, the number of blocks for SPOD was increased, which reduces the duration associated with each individual block (and hence the frequency resolution). For these modes, $T_\zeta = 5$ was chosen, which allows for at least five cycles to be captured while increasing the total number of blocks. As the expected value increases with an increase in the number of realisations in an ensemble, the increase in blocks allows for a better spectral estimate. A higher sampling frequency of $F_s = 125$ was also examined, but no new features were observed in the spectrum.

In addition to using the flow field data from the entire $z = 0$ plane, we also performed an SPOD based only on the variation of instantaneous pressure and skin-friction coefficients and the local Mach number on the monitor curve C5 shown in figure \ref{fig:gridAll}\textit{b}. We found that this further reduced the non-coherent noise that is present in the SPOD modes. We note that the coherent features extracted are not significantly different between the two approaches and thus we have used the second approach exclusively for examining the reconstructed flow field only on the aerofoil surface. 

Flow reconstruction based on the desired SPOD mode was carried out based on equation \ref{eqnReconstr}. However, since the reconstruction is carried out only at a specified frequency $f_0$, it is convenient to represent the reconstructed flow based on the phase of the sinusoidal cycle instead of time, $t$. Thus, we use $\phi = 2\upi f_0 t$ to compute  $\boldsymbol{\tilde{q}}(\boldsymbol{x},\phi)$, with $\phi = 0^\circ$ chosen as when the lift coefficient is the maximum. 

%%%%%%%%%%%%%%%%%%%%%%%%%%%%%%%%%%%%%%%%
\section{Description of flow states}
\label{secResults}
%%%%%%%%%%%%%%%%%%%%%%%%%%%%%%%%%%%%%%%%

We first present a basic description of buffet features for the reference case, following which the results for variation of $M$, $\alpha$, $\Rey$ and $\Lambda$ are reported. For all cases where buffet occurs, the BL remained laminar up to the vicinity of the shock foot, implying laminar buffet, as categorised in \citet{Dandois2018}. Modal decomposition and flow field reconstruction using modes of interest are considered separately in the subsequent section ( \S\ref{secSPOD}).

\subsection{Reference case}
\label{secRef}

\begin{figure} 
\centerline{
\includegraphics[width=\textwidth]{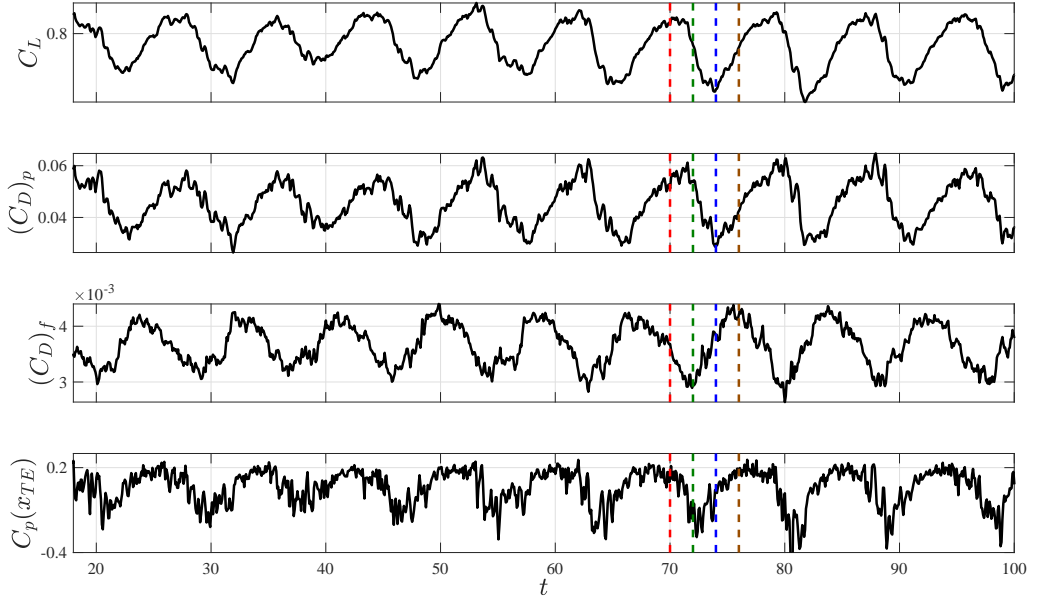}}
\caption{Temporal variation of aerofoil coefficients past initial transients (reference case). Dashed lines correspond to high lift phase (red), low lift (blue), low friction drag (green) and high friction drag (brown).}
    \label{fig:RefClCDCp}
\end{figure}

\begin{figure} 
\centerline{\includegraphics[trim={0.4cm 1.5cm 0cm 2cm},clip,width=.495\textwidth]{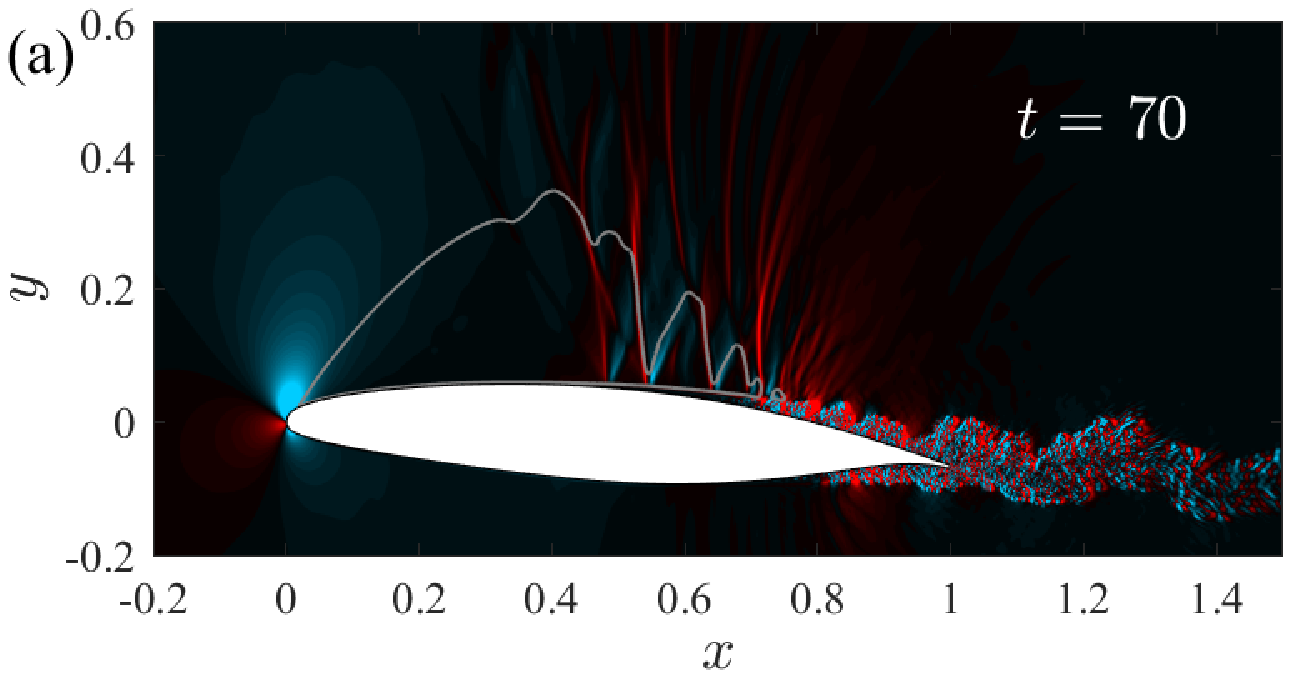}
\includegraphics[trim={0.4cm 1.5cm 0cm 2cm},clip,width=.495\textwidth]{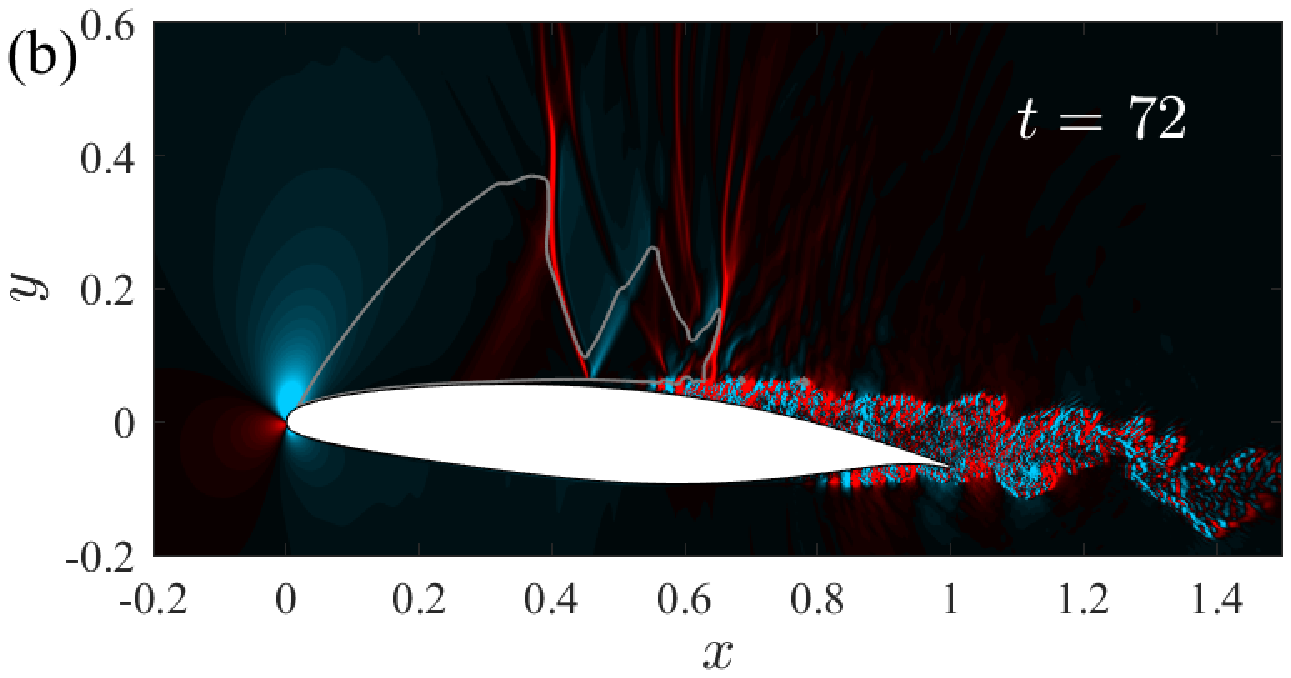}}
\centerline{\includegraphics[trim={0.4cm 1.5cm 0cm 2cm},clip,width=.495\textwidth]{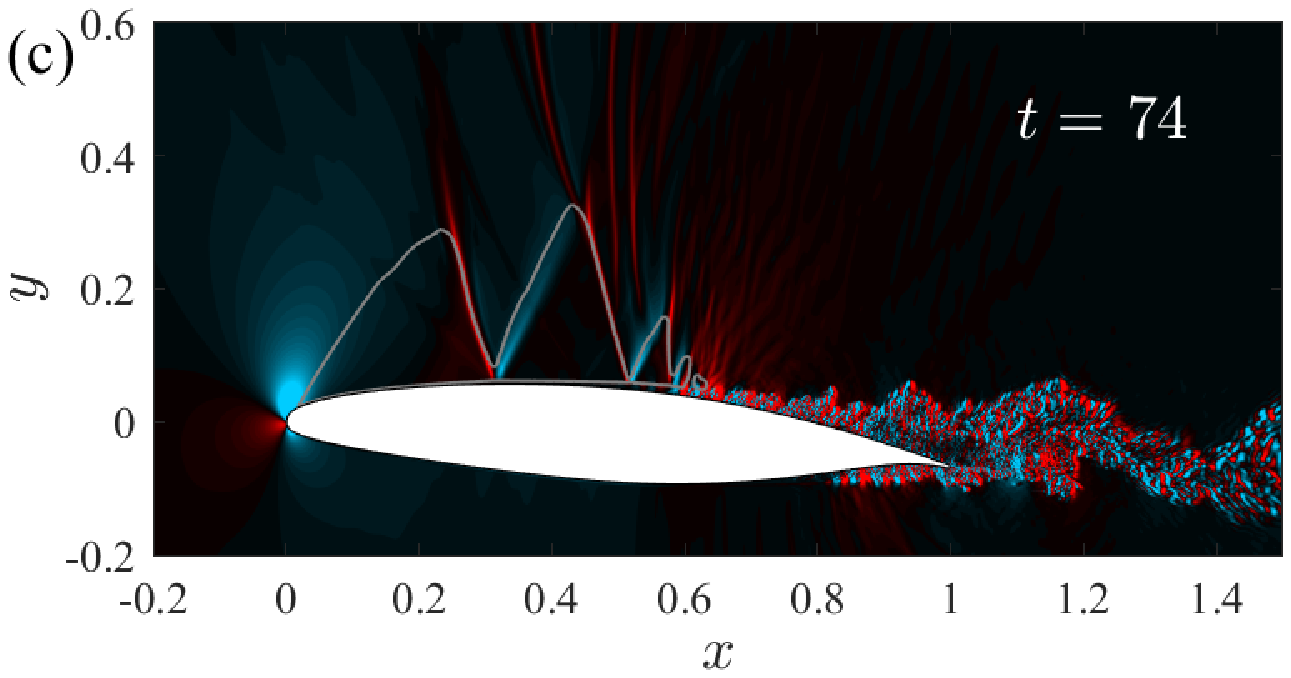}
\includegraphics[trim={0.4cm 1.5cm 0cm 2cm},clip,width=.495\textwidth]{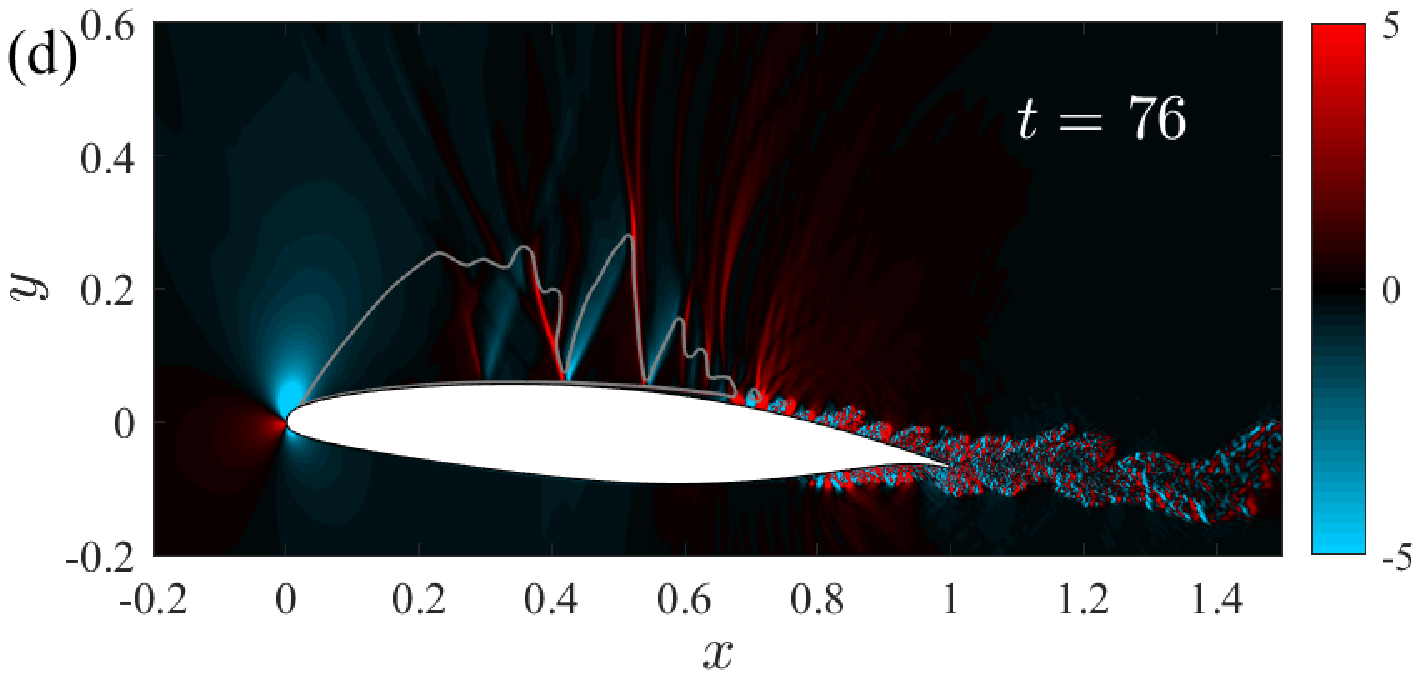}}
\caption{Streamwise density gradient contours on the $x-y$ plane shown at different phases of the buffet cycle for the reference case: (a) high-lift (b) low-skin-friction-drag (c) low-lift and (d) high-skin-friction-drag phases. The sonic line based on the local Mach number is highlighted using a gray curve.}
    \label{fig:RefDensGrad}
\end{figure}

Some aspects of the results for the reference case ($M = 0.7$, $\alpha = 4^\circ$, $\Rey = 5\times 10^5$ and $\Lambda = 0^\circ$)  have already been presented in \citet{Zauner2020}, but it is useful to review these here to aid later comparisons. The temporal variation of various aerofoil coefficients once buffet has been established are shown in figure \ref{fig:RefClCDCp}. Here, $C_L$ and $C_D$ represent the instantaneous span-averaged lift and drag coefficients, with the pressure and skin-friction components of the drag denoted as $(C_D)_p$ and $(C_D)_f$, respectively. Here, the contribution from the blunt trailing edge is neglected when computing the aerodynamic coefficients. The pressure coefficient at the trailing edge, $C_p(x_{TE})$, is also shown. Low-frequency periodic oscillations (period $\tau_\mathrm{LES} = 1/St \approx 8.6$ time units) are apparent for all coefficients shown, with the strong fluctuation of the lift coefficient (fluctuation amplitude $ > 10\%$ of mean $C_L$) clearly indicating the occurrence of buffet. Minor irregularities between cycles can also be seen, which has been reported for supercritical aerofoils \citep{Roos1980} and at moderate $\Rey$ \citep{McDevitt1985}. There are discernible phase differences between all the variables plotted and especially between $C_L$ and $(C_D)_f$. This was observed to be present for all cases simulated here. The time instants when these two coefficients reach their maximum and minimum in a single buffet cycle are highlighted using dashed lines. Contours of streamwise density gradient, $d\rho/dx$, on the $z = 0$ plane at these times are shown in figure \ref{fig:RefDensGrad}. The contour range has been reduced to $[-5,5]$ for clarity while a grey curve is used to delineate the sonic line based on the instantaneous local Mach number, \textit{i.e.} $M_{\mathrm{loc}}=  1$. At all times, the presence of multiple shock waves is evident. This appears to be a characteristic flow feature at moderate $\Rey$, as will be elaborated in  \S\ref{secRe}. For convenience, we will use the term `sonic envelope' to refer to the time-dependent supersonic region, focusing on its downstream edge, beyond which the flow remains subsonic, and the term `shock wave structures' to refer to the multiple shock waves that are present in the flow. We observed that the low-/high-lift phases of the cycle approximately occur when the sonic envelope is at its most upstream/downstream position respectively, while the minimum/maximum $(C_D)_f$ occurs when the sonic envelope is approximately at the mid-point of its upstream/downstream motion, respectively. %We note here that the flow is `most separated' when $(C_D)_f(t)$ attains a local minimum ($t = 72$) and vice-versa ($t = 76$). Evidence for this observation will be provided when examining the reconstructed flow-fields in  \S\ref{secSPOD}, but intuitively, this can be inferred by noting that the velocity gradient at the wall is reduced when the flow is separated.

At $t = 70$ (figure \ref{fig:RefDensGrad}\textit{a}), when the lift is at a maximum and the sonic envelope is at its most downstream position, the shock wave structures appear clustered together at $x \approx 0.6$, while a transitional separation bubble is discernible for $0.4 \leq x \leq 0.7$. Subsequently, as the sonic envelope moves upstream, the number of shock wave structures reduces and they appear strengthened, while the BL separates and transitions at a more upstream position at $t = 72$ (see $x \approx 0.5$ in figure \ref{fig:RefDensGrad}\textit{b}). When the sonic envelope reaches its most upstream position at $t = 74$ (lowest lift, figure \ref{fig:RefDensGrad}\textit{c}), the inclined shock wave structures seem well-separated and the supersonic regions have a triangular shape. The BL reattaches at $x \approx 0.6$, although it remains separated at the trailing edge. At $t = 76$ (figure \ref{fig:RefDensGrad}\textit{d}), at the mid-point of the sonic envelope's motion downstream and when $(C_D)_f(t)$ reaches its maximum, the BL remains mostly attached, with a transitional separation bubble present at $0.5 \leq x \leq 0.7$. A larger supersonic region is present at $t = 70$ and $t = 72$ when compared to $t = 74$ and $t = 76$. 

\begin{figure} 
\centerline{
\includegraphics[width=0.495\textwidth]{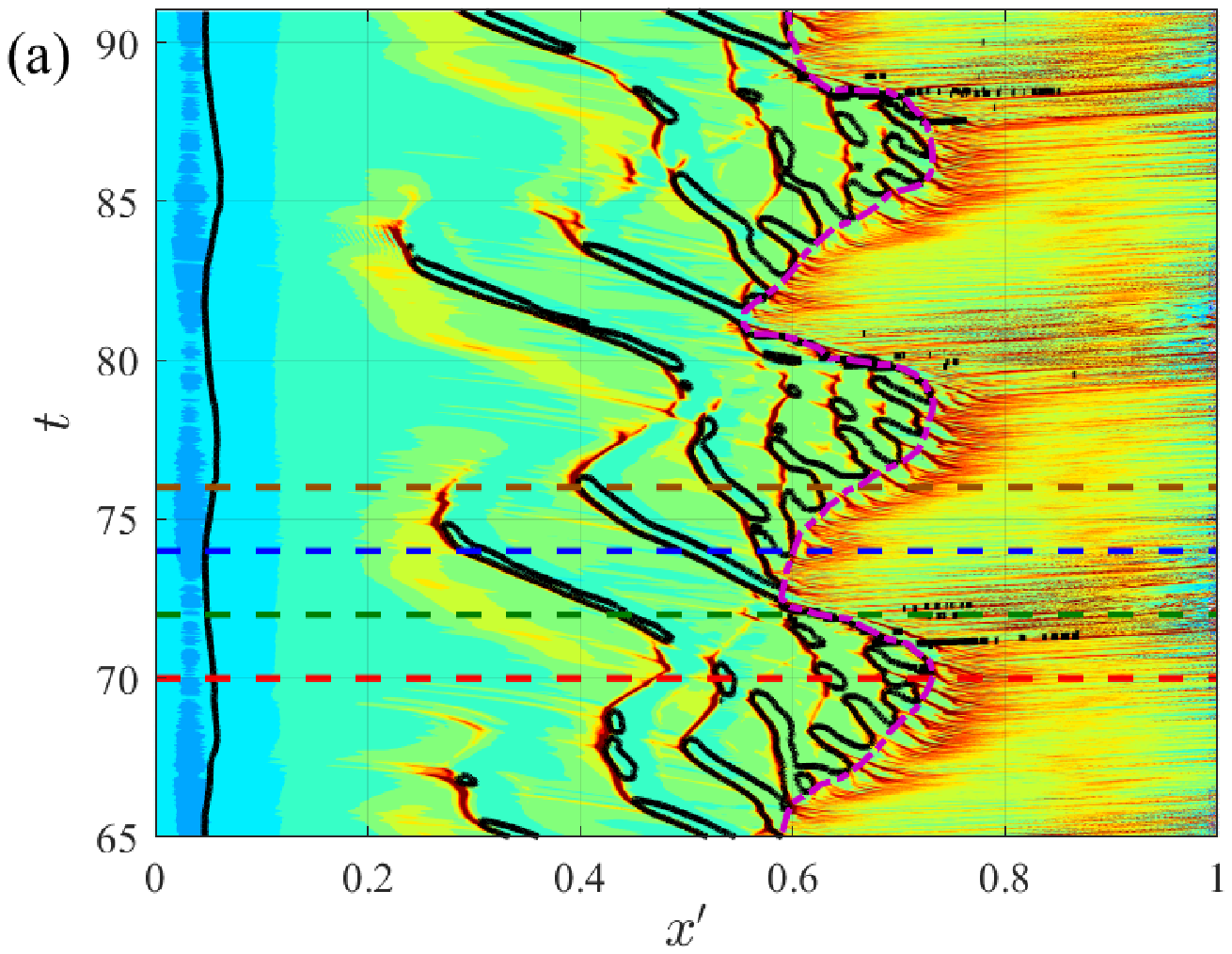}
\includegraphics[width=0.495\textwidth]{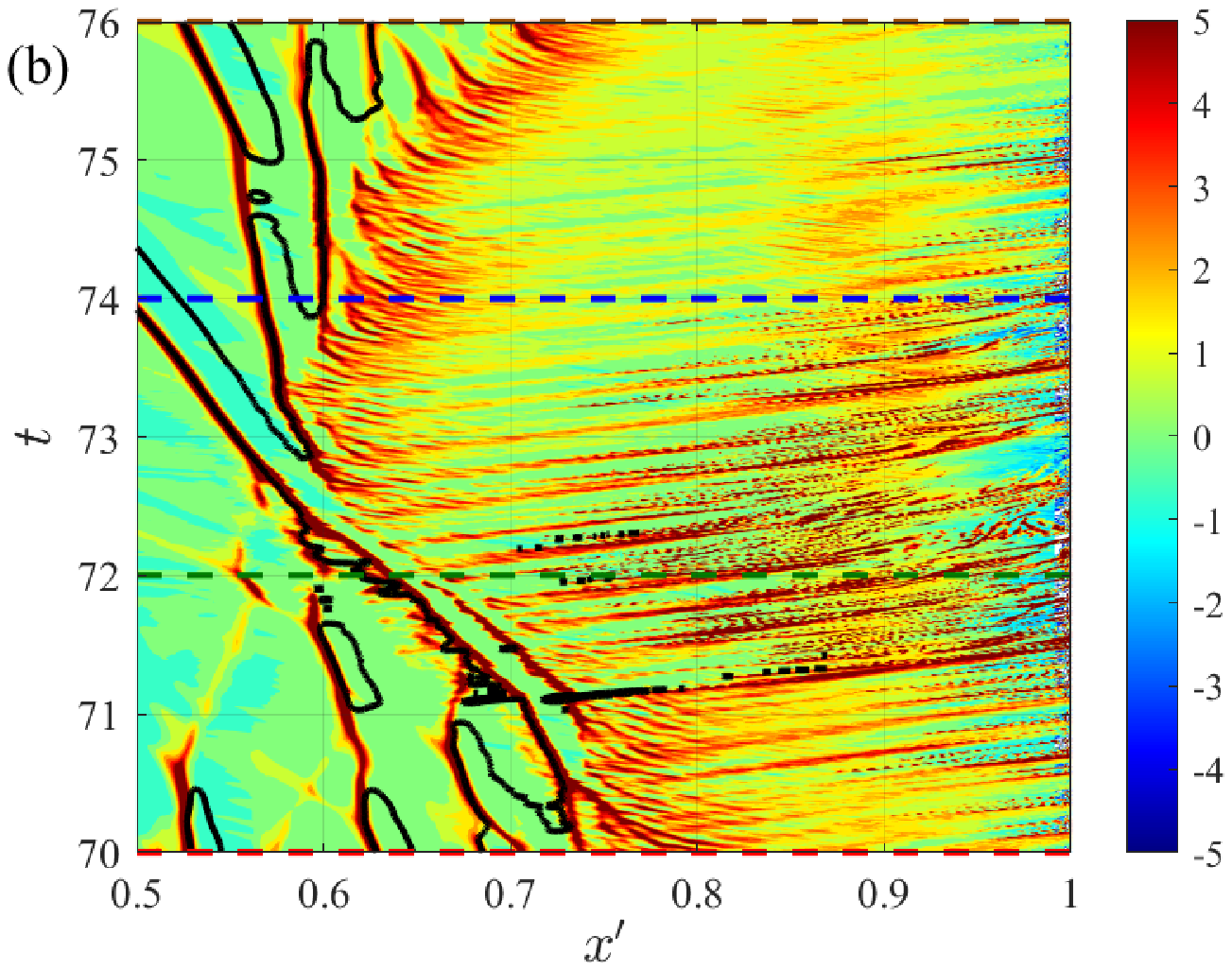}}
\caption{Spatio-temporal variation of streamwise pressure gradient on the suction side of curve C5 for the reference case: (a) entire suction side and (b) close-up. The approximate times associated with different phases of interest -- high lift (red), low skin-friction drag (green), low lift (blue) and high skin-friction drag (brown) -- are highlighted using dashed lines, while the sonic line based on local Mach number is shown using a solid black curve.
}
    \label{fig:RefXTDia}
\end{figure}

The shock wave motion can be further understood by examining the flow features on the monitor curve C5 highlighted in figure \ref{fig:gridAll}\textit{b}. This curve is at a wall-normal distance of 0.05 from the aerofoil's surface and is above the BL in the supersonic regions of the flow (cf. figure \ref{fig:RefDensGrad}). The spatio-temporal variation of the streamwise pressure gradient, $dp/dx$, on the suction side of this curve is shown in figure \ref{fig:RefXTDia} (figure \ref{fig:RefXTDia}\textit{b} shows a close-up view of the sonic envelope for a single buffet cycle). The sonic line based on the local Mach number is overlaid as a black curve, while the pink dashed curve in figure \ref{fig:RefXTDia}\textit{a} approximates the sonic envelope. Here, the horizontal axis represents the variation in the chordwise direction, while the vertical axis represents time. Thus, features with a positive slope represent downstream moving structures and vice versa. The time instants associated with the extrema discussed above are highlighted using dashed horizontal lines. We emphasise that what is shown are flow features on a single curve at a specified wall-normal distance, so the changes, especially of the sonic line, must be interpreted cautiously. 

The sonic envelope (pink dashed curve) is approximately at its most downstream and upstream position at $t = 70$ (high-lift) and 74 (low-lift), respectively. Emanating from the sonic envelope, we see streaks in the sonic line (black curve) oriented with a negative slope, \textit{i.e.}, propagating upstream. These streaks are due to the shock wave structures and are seen to be only generated during the downstream motion of the envelope. These shock wave structures weaken as they travel upstream and eventually seem to degenerate into downstream propagating pressure waves. In the subsonic region downstream of the sonic envelope, linear streaks in the pressure gradient contour at a positive slope can be seen (figure \ref{fig:RefXTDia}\textit{b}). The phase speed of the waves associated with these streaks can be computed from the figure as the inverse of the slope. This falls in the range $0.65-0.7U_\infty$. Lower convection speeds of $\approx 0.55-0.65 U_\infty$ were observed for other cases studied here. This is similar to the convection speeds, $0.7U_\infty$ and  $0.41U_\infty$, of Kelvin-Helmholtz (K-H) instabilities that are reported to accompany buffet in \citet{Dandois2016} and \citet{Dandois2018}, respectively. 

The model proposed in \cite{Lee1990} assumes that pressure waves are generated at the shock foot and convect downstream within the BL until they reach the TE (see \S\ref{secIntro}). Here, the streaks seen in  figure \ref{fig:RefXTDia}\textit{b} represent the dominant pressure waves within the BL that reach the TE starting from the sonic envelope. The approximate time required for these waves to do so can be inferred from the figure to vary in the range $0.4 \leq t_{\mathrm{down}} \leq 0.6$ (based on sonic envelope's position). Additionally, as shown later in \S\ref{subSecSPODModeFeatures}, upstream propagating waves outside the BL can be identified using SPOD. The buffet frequency predicted by Lee's model based on these upstream and downstream propagating waves is considered there.

%%%%%%%%%%%%%%%%%%%%%%%%%%%%%%%%%%%%%%%%
\subsection{Effect of Mach number}
\label{secMach}
%%%%%%%%%%%%%%%%%%%%%%%%%%%%%%%%%%%%%%%%

When the freestream Mach number alone is varied (with $\alpha = 4^\circ$, $\Rey = 5\times10^5$ and $\Lambda = 0^\circ$), the lowest value at which a small pocket of supersonic flow develops in the flow field (approximate critical Mach number) was found to be $M = 0.6$. Increasing $M$ further, buffet was observed in the range $0.7 \leq M \leq 0.85$, above which it was absent. However, for $M \geq 0.8$ shock wave structures were observed on both the suction and pressure sides of the aerofoil. Buffet features in the former range are reported first, while the cases of higher $M$ are presented subsequently.

%%%%%%%%%%%%%%%%%%%%%%%%%%%%%%%%%%%%%%%%
\subsubsection{Buffet onset and deep buffet}
\label{subsecLowMach}
%%%%%%%%%%%%%%%%%%%%%%%%%%%%%%%%%%%%%%%%
\begin{figure} 
\centerline{
\includegraphics[width=0.495\textwidth]{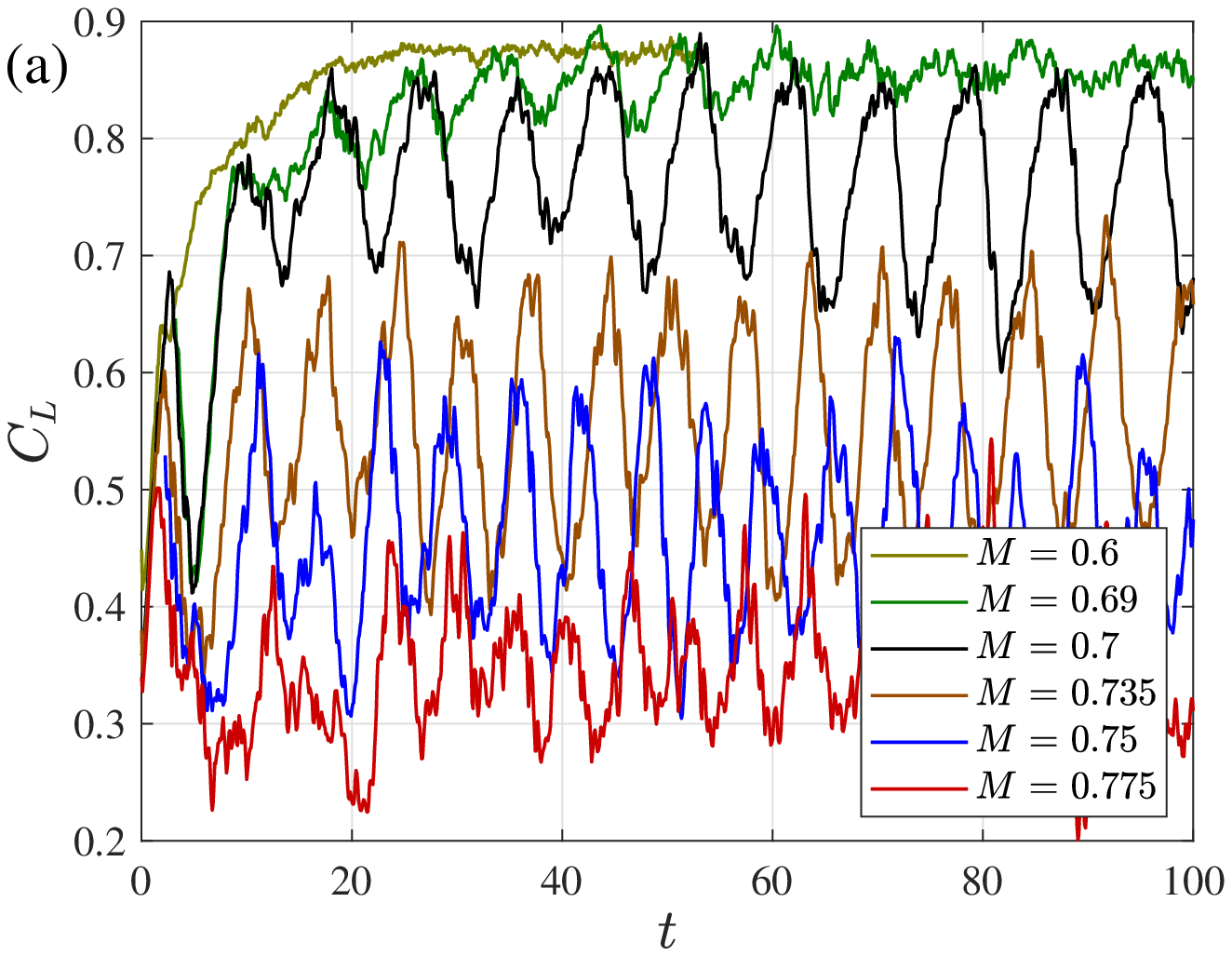}
\includegraphics[width=0.495\textwidth]{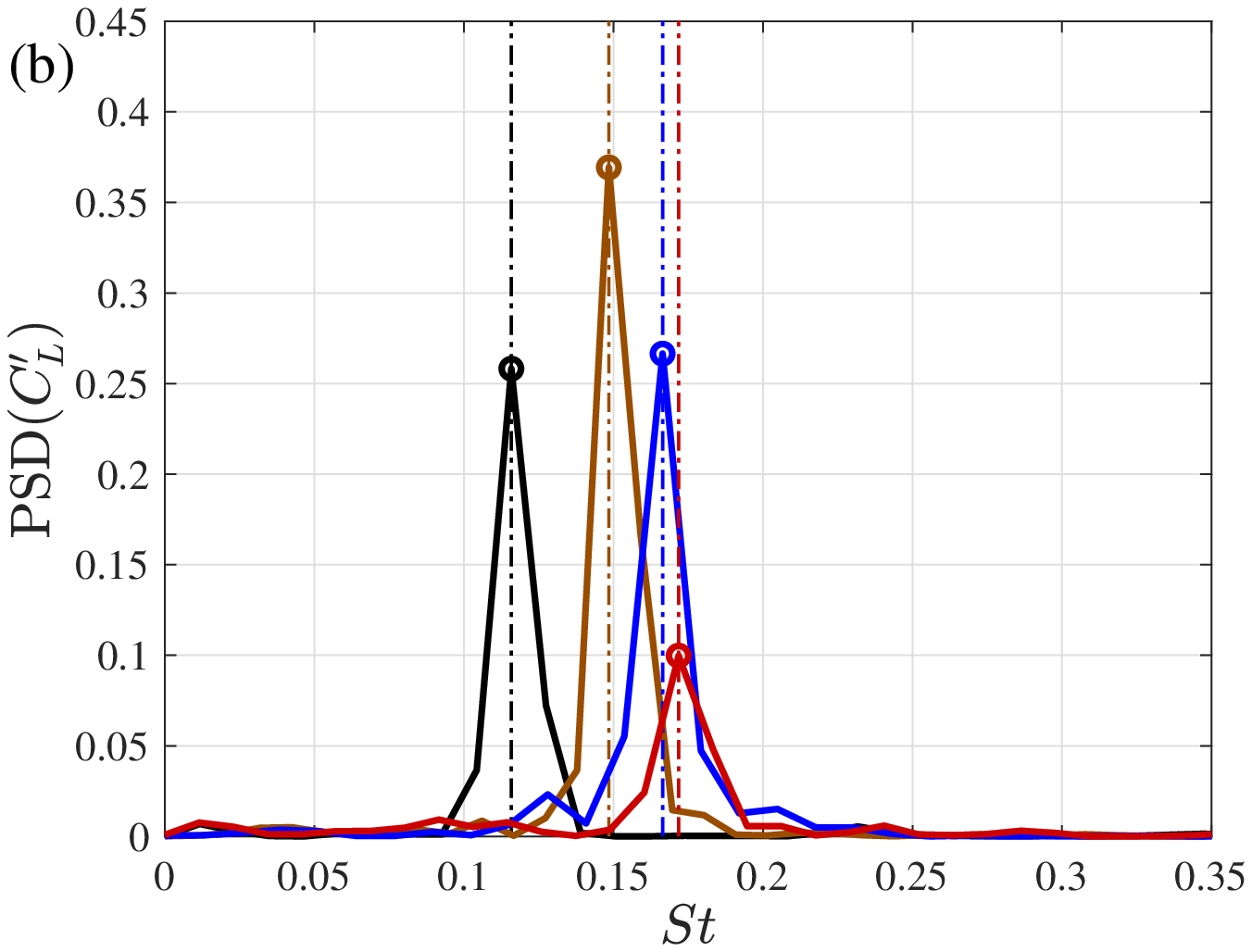}}
\caption{Effect of varying freestream Mach number: (a) temporal variation of lift coefficient and (b) power spectral density of its fluctuating component as a function of the Strouhal number. The reference case of $M = 0.7$ is shown using a black curve.}
\label{fig:ClMach}
\end{figure}

The temporal evolution of $C_L$ for a few select cases of $M$ in the range $0.5 \leq M\leq 0.775$ is shown in figure \ref{fig:ClMach}\textit{a}. For all cases simulated in the range $0.5 \leq M \leq 0.68$ (see Table \ref{tableCases}), the variation of $C_L$ is similar to that shown for $M =0.6$ in the figure, with the equilibrium values attained past transients being approximately the same. This occurs even though the flow is entirely subsonic for $M = 0.5$ and becomes transonic at and above the approximate critical Mach number of $M = 0.6$. Interestingly, at $M = 0.69$, sinusoidal oscillations can be discerned during the initial evolution ($t < 50$), but they slowly dampen with time, with the $C_L$ stabilising at a value close to that of $M = 0.6$. By contrast, a small increase in $M$ to 0.7 (reference case) leads to relatively large-amplitude sustained oscillations implying an abrupt onset of buffet with $M$, which is similar to the results reported for turbulent buffet in \cite{Giannelis2018}. This trend suggests that a marginally stable eigenmode becomes unstable as the parameter, $M$, is increased from $M=0.69$ to the onset value of $M = 0.7$. This is characteristic of a supercritical Hopf bifurcation  and confirms to the global linear instability model proposed for buffet \citep{Crouch2007}. Interestingly, the maximum of $C_L(t)$ at $M= 0.7$ coincides with that of the approximately steady value attained for pre-onset conditions examined. Also, a sharp drop in mean lift coefficient past onset is apparent, with $\overline{C}_L$ (the time average of $C_L$) dropping almost linearly with $M$ in a small range of $0.7 \leq M \leq 0.775$ from approximately 0.85 to $0.3$. 

The power spectral density estimate (PSD) of the fluctuating component of the lift coefficient, $C_L'$, (computed as a periodogram of the signal using a Hamming window function) is shown in figure \ref{fig:ClMach}\textit{b} for cases where buffet is observed. The dominant peaks and their corresponding $St$ are highlighted using circles and broken vertical lines, respectively, with the latter indicating the buffet frequencies. It is evident from the figure that the buffet frequency increases monotonically in the range shown. Such a monotonic increase in frequency with $M$ for turbulent buffet has been previously reported for aerofoils \citep{Dor1989, Jacquin2009, Giannelis2018, Brion2020} and swept wings \citep{Dandois2016}. 

\begin{figure} 
\centerline{
\includegraphics[trim={0.3cm 1.5cm 0cm 2cm},clip,width=.495\textwidth]{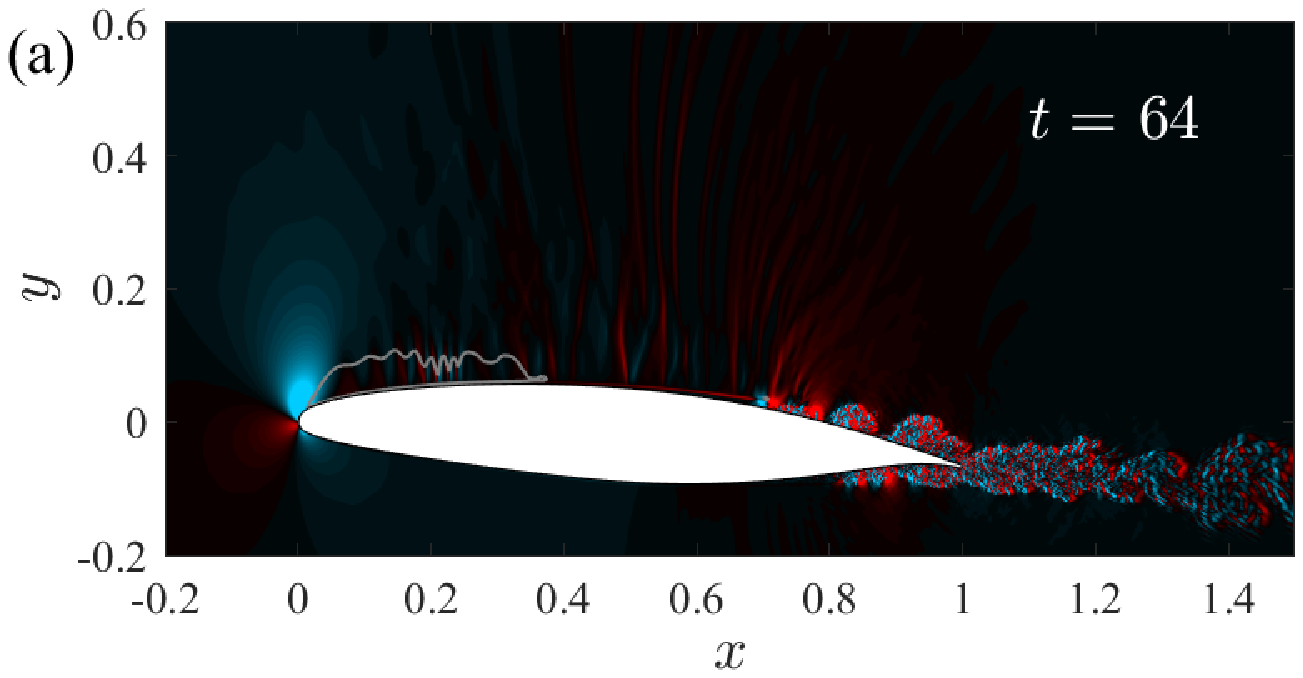}
\includegraphics[trim={0.3cm 1.5cm 0cm 2cm},clip,width=.495\textwidth]{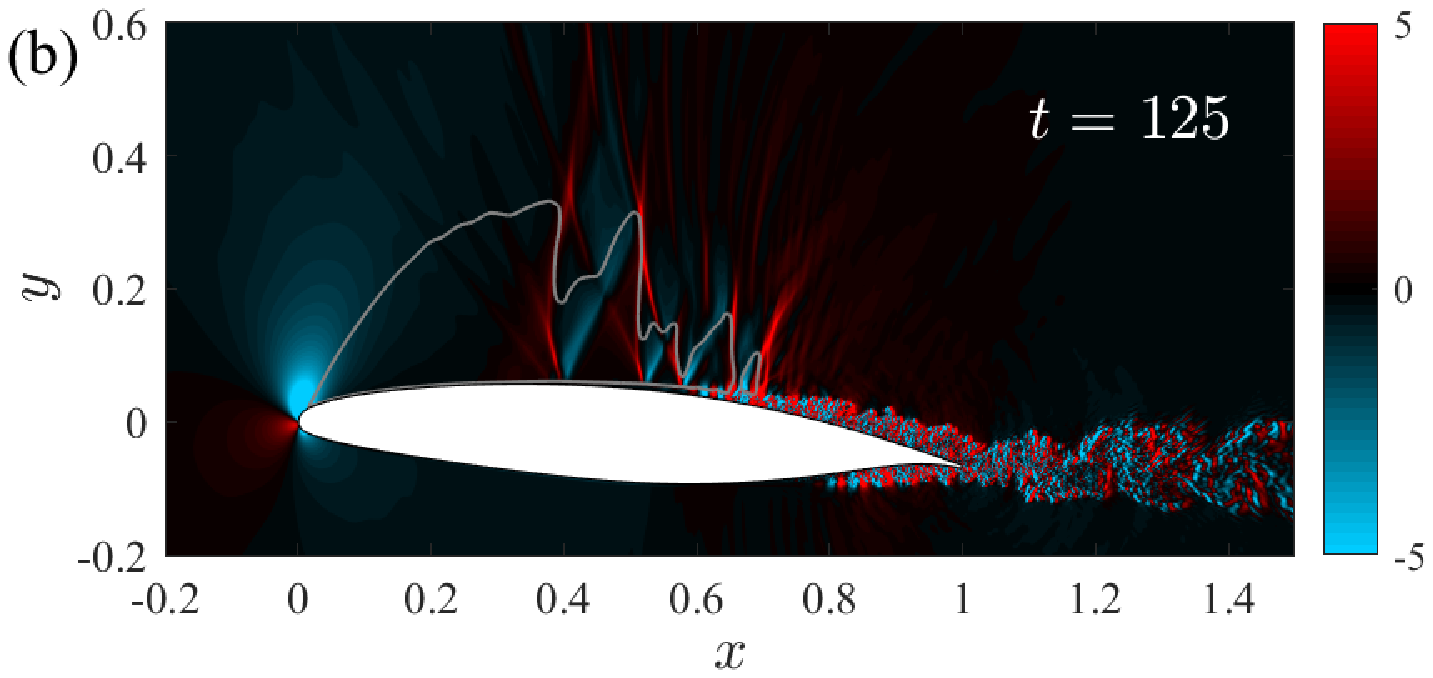}}
\caption{Streamwise density gradient contours on the $x-y$ plane shown for long-time flow states for freestream Mach numbers below buffet onset: (a) $M = 0.65$ and (b) $M = 0.69$.}
    \label{fig:M65N69DensGrad}
\end{figure}

The flow features of pre-onset cases of $M$ are shown in figure \ref{fig:M65N69DensGrad} using streamwise density gradient contours. At $M = 0.65$, a small elongated pocket of supersonic region is present. While there is no shock wave terminating the supersonic region, multiple weak pressure waves can be discerned in the flow at $x \approx 0.5$. Similar ``wavelets" were reported in \citet{Hilton1947} for low $M$ at which no shock waves were present. In contrast to $M=0.65$, a shock system is observed at $M = 0.69$, with the pressure waves now strengthening sufficiently to form oblique shock waves. Note that the time instant shown for $M = 0.69$ is well past the initial transient evolution and after the buffet mode is fully damped (cf. figure \ref{fig:ClMach}\textit{a}). Interestingly, these long-time equilibrium flow features of $M = 0.69$ resemble those seen for the reference case ($M = 0.7$) in the high-lift phase (see figure \ref{fig:RefDensGrad}\textit{a}), which is also indicated by the evolution of $C_L$ for these cases. 

\begin{figure} 
\centerline{
\includegraphics[trim={0.3cm 1.5cm 0cm 2cm},clip,width=0.495\textwidth]{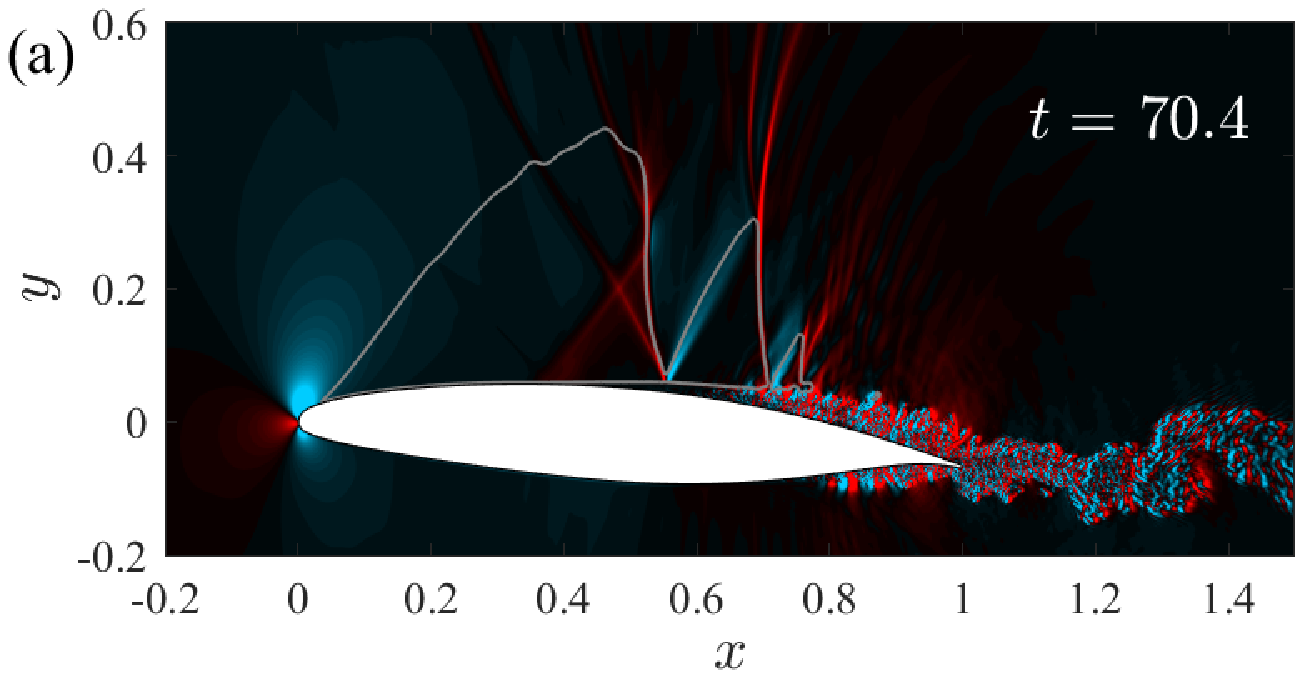}
\includegraphics[trim={0.3cm 1.5cm 0cm 2cm},clip,width=0.495\textwidth]{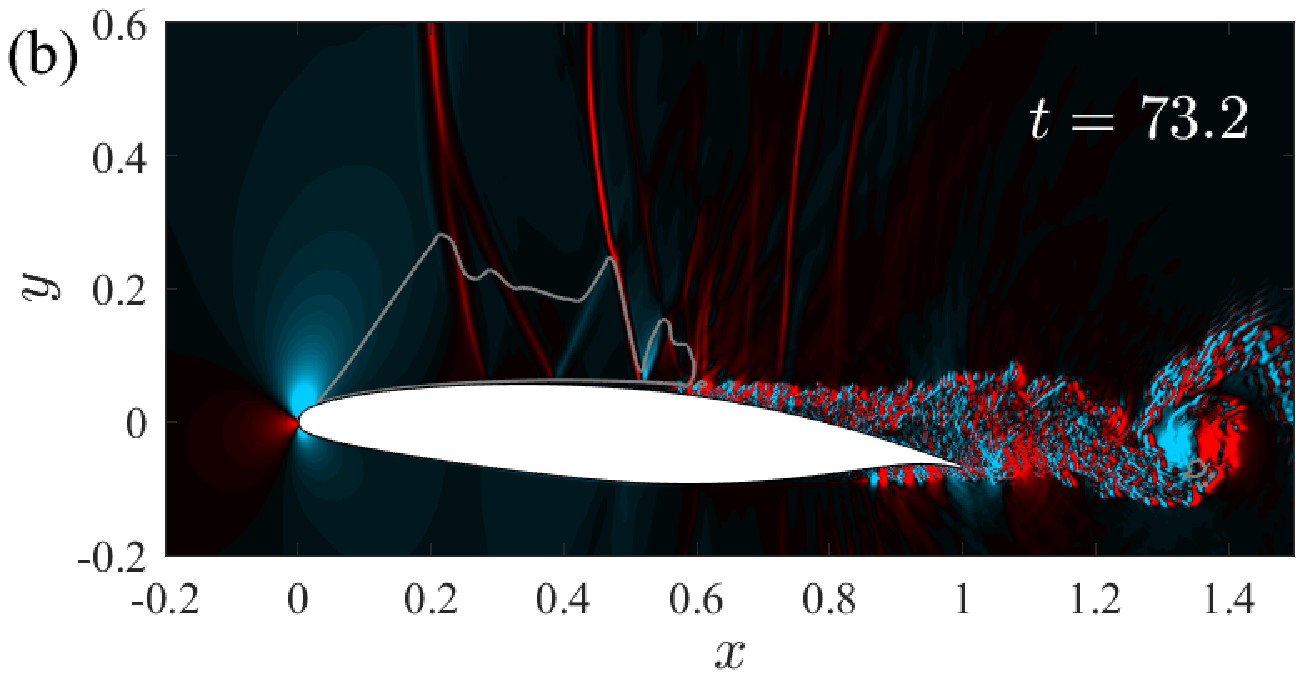}}
\centerline{
\includegraphics[trim={0.3cm 1.5cm 0cm 2cm},clip,width=0.495\textwidth]{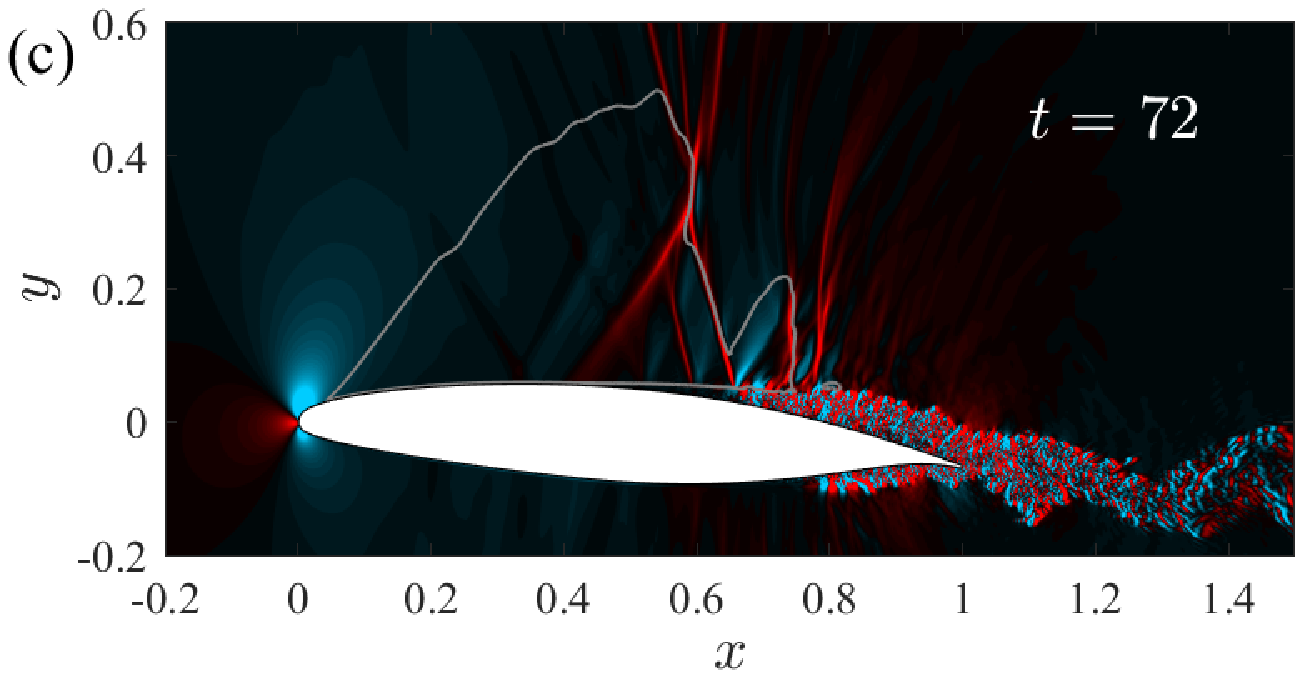}
\includegraphics[trim={0.3cm 1.5cm 0cm 2cm},clip,width=0.495\textwidth]{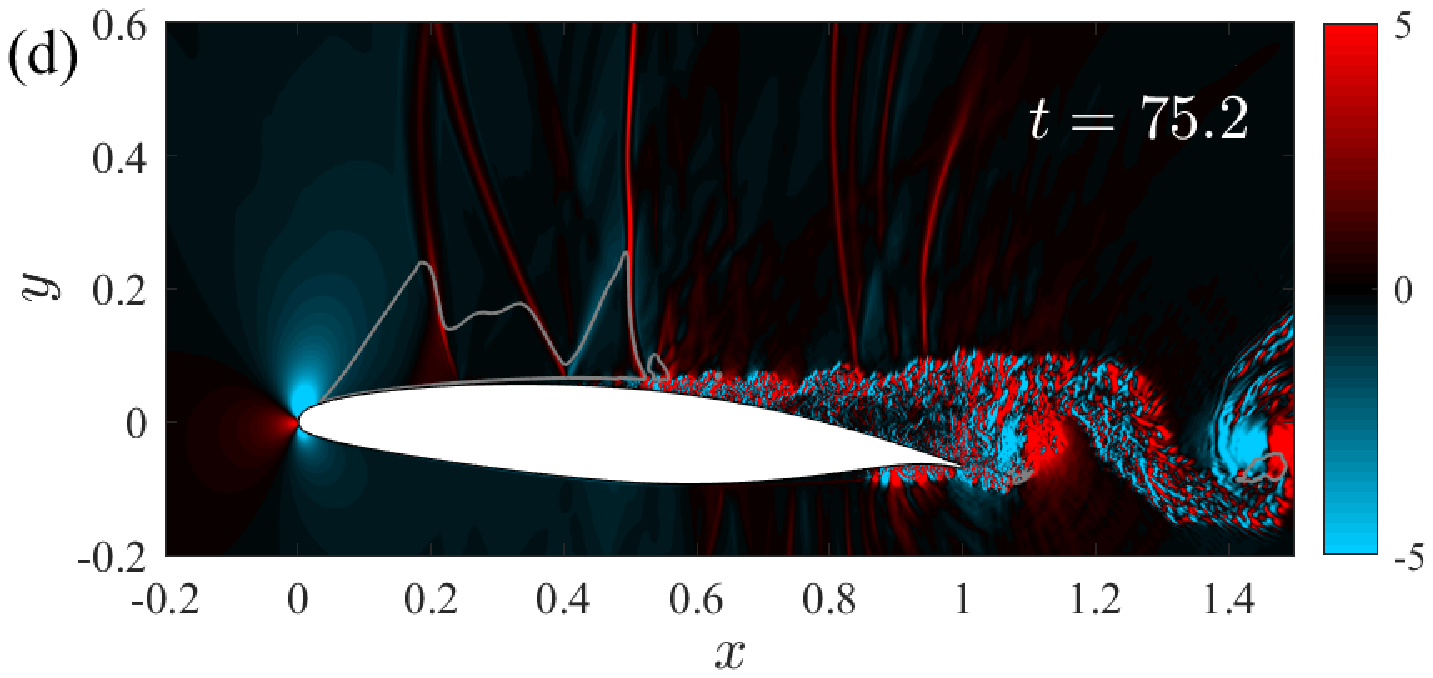}}
\caption{Streamwise density gradient contours on the $x-y$ plane shown for the high (left) and low (right) lift phases of the buffet cycle for the cases $M = 0.735$ (top) and 0.75 (bottom).}
    \label{fig:M735DensGrad}
\end{figure}

Flow features above the onset $M$ are shown for $M = 0.735$ (top) and $M = 0.75$ (bottom) in figure \ref{fig:M735DensGrad} at high (left) and low lift (right) phases of the buffet cycle. Note that the maximum fluctuation energy of $C_L$ is achieved for $M = 0.735$ (cf. figure \ref{fig:ClMach}\textit{a}). For both cases, the flow features far from the aerofoil surface change drastically in a given cycle indicating that buffet can affect the entire supersonic envelope. Strong vortices that shed into the wake can be seen in the latter phase. Indeed, for all cases where buffet is observed in this study, the vortices were found to be strongest when the sonic envelope is at its most upstream position (shown later in figures \ref{fig:M8DensGrad}, \ref{fig:AoA5DensGrad} and \ref{fig:ReDensGrad}). This is also supported by the results based on a time-frequency analysis reported later in \S\ref{subsubsecSPODTemporal}. This contradicts the model proposed in \citet{Hartmann2013}, which requires that the vortices that reach the TE are most intense when the shock wave is at its most downstream position and  vice versa (see their figure 15, p. 14). 

\begin{figure} 
\centerline{
\includegraphics[width=.495\textwidth]{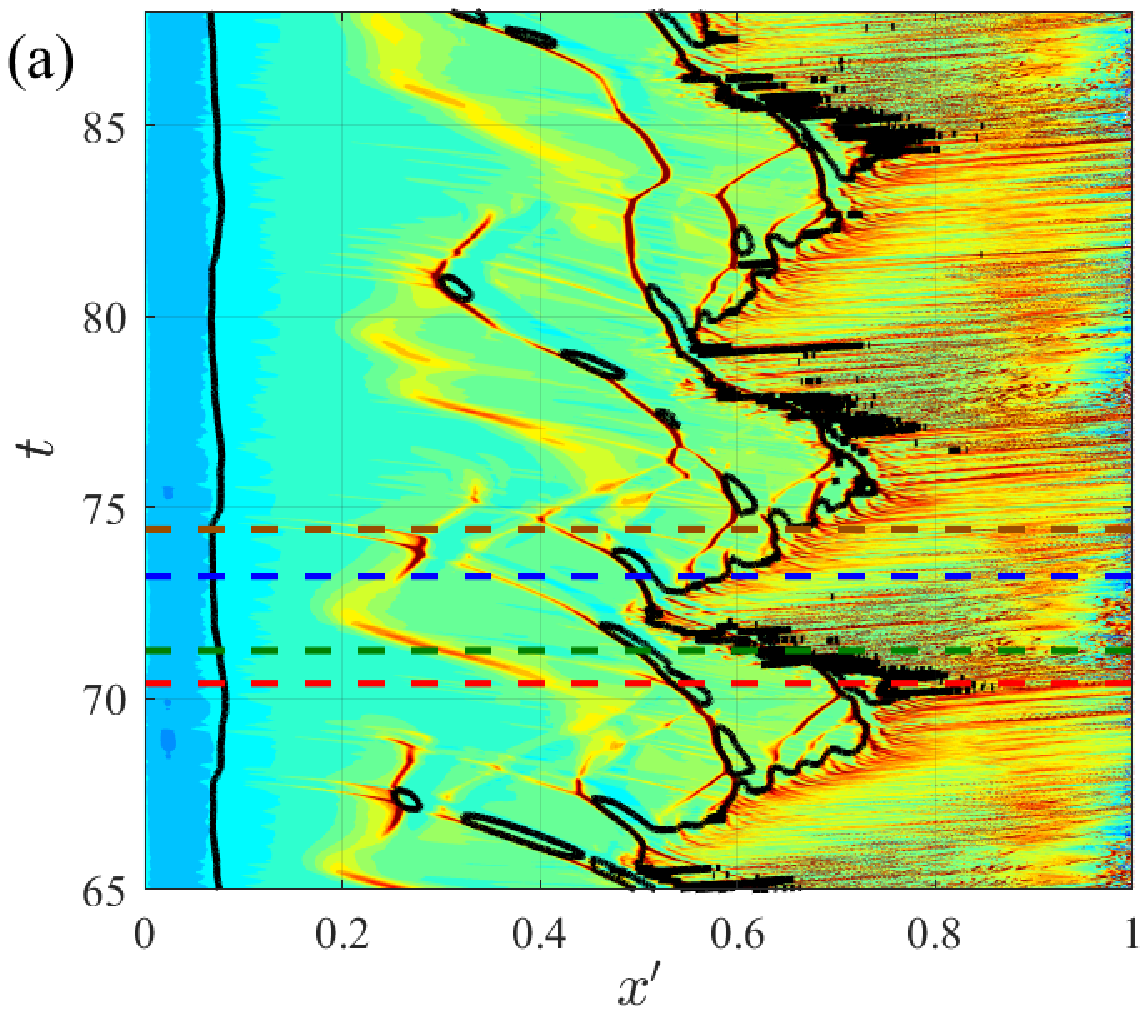}
\includegraphics[width=.495\textwidth]{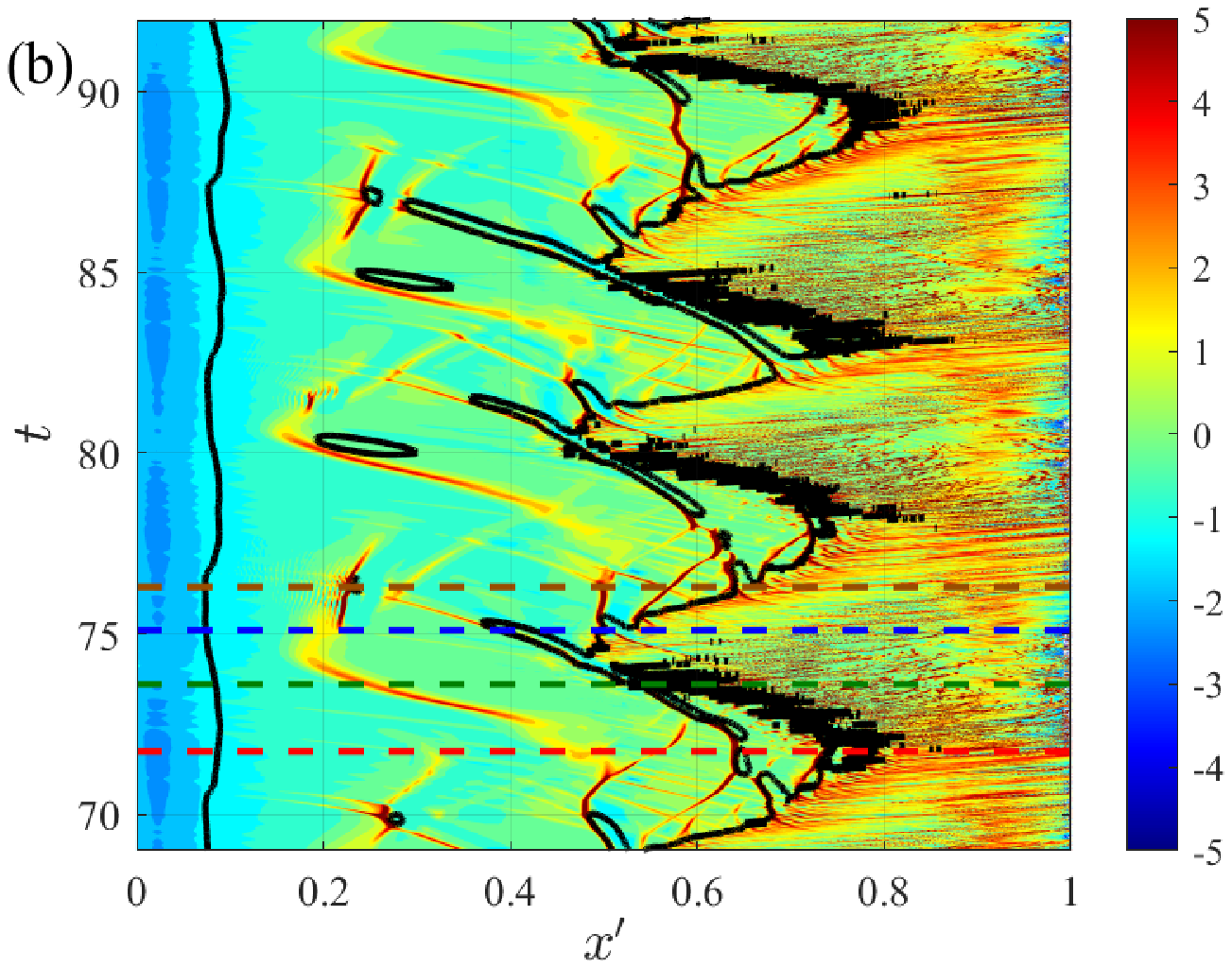}}
\caption{Spatio-temporal variation of streamwise pressure gradient on the suction side of C5 for (a) $M = 0.735$ and (b) $M = 0.75$.}
    \label{fig:M0p735XTDia}
\end{figure}

The spatio-temporal variation of the pressure gradient along the C5 monitor curve is shown in figure \ref{fig:M0p735XTDia}. Multiple transitions between locally supersonic and subsonic regions that were observed for $M =0.7$ as the sonic envelope moves downstream  (figure \ref{fig:RefXTDia}) are reduced here with pressure waves present in place of shock wave structures. The approximate times associated with the local extrema of $C_L$ and $(C_D)_f$ are highlighted in the figure using dashed vertical lines (see figure \ref{fig:RefXTDia} for colour scheme). As expected, the maximum (minimum) $C_L$ occurs when the sonic envelope is at its most downstream (upstream) position. Interestingly, similar to the reference case, the minimum (maximum) $(C_D)_f$ occurs when the sonic envelope is approximately in the mid position of its upstream (downstream) motion. Downstream propagating structures in the BL, similar to those seen for the reference case, can also be observed. The convection velocities of these were observed to be slightly lower than in the reference case, with the approximate range of their phase speed being $0.55-0.65U_\infty$, while the time required for them to reach the TE is in the range $0.5 \leq t_\mathrm{down} \leq 1.5$.

\begin{figure} 
\centerline{
\includegraphics[width=0.49\textwidth]{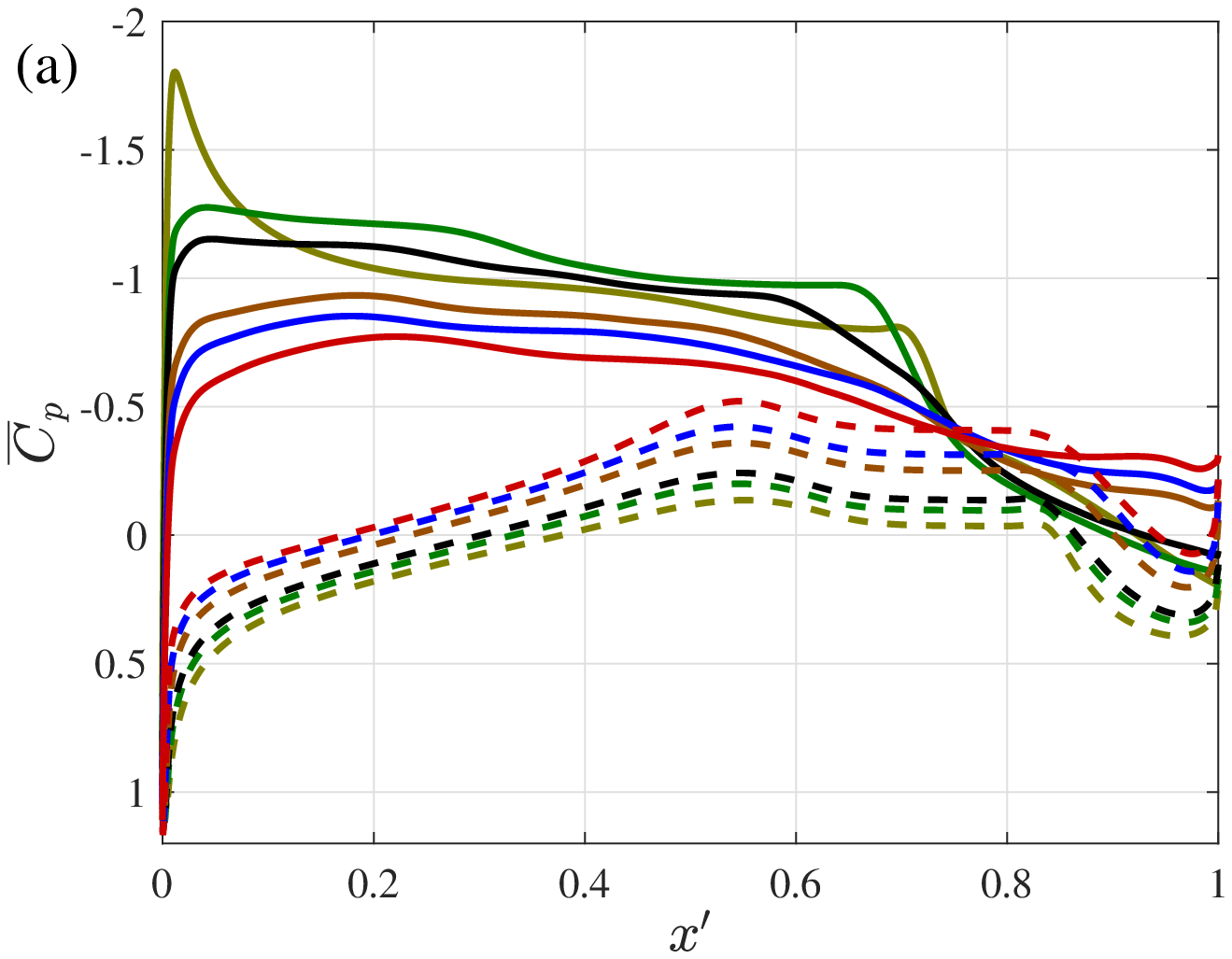}
\includegraphics[width=0.49\textwidth]{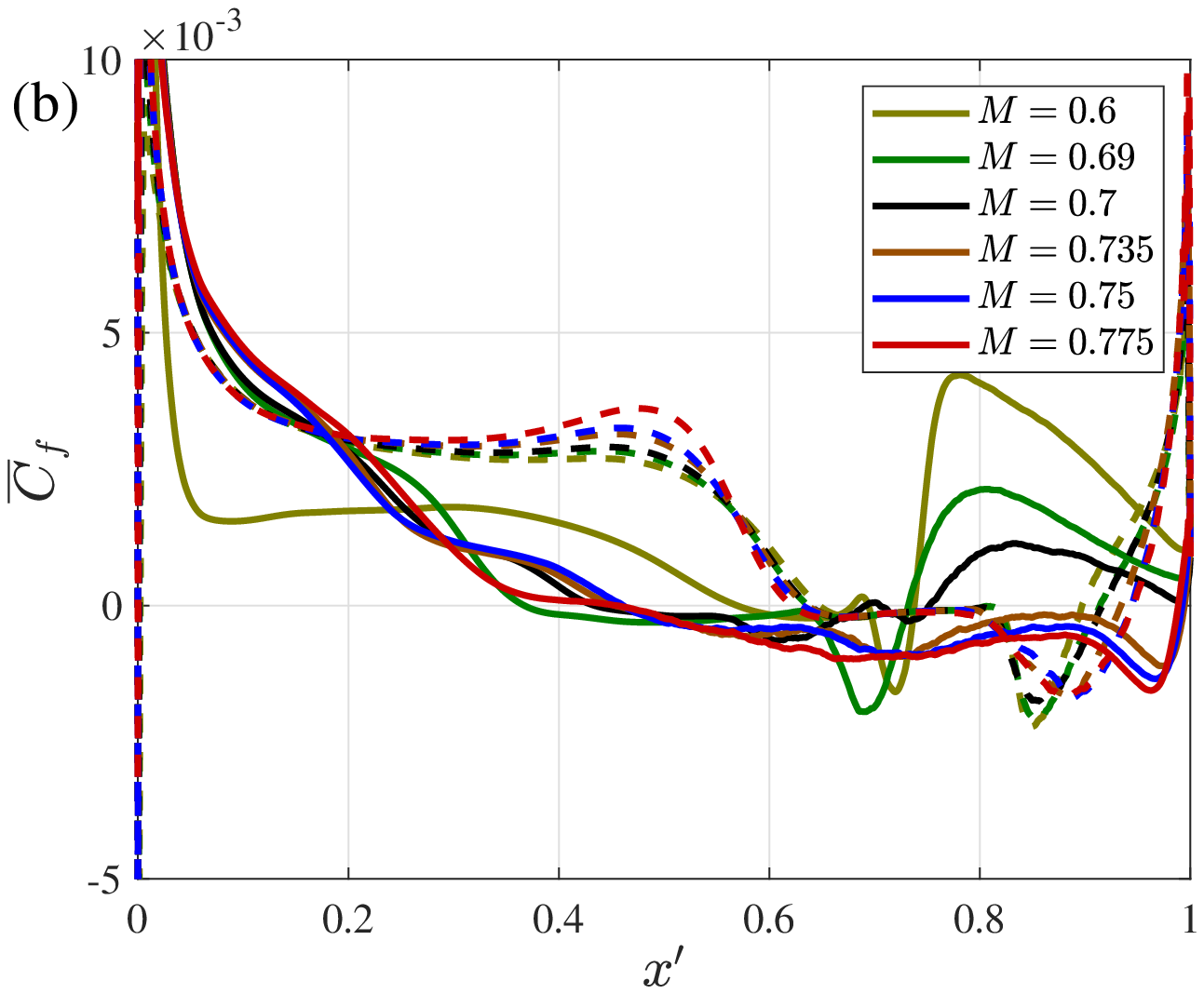}}
\caption{Variation of mean (a) pressure and (b) skin friction coefficients along suction (solid curves) and pressure (dashed curves) surfaces for various Mach numbers.}
    \label{fig:StatCompareCpMach}
\end{figure}

Mean pressure and skin friction coefficients  $\overline{C}_p$ and $\overline{C}_f$ (computed by time- and span-averaging the flow field) are compared for various $M$ in figure \ref{fig:StatCompareCpMach}. As $M$ increases, we see an increase of $\overline{C}_p$ on the suction surface, and a reduction on the pressure surface. This accounts for the monotonic reduction in mean lift with $M$ seen in figure \ref{fig:ClMach}\textit{a}. The presence of large-amplitude buffet causes what has been referred to as shock smearing \citep{Giannelis2017}, where the mean pressure reduces along the aerofoil surface in a gradual fashion in contrast to the abrupt drop present when a stationary shock wave occurs \citep{Jacquin2009}. This can be inferred by comparing the pre-onset cases of $M \leq 0.69$ with the rest. Examining the mean flow field, the mean shock position was observed to be approximately a constant at $x \approx 0.63$ in the range $0.7 \leq M \leq 0.8$ (see figures \ref{fig:RefSPODReconXT} and \ref{fig:SPODM8ShockPosn}). From figure \ref{fig:StatCompareCpMach}\textit{b}, we see that the mean skin-friction remains approximately a constant on the pressure surface for all $M$, but changes significantly on the suction surface when $M$ is increased above onset. For $M = 0.6$, the presence of a short separation bubble ($C_f < 0$) can be inferred in $0.64 \leq x' \leq 0.66 $, while for $M = 0.69$ and 0.7, there is another region of flow separation seen slightly upstream to this bubble, with the two merging for $M\geq0.735$ to form a separation zone that extends from upstream of the shock foot to close to the TE. Note that although this merged separation zone is present in the mean flow, there are phases in the buffet cycle (high-lift phase) when there is flow reattachment (see \S\ref{secSPOD}).

%%%%%%%%%%%%%%%%%%%%%%%%%%%%%%%%%%%%%%%%
\subsubsection{High Mach number features and buffet offset}
\label{subsechighMach}
%%%%%%%%%%%%%%%%%%%%%%%%%%%%%%%%%%%%%%%%

\begin{figure} 
\centerline{
\includegraphics[width=.495\textwidth]{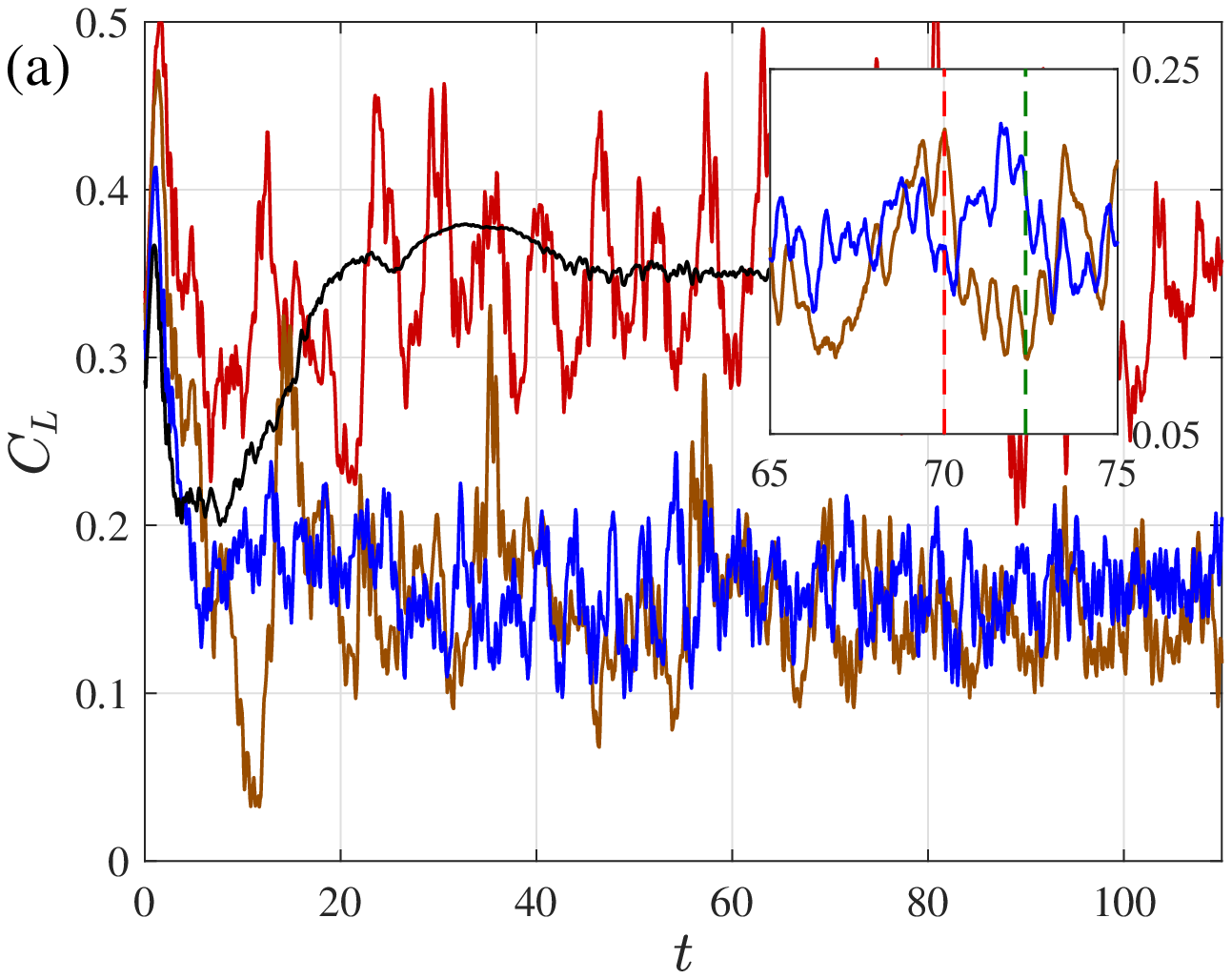}
\includegraphics[width=.495\textwidth]{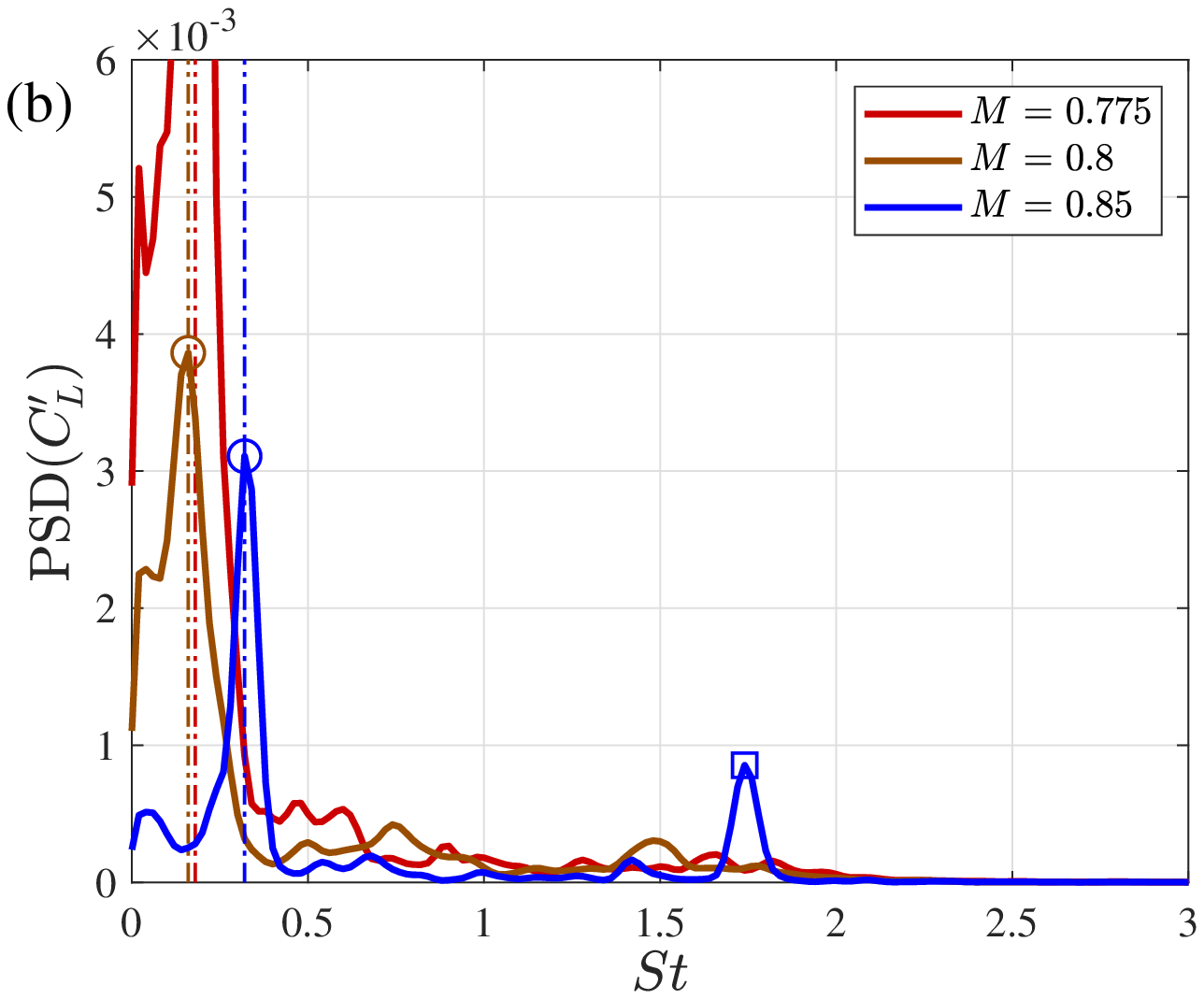}}
\caption{(a) Temporal variation of lift coefficient for various freestream Mach number and (b) the power spectral density of its fluctuating component.}
    \label{fig:clVsT_HighM}
\end{figure}
For $M \geq 0.8$, shock wave structures are present on both aerofoil surfaces. The temporal variation of $C_L$ for these cases and the PSD of the corresponding fluctuating component are shown in figure \ref{fig:clVsT_HighM}. The case of $M = 0.775$, for which shock wave structures are observed only on the suction side, is additionally shown for reference. Buffet offset is seen to occur at $M = 0.9$, with $C_L$ approximately a constant. At this $M$ the mean lift is seen to be significantly higher than that of $M = 0.85$. The PSD at the buffet frequency (highlighted using circles) reduces by an order of magnitude as $M$ is increased from $0.775$ to $0.8$. 

\begin{figure} 
\centerline{
\includegraphics[trim={0cm 0cm 0cm .4cm},clip,width=.495\textwidth]{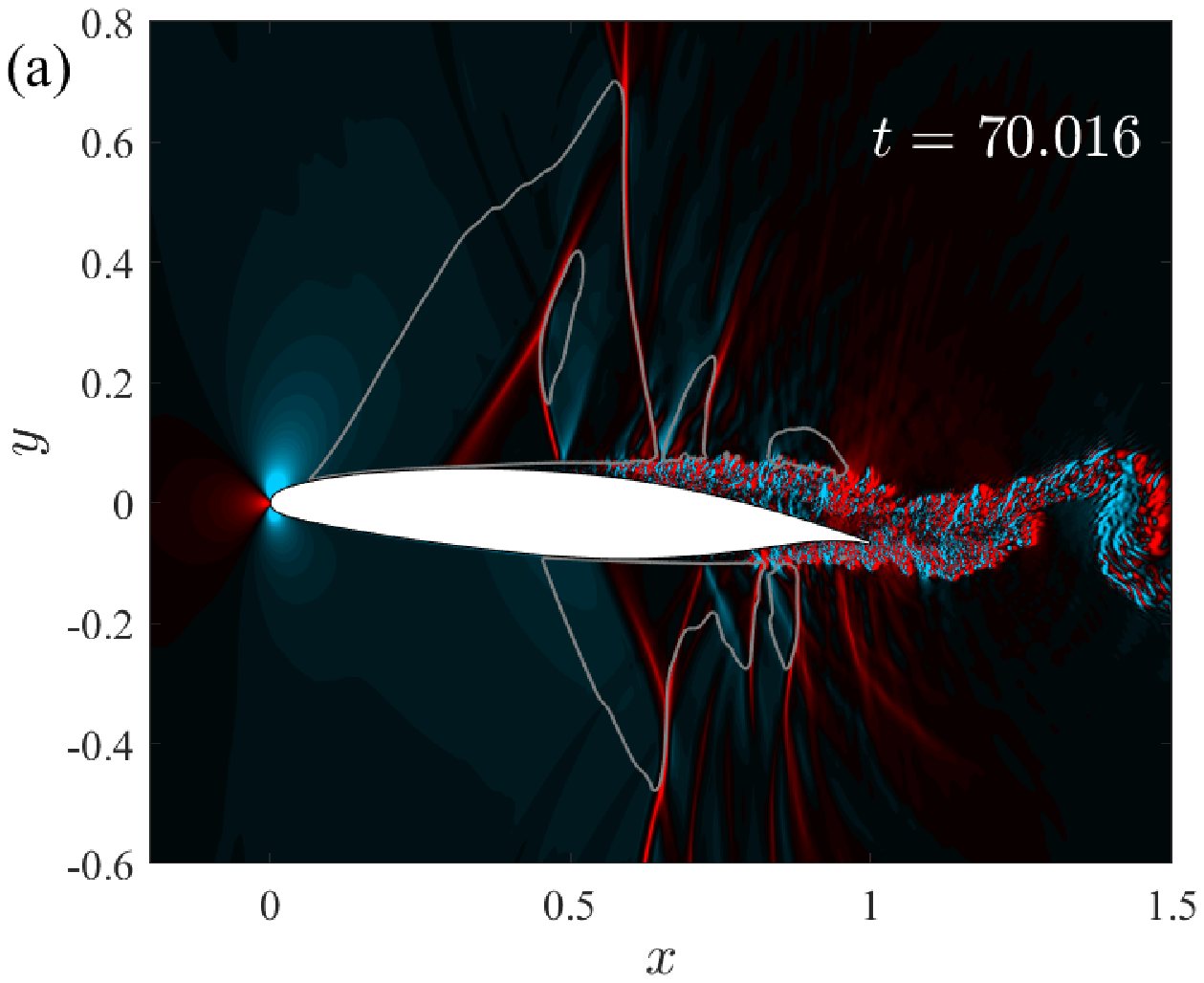}
\includegraphics[trim={0cm 0cm 0cm .4cm},clip,width=.495\textwidth]{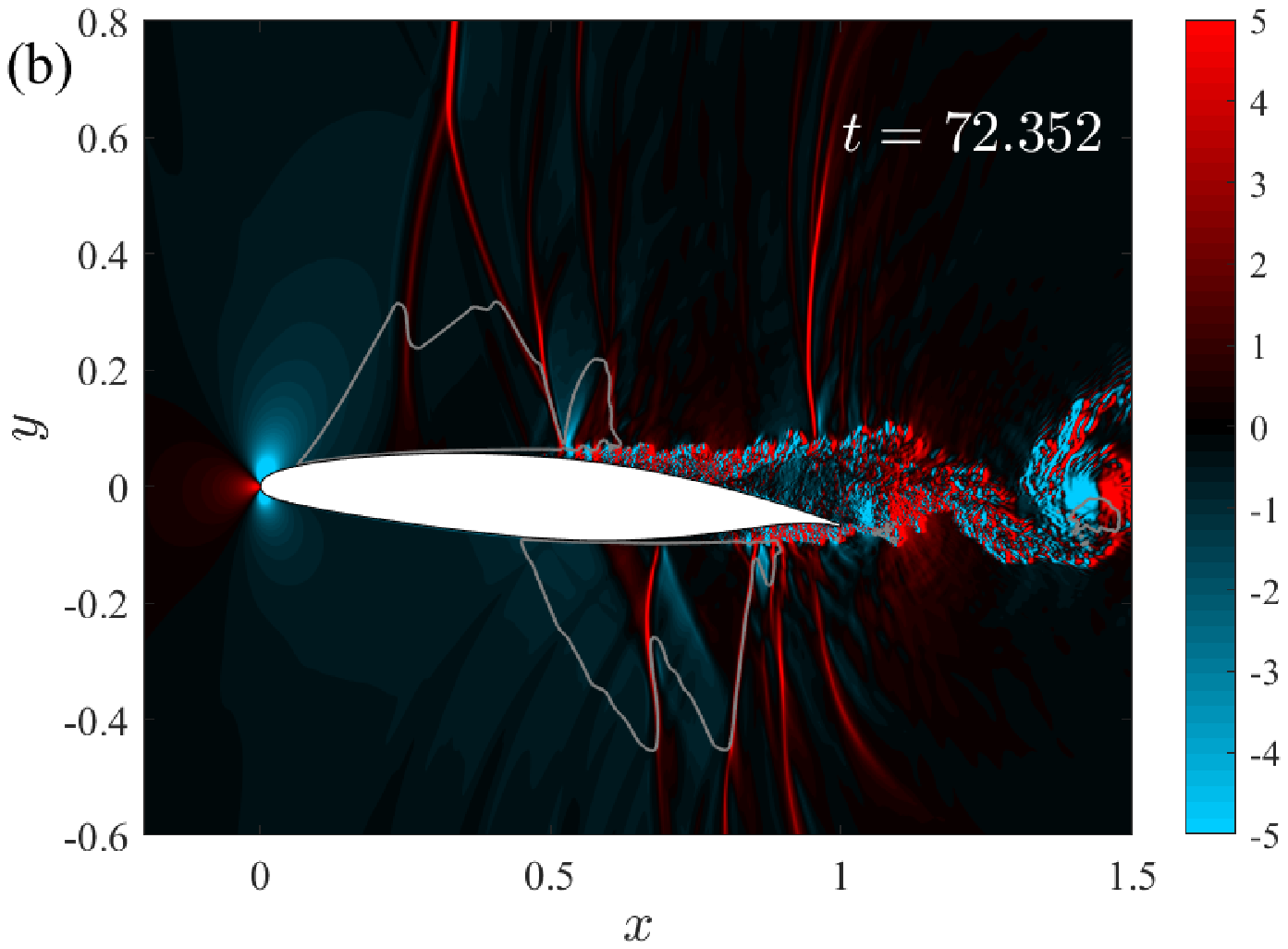}}
\caption{Streamwise density gradient contours on the $x-y$ plane shown for the (a) high and (b) low lift phases of the buffet cycle for $M = 0.8$.}
    \label{fig:M8DensGrad}
\end{figure}

The inset in figure \ref{fig:clVsT_HighM}\textit{a} shows the variation of $C_L$ for a shorter duration past transients for $M = 0.8$ and 0.85. For $M = 0.8$, high and low lift phases (red and green dashed lines) that occur at a low frequency can be discerned, but are now accompanied intermittently by high frequency oscillations of significant amplitude. The instantaneous flow features at these phases are shown in figure \ref{fig:M8DensGrad}. Both the suction and pressure sides have substantial regions of supersonic flow. This development of a supersonic region on the pressure side was observed with increasing $M$ first for $M = 0.775$, with a small ($< 0.1c$) pocket of supersonic flow occurring on the pressure side close to mid-chord in part of the buffet cycle, but without any shock wave present. From figures \ref{fig:M8DensGrad} and \ref{fig:clVsT_HighM}, we can see that although the temporal fluctuations in $C_L$ are low for this $M$, there are significant changes in the sonic envelope positions and the wall-normal extent of the supersonic region. Indeed, the qualitative flow features on the suction side resemble those seen for the other cases where buffet occurs (cf. figure \ref{fig:M735DensGrad}).  

\begin{figure} 
\centerline{
\includegraphics[trim={1cm 0cm 0cm .5cm},clip,width=.495\textwidth]{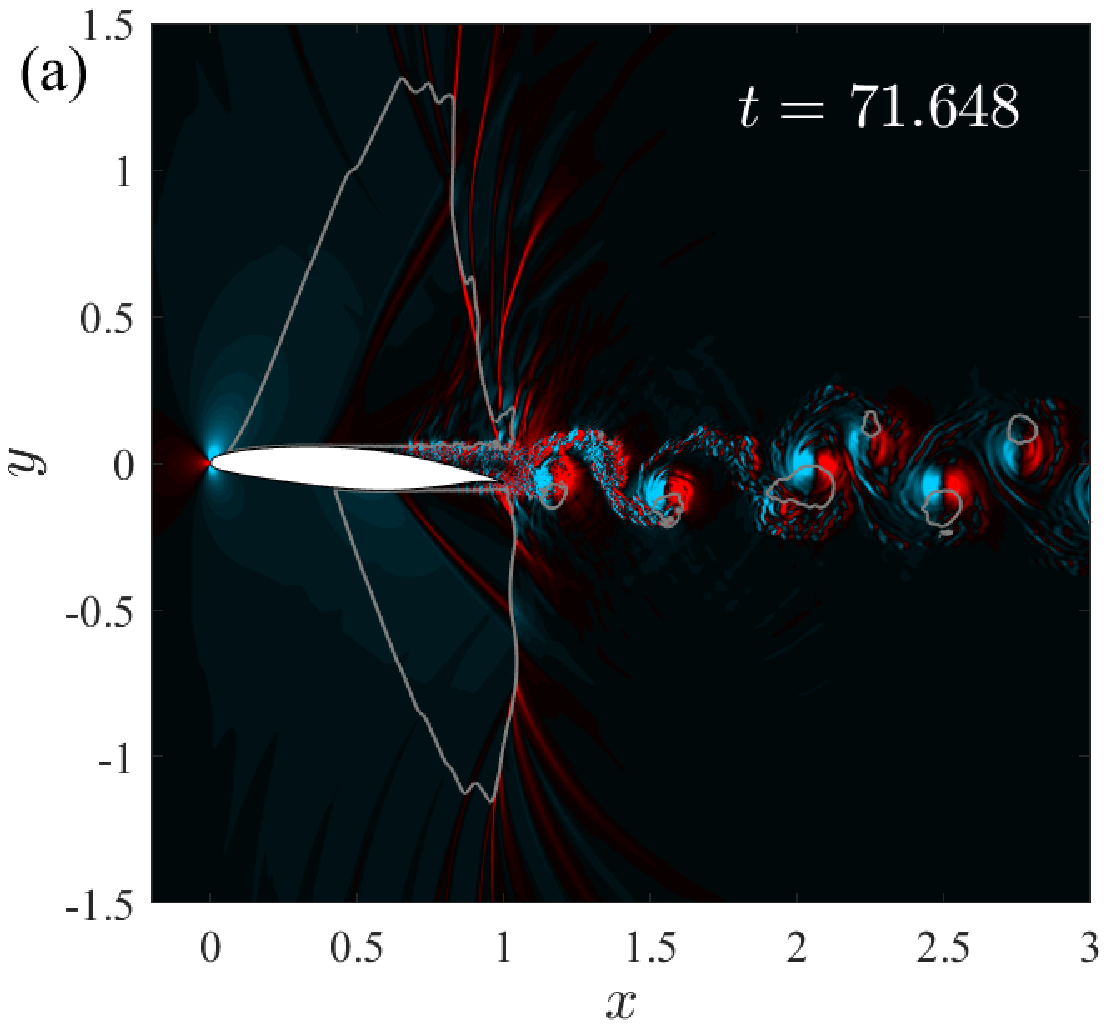}
\includegraphics[trim={1cm 0cm 0cm .5cm},clip,width=.495\textwidth]{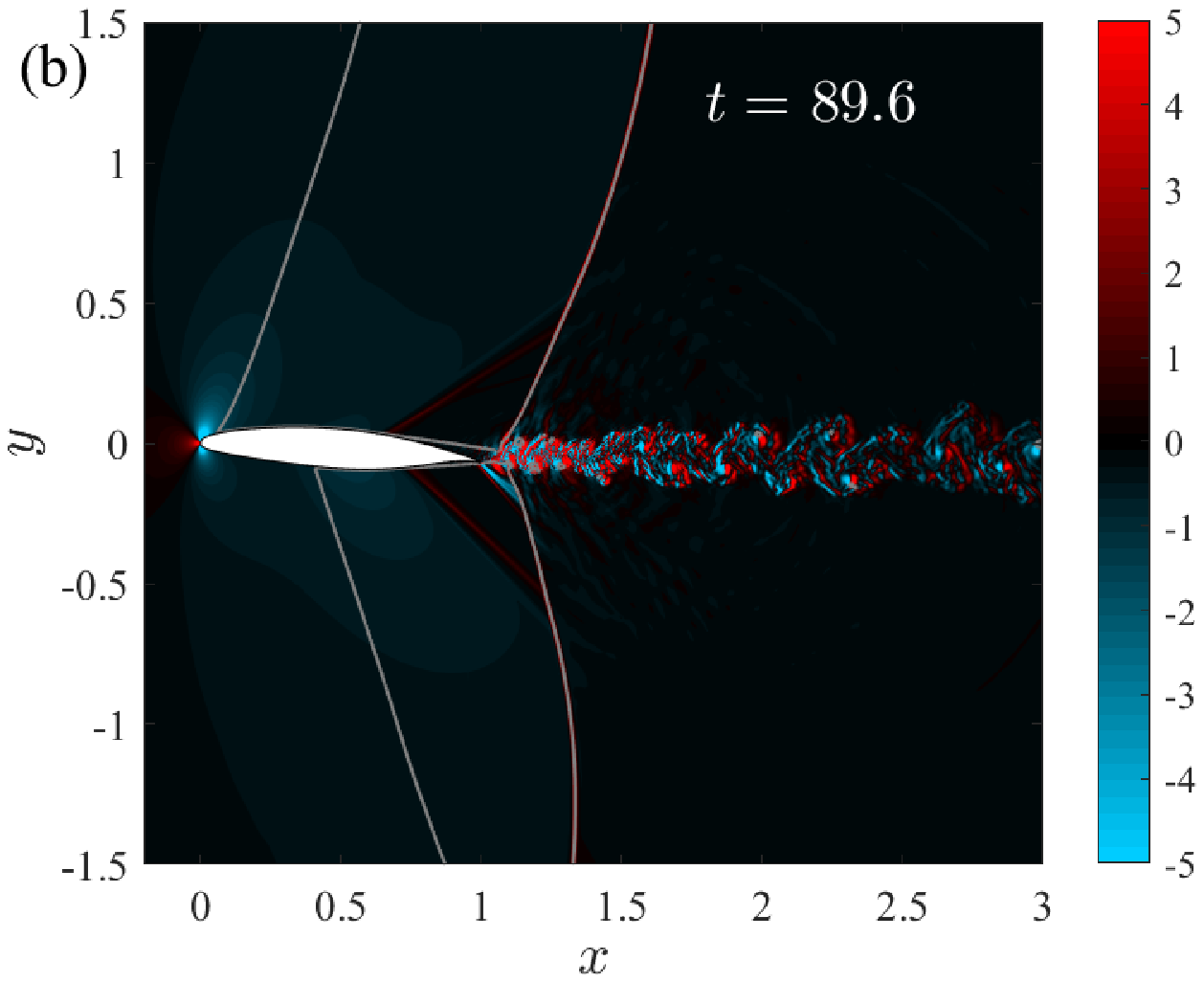}}
\caption{Instantaneous streamwise density gradient contours on the $x-y$ plane shown at (a) $M = 0.85$ and (b) $M = 0.9$.}
    \label{fig:M85N9DensGrad}
\end{figure}

For $M = 0.85$, as shown in figure \ref{fig:clVsT_HighM}\textit{b}, a peak at $St \approx 1.7$ in the PSD is present (highlighted by the square symbol) in addition to that at the buffet frequency. The former arises from a regular vortex shedding and is shown in figure \ref{fig:M85N9DensGrad}\textit{a} at an arbitrary instant. For this $M$, a single shock wave exists at the TE on each aerofoil surface. The shock wave positions on both sides were observed not to change significantly from that shown, although buffet could still be discerned. This  was also confirmed using SPOD (not shown). Strongly coherent vortices in the wake, indicating a K\'arm\'an vortex street, are apparent from the figure. By contrasting with the reference case, where large amplitude shock wave motion is observed but the vortical structures are not dominant, we can make an important conclusion that vortex shedding does not \textit{directly} influence buffet. That is, an increase in vortex shedding strength does not necessarily translate to an increase in buffet amplitude. 

The flow features for the offset freestream Mach number, $M = 0.9$, are shown in figure \ref{fig:M85N9DensGrad}\textit{b}. Unlike all the other cases simulated, the BL is now laminar up to the TE, although BL separation still occurs just upstream of the TE on the suction surface. The increase in $C_L$ observed for $M = 0.9$ as compared to $M = 0.85$ (figure \ref{fig:clVsT_HighM}\textit{a}) is likely due to this separation being relatively aft for the former. Vortices in the wake are also smaller. Similar results of subdued vortex shedding have been observed when $M$ is increased close to unity for transonic flows over circular cylinders \citep{Murthy1978} while the shock wave moving to the TE has been seen for the NACA 0012 aerofoil at zero incidence \citep{Bouhadji2003}. A fish-tail structure was reported in the latter flow, similar to the one that can be discerned here in the aft of the aerofoil, starting from $x \approx 0.6$. It is interesting to contrast these results with those reported in \citet{Giannelis2018}. They reported offset to occur abruptly, which is not the case here. As shown above, weak shock wave motion is present at $M = 0.85$ implying that buffet fades off gradually as $M$ is increased from 0.8 to 0.9. 

\subsubsection{Buffet at zero incidence on a supercritical aerofoil}
\label{subSecZeroAoA}

 Possibly based on distinctions made previously \citep{Lee2001,Iovnovich2012}, buffet has been classified in \citet{Giannelis2017} as either Type I or Type II. The former is typically associated with buffet on biconvex or symmetric aerofoils at zero incidences and is characterised by the presence of shock waves on both aerofoil surfaces. The latter is observed on supercritical aerofoils at relatively high incidence, with shock waves present only on the suction side. Based on the differences in the models for Type I buffet in \citet{Gibb1988} and Type II buffet in \citet{Lee1990}, it was proposed in \citet{Lee2001} that the two types are sustained by distinct mechanisms. This distinction was also noted in \citet{Iovnovich2012}, where the two were referred to as type one and two.

\begin{figure} 
\centerline{
\includegraphics[width=.9\textwidth]{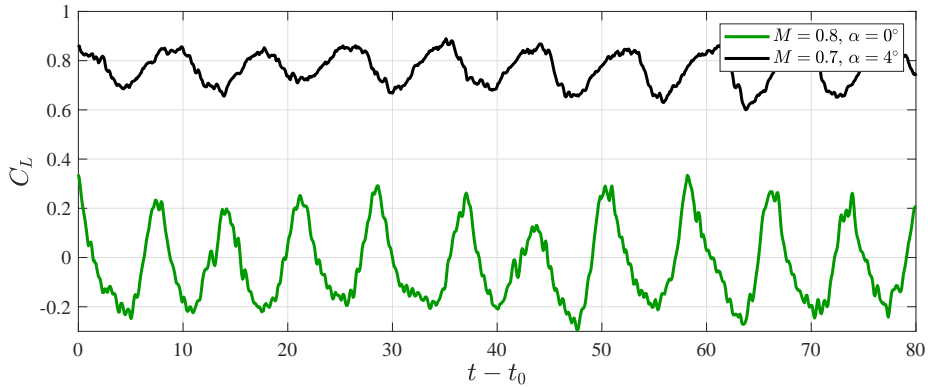}}
\caption{Temporal variation of lift coefficient for buffet at zero and non-zero incidence angles.}
    \label{fig:clVsT_0AoA}
\end{figure}

\begin{figure} 
\centerline{
\includegraphics[trim={0cm 0cm 0cm .5cm},clip,width=.495\textwidth]{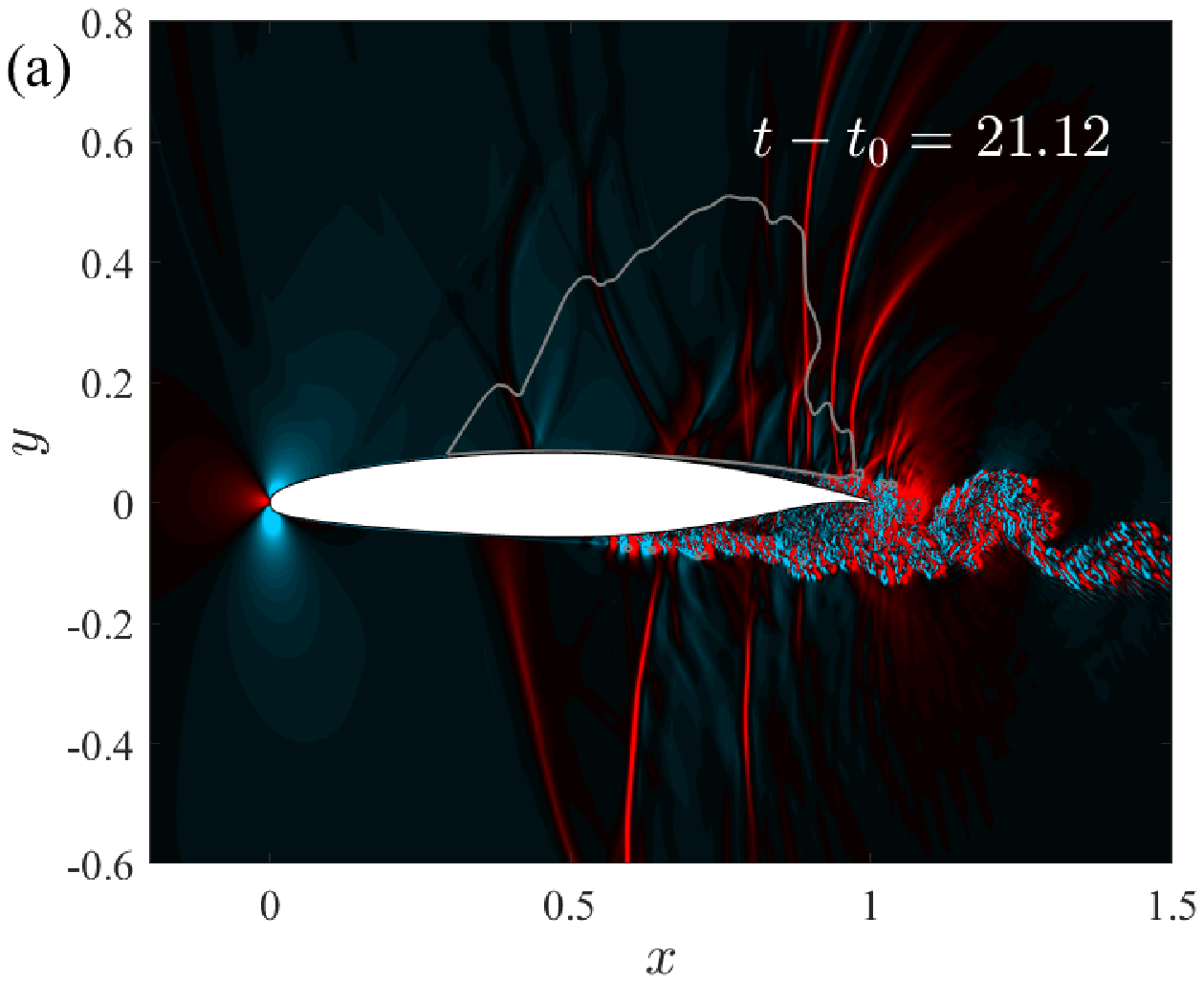}
\includegraphics[trim={0cm 0cm 0cm .5cm},clip,width=.495\textwidth]{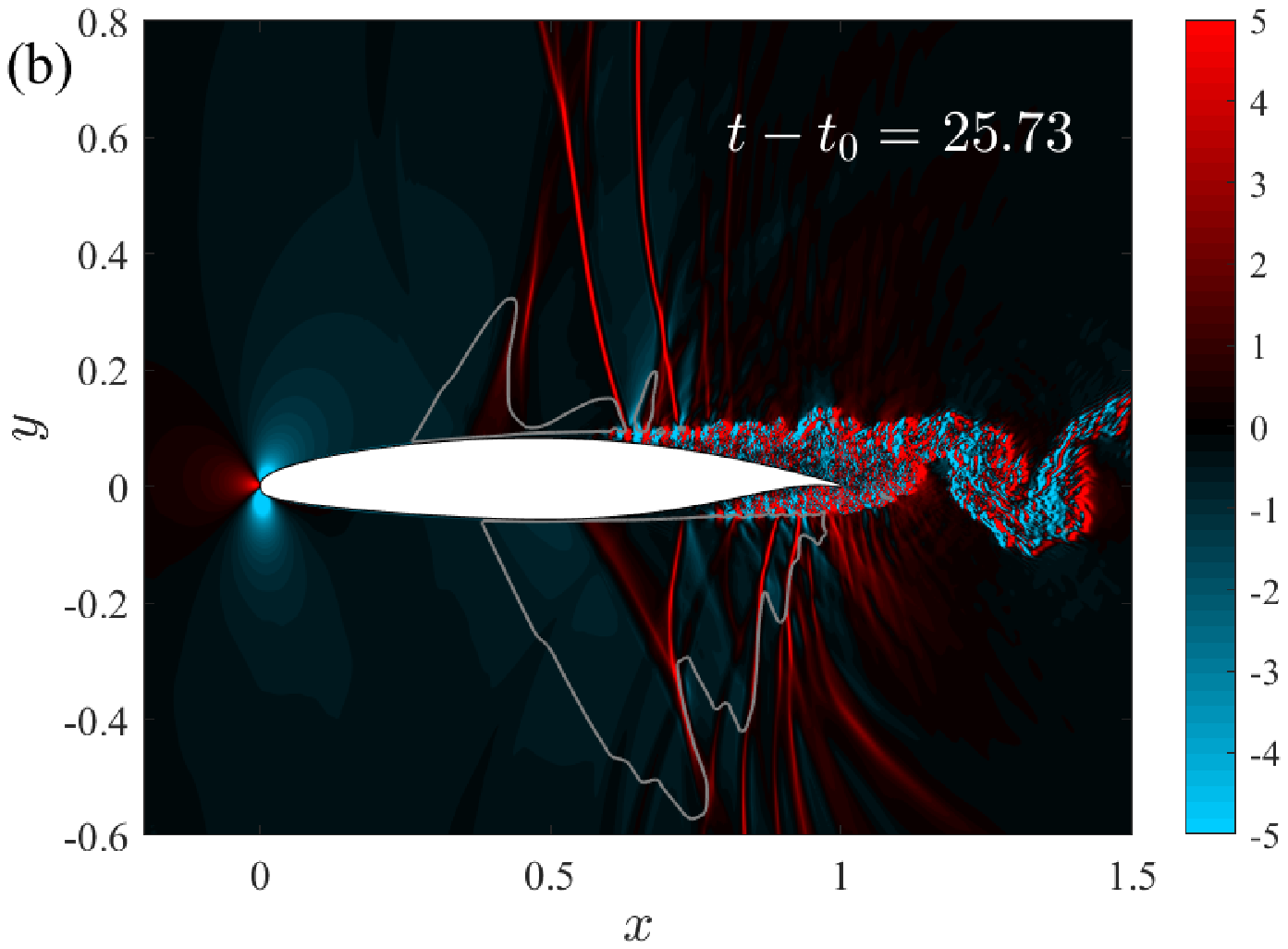}}
\caption{Streamwise density gradient contours on the $x-y$ plane shown for the (a) high and (b) low lift phases of the buffet cycle for $M = 0.8$ at $\alpha = 0^\circ$.}
    \label{fig:0AoAM8DensGrad}
\end{figure}

\begin{figure} 
\centerline{
\includegraphics[width=.495\textwidth]{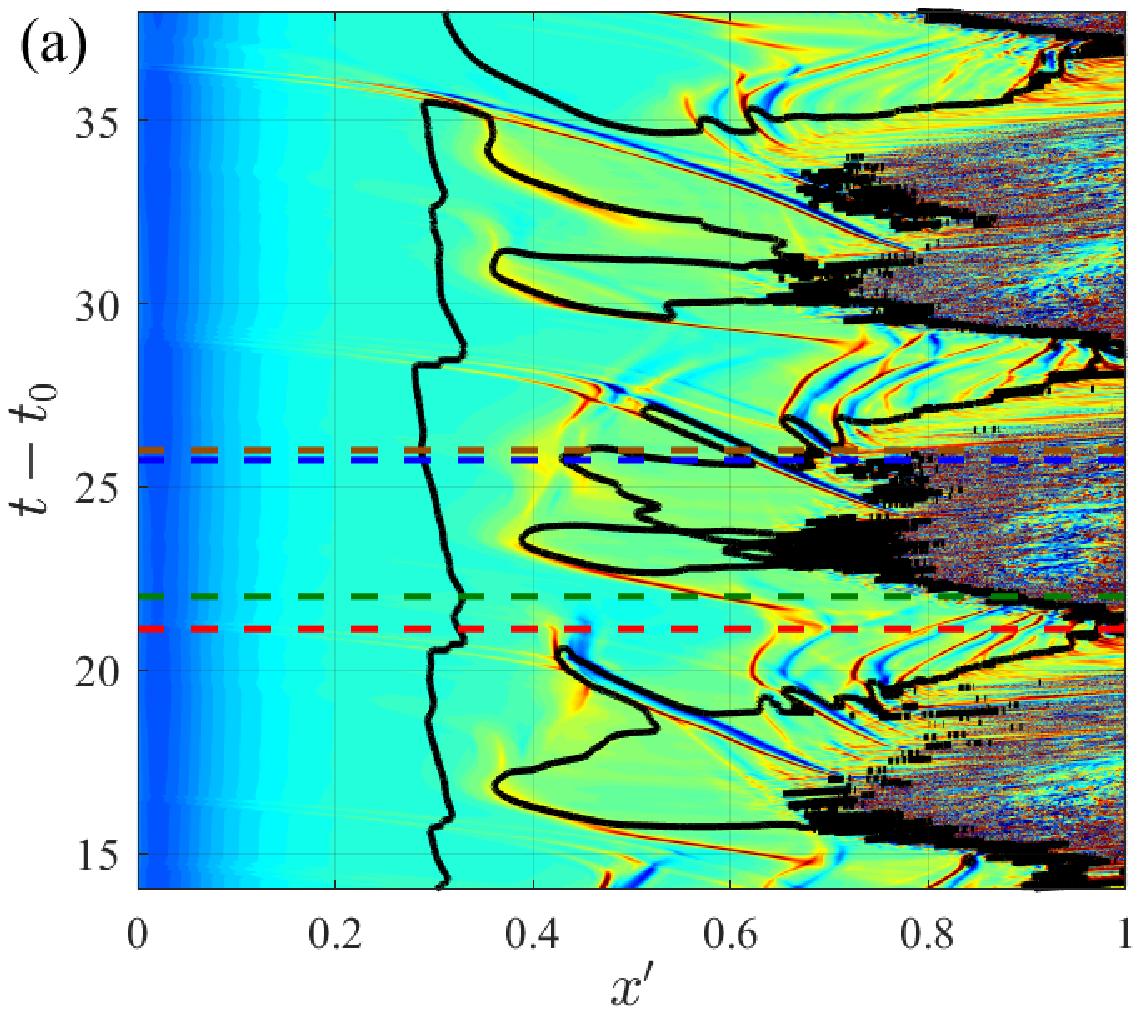}
\includegraphics[width=.495\textwidth]{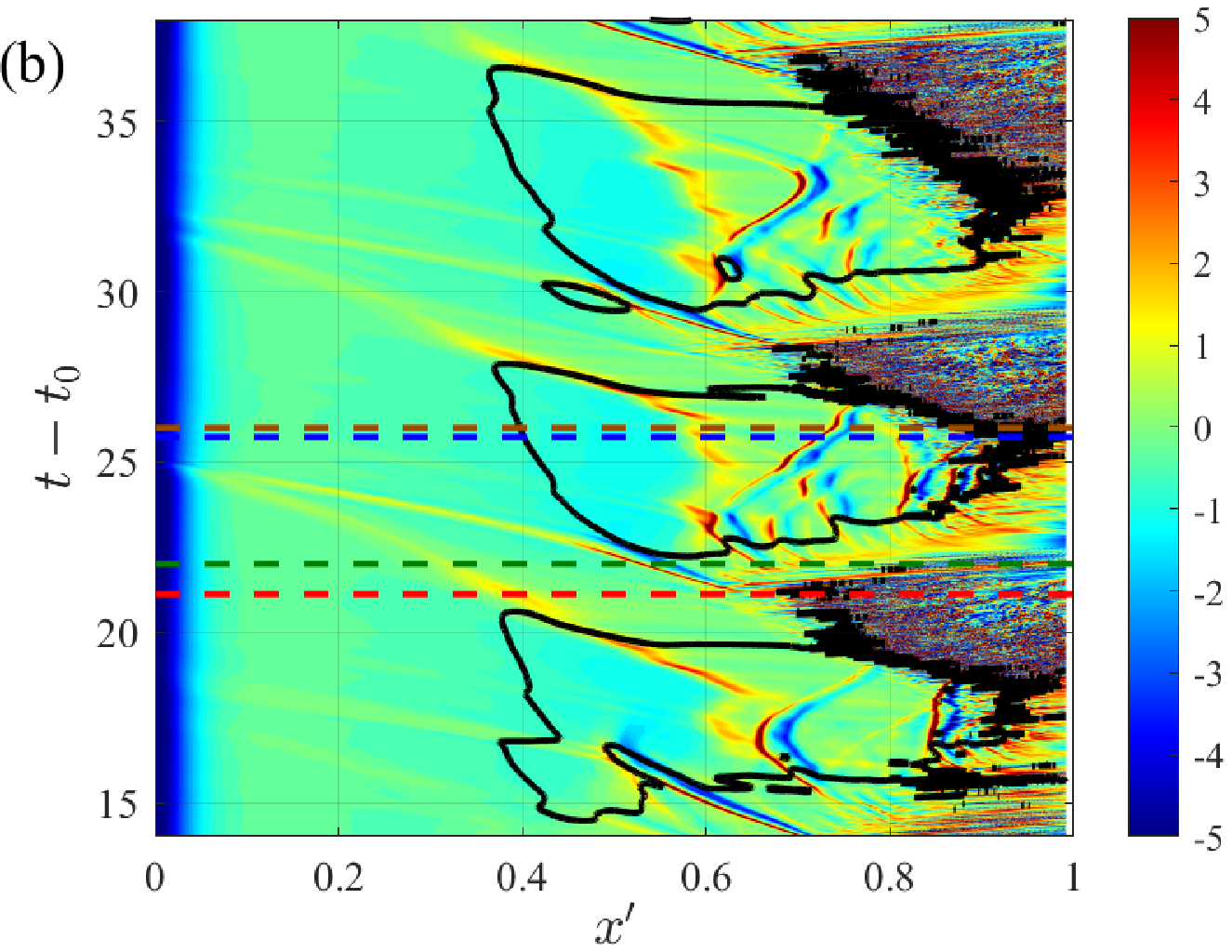}}
\caption{Spatio-temporal variation of streamwise pressure gradient on the (a) suction and (b) pressure sides of curve C5.}
    \label{fig:0AoAXTDia}
\end{figure}

 Here, for $0.7 \leq M \leq 0.775$, the buffet observed is clearly Type II. However, for $0.8 \leq M \leq 0.85$, we have buffet with shock waves appearing on both surfaces, albeit at a high incidence angle of $\alpha = 4^\circ$. To check if similar buffet features can be observed at zero incidence, we carried out simulations for an additional case of $M = 0.8$ and $\alpha = 0^\circ$. The temporal variation of $C_L$ past transients is shown in Fig \ref{fig:clVsT_0AoA}, with the reference case result also provided for comparison. The frequency of the oscillations seen was computed as $St \approx 0.14$ for the zero-incidence case, which is slightly higher than that of the reference case ($St \approx 0.12$), but lower than that of $M = 0.8$ case at $\alpha = 4^\circ$ ($St \approx 0.16$). Interestingly, the mean lift can be inferred to be zero for the zero-incidence case, despite the aerofoil not being symmetric. The density gradient contours at the high and low lift phases are shown in figure \ref{fig:0AoAM8DensGrad}, while the spatio-temporal variations of the streamwise pressure gradient on both the suction and pressure sides of the curve C5 are shown in Fig. \ref{fig:0AoAXTDia}. Pockets of supersonic flow and shock wave structures are seen on both surfaces for parts of the buffet cycle, although in other parts of the buffet cycle the flow becomes mostly or entirely subsonic on one of the surfaces. The high-lift (red dashed line) and low-lift (blue dashed line) phases are characterised by the shock wave structures being at their most downstream position on the suction and pressure sides, respectively. Furthermore, the extent of the supersonic regions on either surface varies approximately 180$^\circ$ out of phase with the other surface which can be inferred by comparing the sonic envelope's positions at the high and low-lift phases in Fig. \ref{fig:0AoAXTDia}. This suggests that the present case is of a Type I buffet, albeit on a supercritical aerofoil (and asymmetrical flow conditions). This is further explored using SPOD in \S\ref{subSecSPODHighM}.

%%%%%%%%%%%%%%%%%%%%%%%%%%%%%%%%%%%%%%%%
\subsection{Effect of angle of attack}
\label{secAoA}
%%%%%%%%%%%%%%%%%%%%%%%%%%%%%%%%%%%%%%%%

In this section, we examine the effect of incidence angles on buffet. As noted previously, the highest angle reported here is $\alpha = 6^\circ$, although buffet was also observed for $\alpha = 7^\circ$, albeit accompanied by minor but persistent grid-level oscillations. At a higher incidence of $\alpha = 8^\circ$, preliminary simulations showed that the aerofoil stalls with separation beginning at the leading edge, causing the simulations to fail in the current grid. As the grid requirements for capturing leading edge stall would exceed current computational resources, we were unable to determine if buffet persists in the presence of stall, although this seems unlikely based on previous studies \citep{Iovnovich2012,Giannelis2018}. %Indeed, several studies have noted that offset with $\alpha$ occurs when the flow stalls for turbulent buffet. 

\begin{figure} 
\centerline{
\includegraphics[width=.495\textwidth]{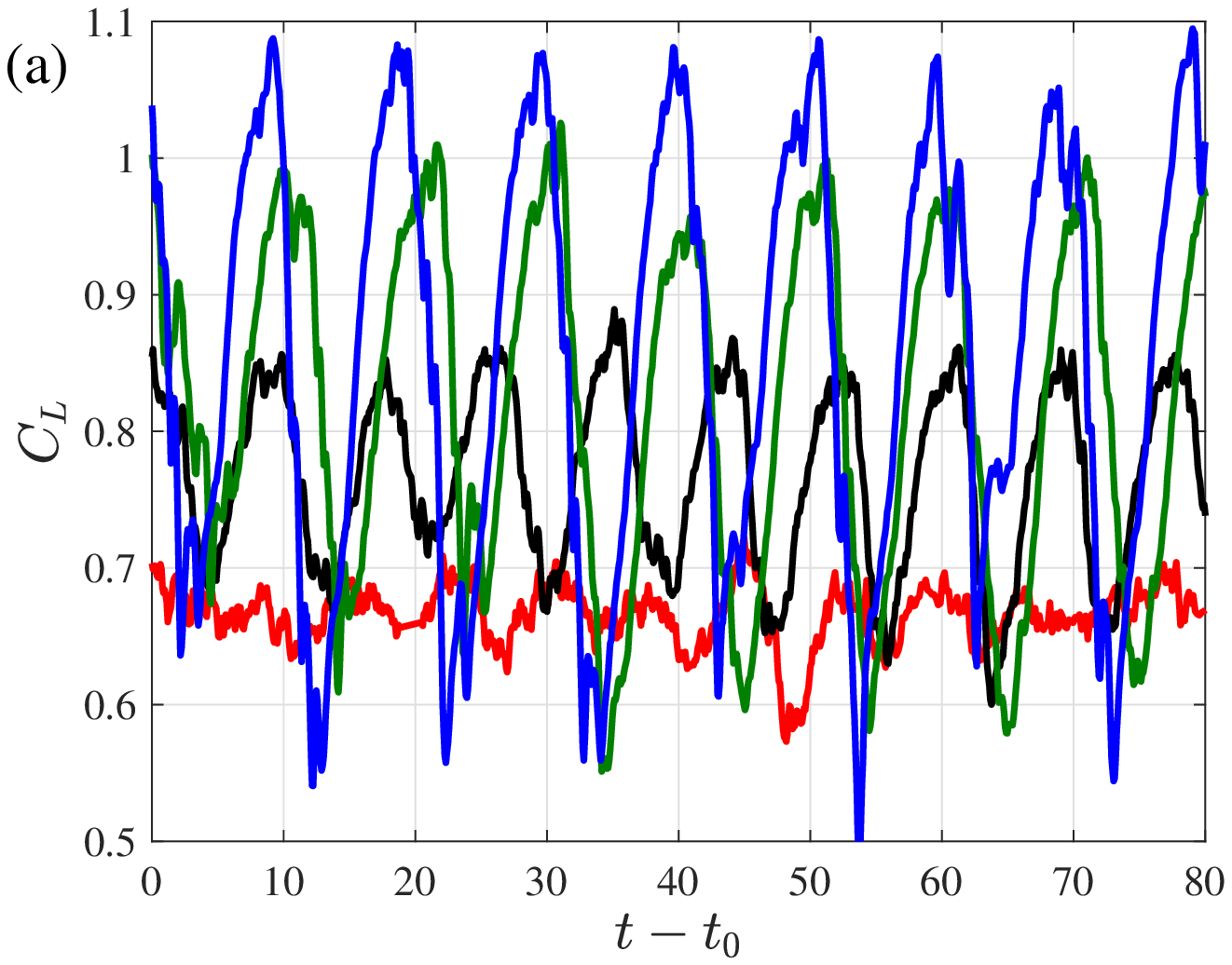}
\includegraphics[width=.495\textwidth]{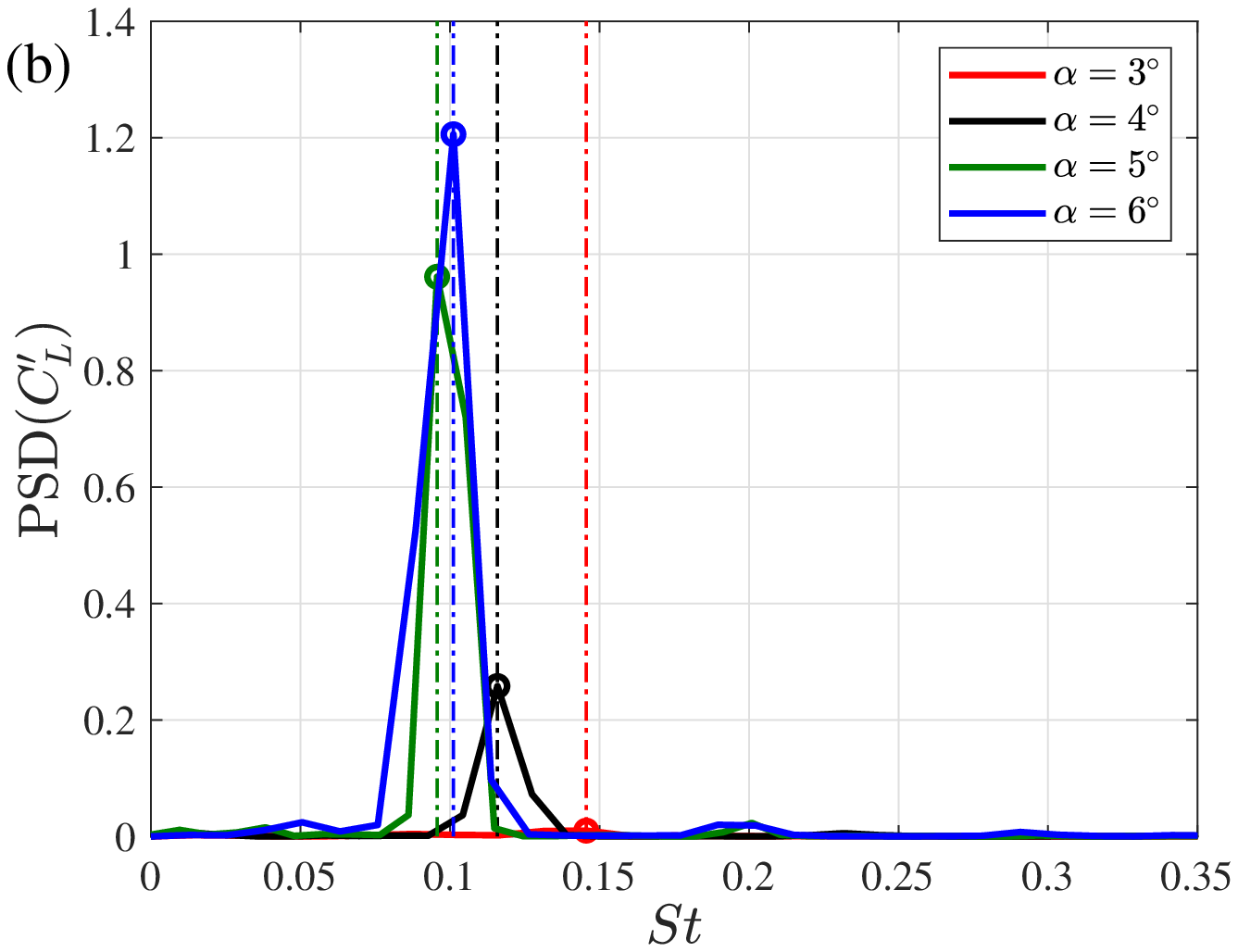}}
\caption{(a) Temporal variation of the lift coefficient past initial transients for various incidence angles and the (b) power spectral density of its fluctuating component.}
\label{fig:clVsT_AoA}
\end{figure}

The variation of $C_L$ with $\alpha$ and the PSD of its fluctuating component are shown in figure \ref{fig:clVsT_AoA}. Note that the time offset, $t_0$, is chosen approximately to coincide with the high-lift phase of buffet as well as being large enough to remove transient effects. For $\alpha = 3^\circ$, irregular temporal variations are observed, although footprints of buffet can also be discerned. This was confirmed using SPOD (not shown for brevity). For the same flow settings, \citet{Zauner2020a} have performed DNS in a wider domain and observed weak buffet indicating onset at this $\alpha$. Based on the observation of \citet{Zauner2020} that a reduction in span can lead to more irregularity in the buffet cycle, we conclude that buffet onset occurs at $\alpha = 3^\circ$.

With increasing $\alpha$, we see a substantial increase in the fluctuation energy of lift. The buffet frequency decreases when $\alpha$ is increased from 3$^\circ$ to 5$^\circ$, but remains approximately a constant when $\alpha$ is raised to 6$^\circ$. For turbulent buffet, a similar increase in buffet amplitude is commonly reported \citep{Jacquin2009,Giannelis2018}. However, in contrast to the present results, the frequency is reported to be approximately a constant \citep{Jacquin2009} or increase \citep{Dor1989,Brion2020} for turbulent buffet.

%However, the effect of $\alpha$ on frequency is somewhat unclear. For example, the computational study of \citet{Giannelis2018} noted a 76\% increase in turbulent buffet frequency when changing $\alpha$ from $5^\circ$ to $7^\circ$ for low $M$, but only around a $4\%$ increase at high $M$ in the buffet regime indicating that the variation with $\alpha$ can be crucially dependent on $M$. The experimental investigations reported in \citet{Jacquin2009} indicate that $\alpha$ negligibly affects buffet frequency, while others \citep{Dor1989,Brion2020} report an increase of frequency with $\alpha$. Thus, from the available data, it is unclear whether there are significant differences in the effect of $\alpha$ on laminar and turbulent buffet's frequency. 

\begin{figure} 
\centering
\includegraphics[trim={0.3cm 1.5cm 0cm 2cm},clip,width=.495\textwidth]{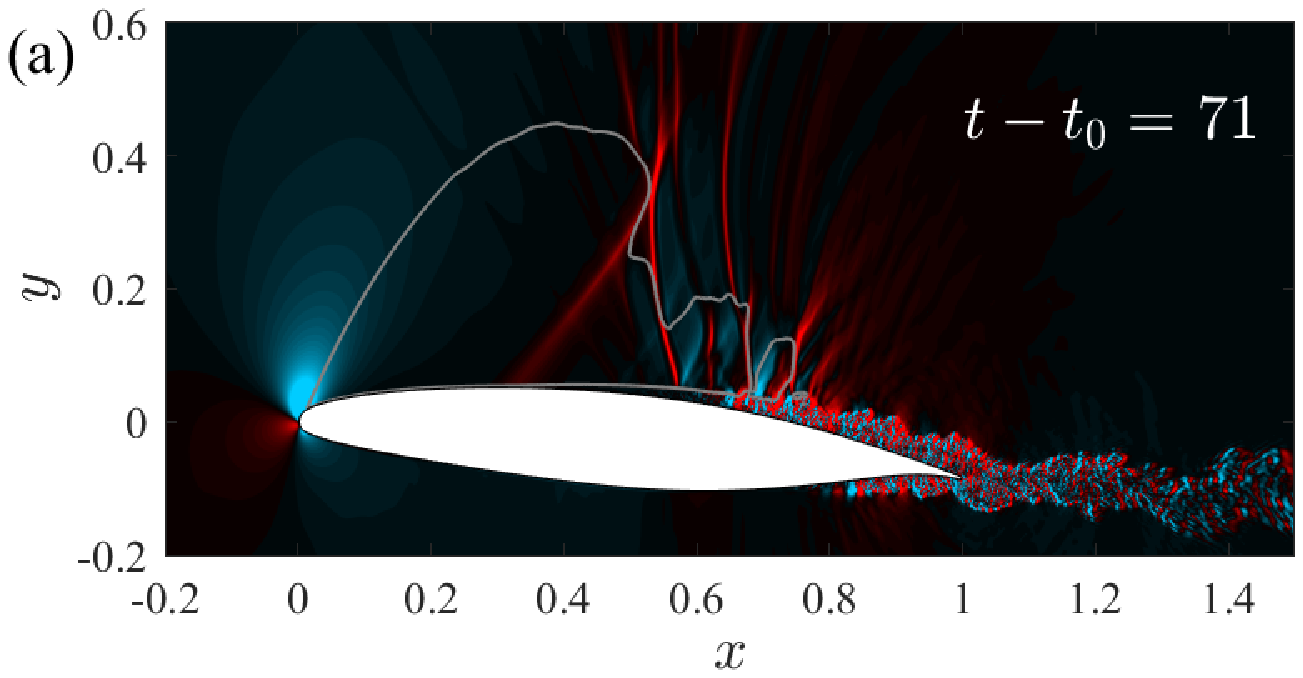}
\includegraphics[trim={0.3cm 1.5cm 0cm 2cm},clip,width=.495\textwidth]{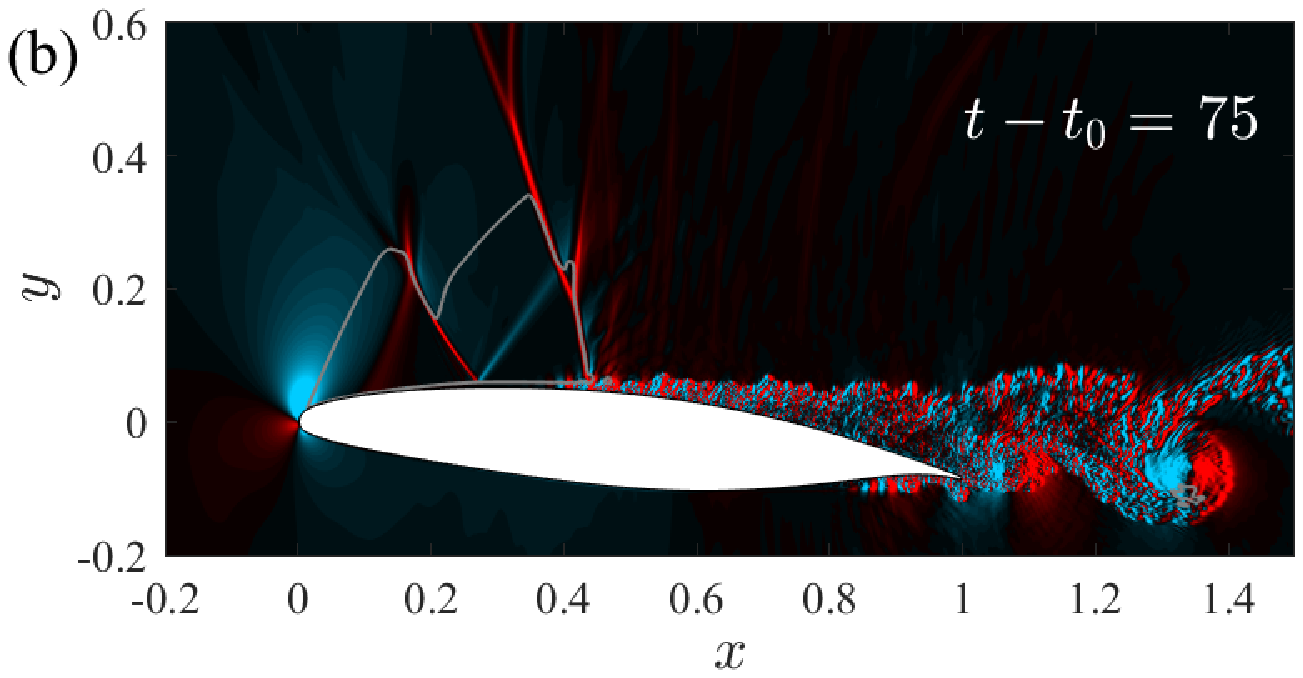}
\includegraphics[trim={0.3cm 1.5cm 0cm 2cm},clip,width=.495\textwidth]{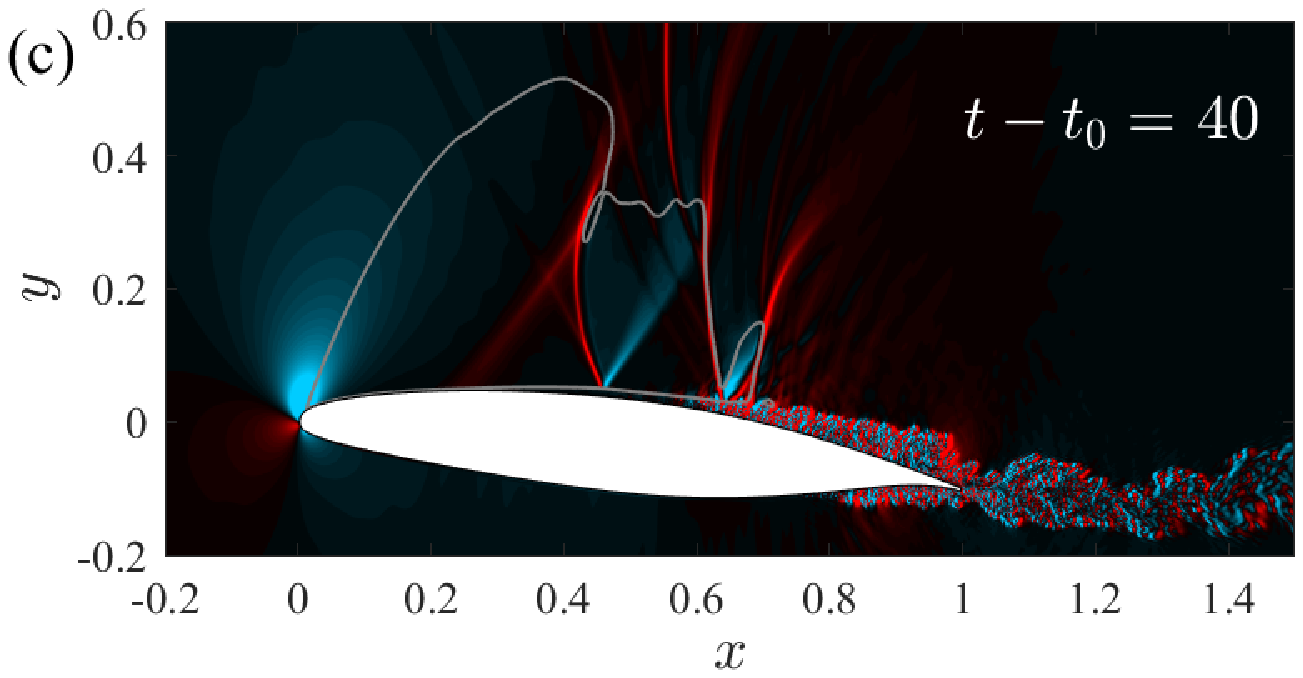}
\includegraphics[trim={0.3cm 1.5cm 0cm 2cm},clip,width=.495\textwidth]{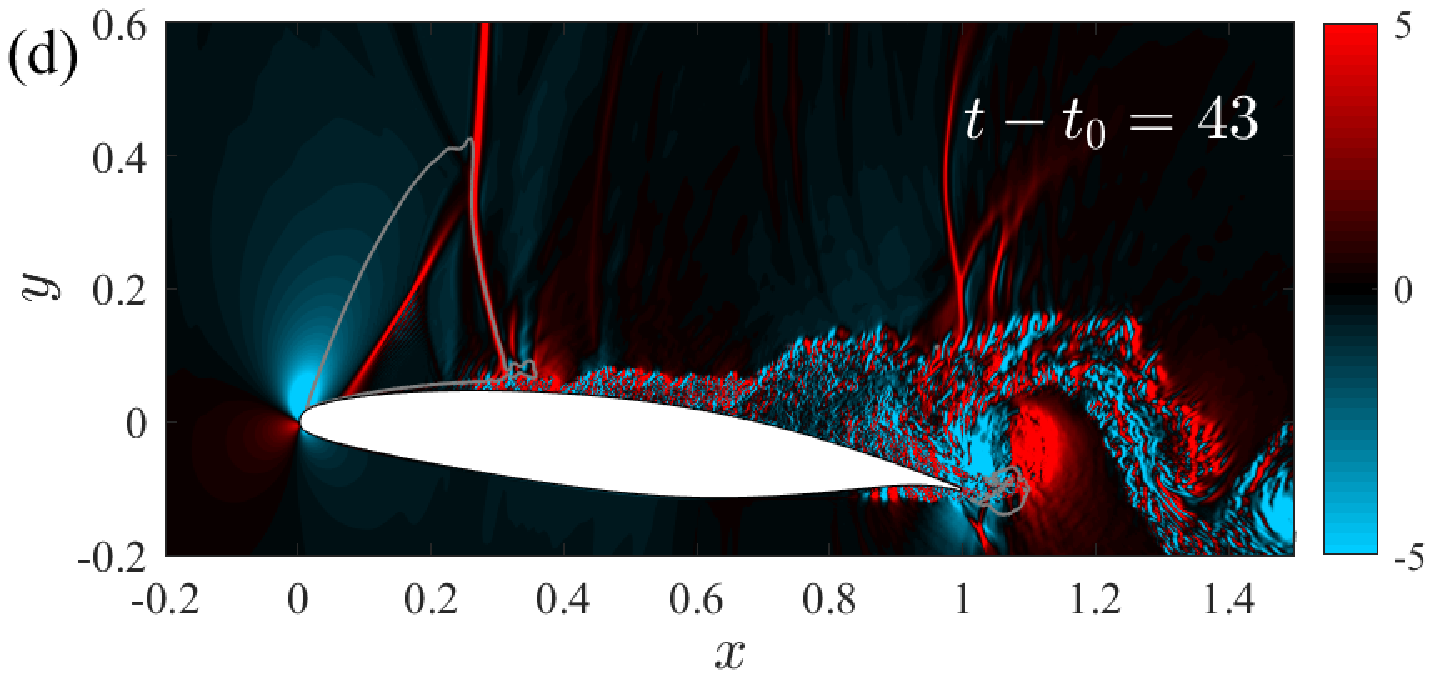}
\caption{Streamwise density gradient contours on the $x-y$ plane shown for the high (left) and low (right) lift phases of the buffet cycle for the cases, $\alpha = 5^\circ$ (top) and $6^\circ$ (bottom).}
    \label{fig:AoA5DensGrad}
\end{figure}

No qualitative differences in mean aerofoil coefficients ($\overline{C}_p$ and $\overline{C}_f$) were observed with variations in $\alpha$ and thus, these results are not shown for brevity. The instantaneous flow features highlight some interesting quantitative differences, as shown in figure \ref{fig:AoA5DensGrad} using streamwise density gradient contours. For higher $\alpha$, we see a large supersonic region in the high lift phase (left) which reduces considerably in size in the low-lift phase (right). This is accompanied by a large upstream excursion of the shock structures, which is especially evident for $\alpha = 6^\circ$. Similar to cases of buffet at high $M$, large-scale vortices are observed in the low-lift phase. 

\begin{figure} 
\centerline{\includegraphics[trim={0cm 0cm 0.5cm 0cm},clip,width=.495\textwidth]{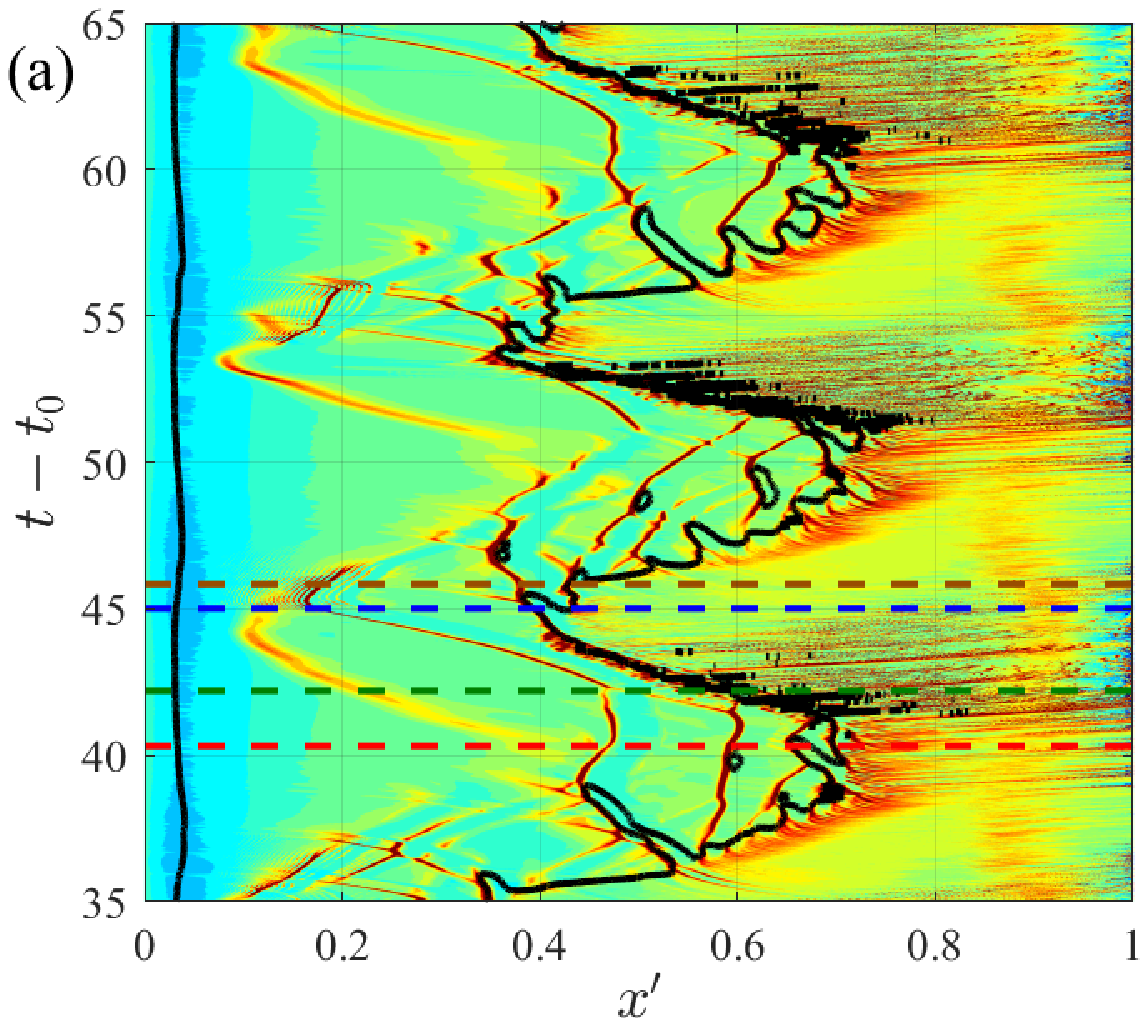}
\includegraphics[trim={0cm 0cm 0.5cm 0cm},clip,width=.495\textwidth]{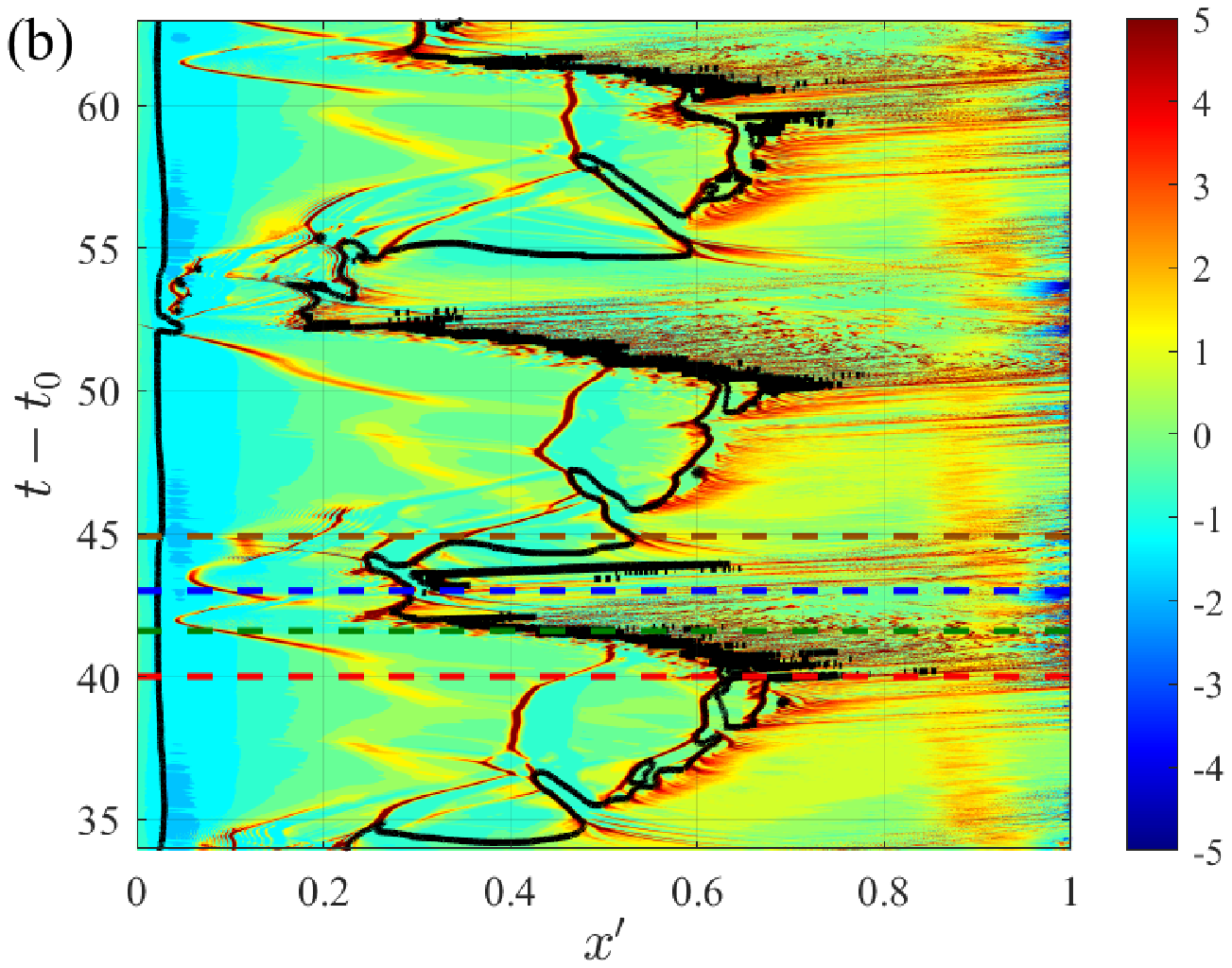}}
\caption{Spatio-temporal variation of streamwise pressure gradient on the suction side of C5 for (a) $\alpha = 5^\circ$ and (b) $\alpha = 6^\circ$.}
    \label{fig:AoA6XTDia}
\end{figure}

The spatio-temporal variation of the pressure gradient on the suction side of curve C5 is shown for $\alpha=5^\circ$ and $\alpha=6^\circ$ in figure \ref{fig:AoA6XTDia}. The sonic envelope's minimum and maximum position is, as with the other cases, seen to approximately align with the time at which the lift is lowest and highest (horizontal blue and red dashed lines), respectively. However, the motion is no longer symmetrical about the mean position, with the sonic line (black curve) resembling an inverse sawtooth wave at $\alpha = 6^\circ$. Based on the slopes observed, we can infer that the sonic envelope moves upstream rapidly, while its downstream excursion contains phases in which it is substantially slower. Comparing with the reference case (figure \ref{fig:RefXTDia}), we find that the mean shock position moves upstream with increasing $\alpha$. The sonic envelope moves more rapidly compared to the reference case, which is expected given the strong increase in amplitude, as opposed to the moderate reduction in frequency of the buffet cycle seen at higher $\alpha$ (figure \ref{fig:clVsT_AoA}\textit{b}). Additionally, the number of shock wave structures is reduced. The time required for the pressure waves to reach the TE was found to be in the range $0.6 \leq t_\mathrm{down} \leq 1.1$. 

%%%%%%%%%%%%%%%%%%%%%%%%%%%%%%%%%%%%%%%%
\subsection{Effect of Reynolds number}
\label{secRe}
%%%%%%%%%%%%%%%%%%%%%%%%%%%%%%%%%%%%%%%%

\begin{figure} 
\centering
\includegraphics[width=.495\textwidth]{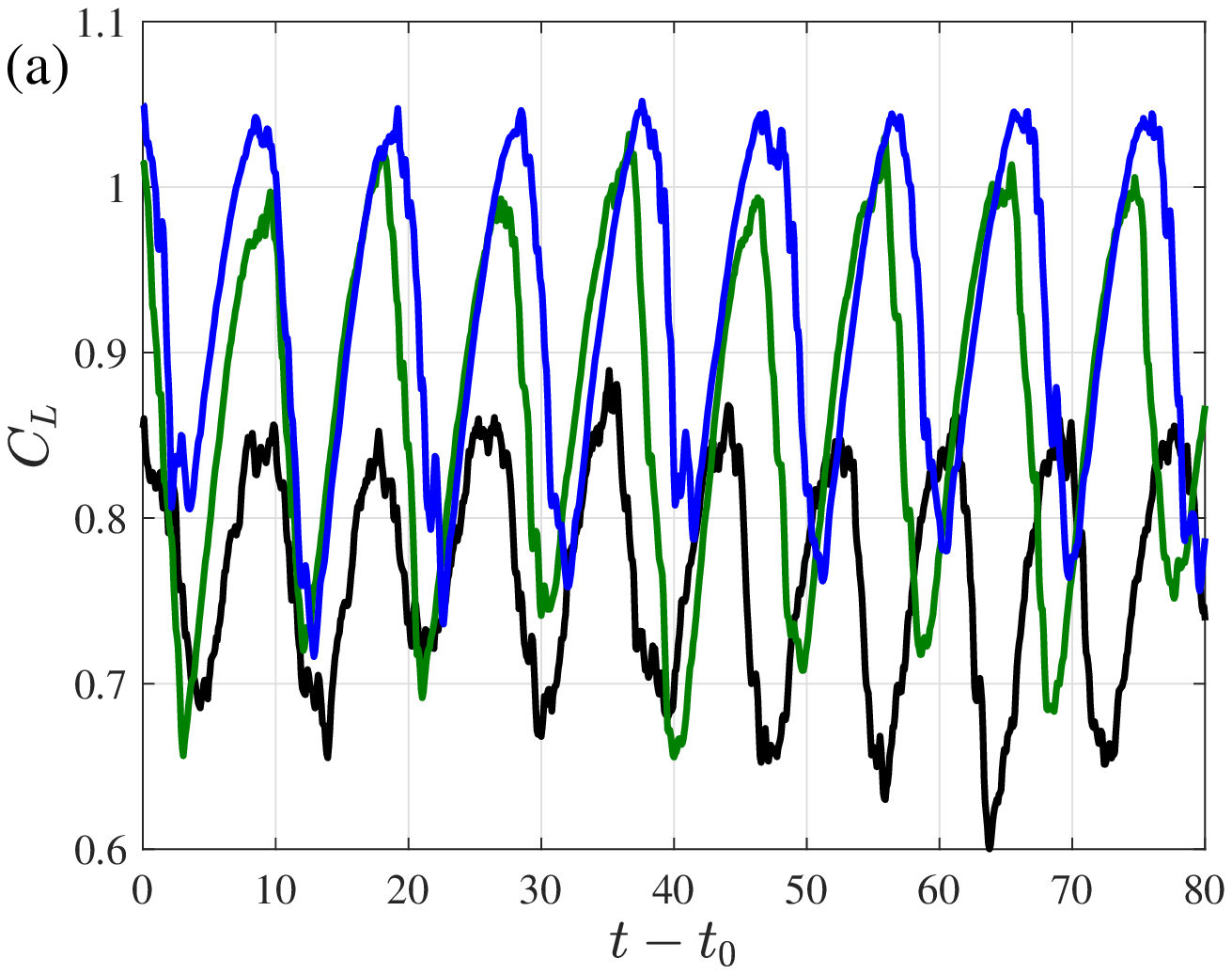}
\includegraphics[width=.495\textwidth]{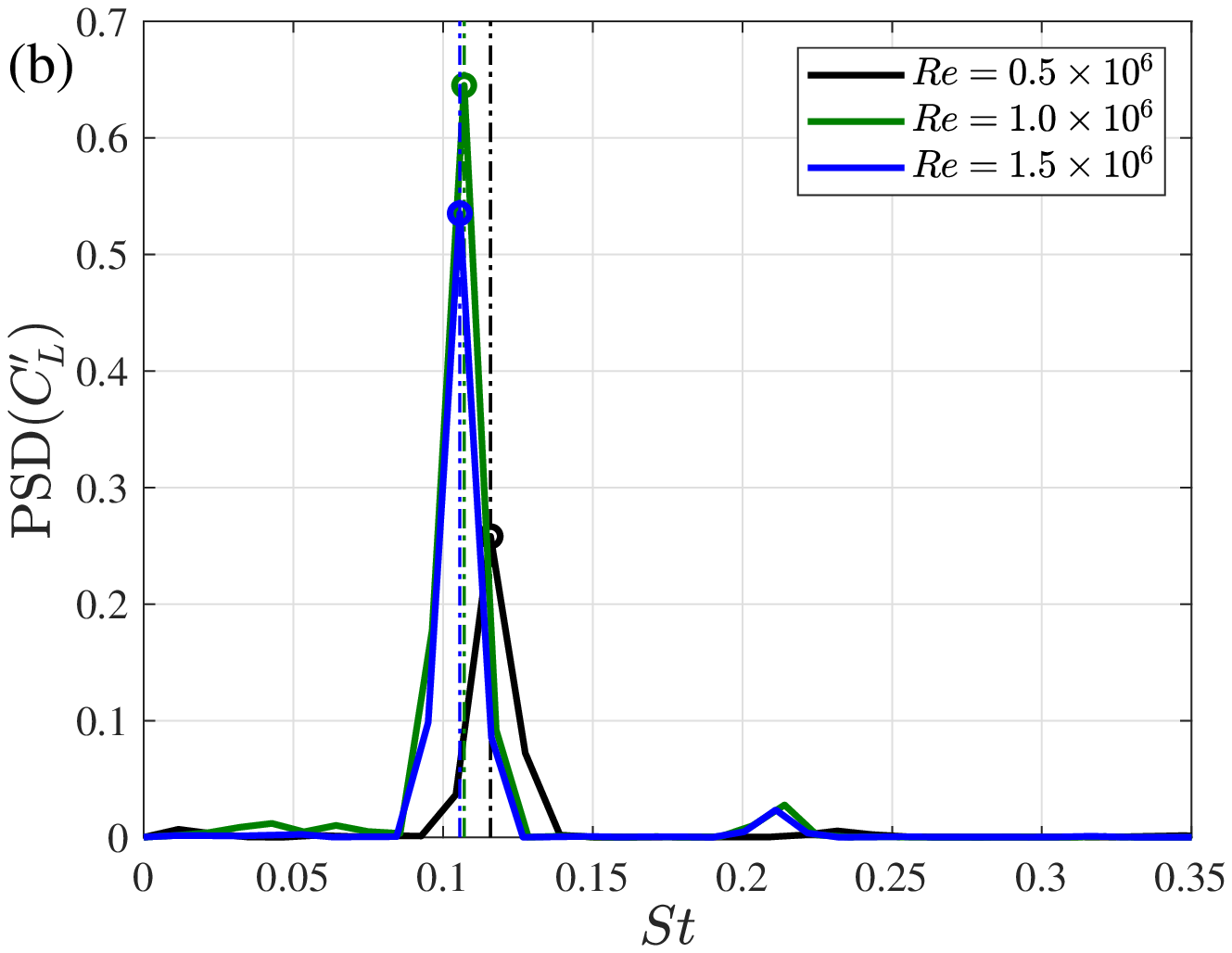}
\caption{(a) Temporal variation of lift coefficient past initial transients for various freestream Reynolds numbers and (b) the PSD of its fluctuating component.}
    \label{fig:clVsT_Re}
\end{figure}

The effect of varying $\Rey$ is reported here. At the lowest value of $\Rey = 2\times10^5$ simulated, buffet is absent and only BL separation leading to stall was observed. This case is not shown here and we focus only on cases in the range $5\times10^5 \leq \Rey \leq 1.5\times10^6$. One of the motivating factors for this is the presence of multiple shock wave structures at $\Rey = 5\times 10^5$. Most other studies, which are usually at higher $\Rey$, report a single shock wave and not the shock system seen here. The temporal variation of $C_L$ and the PSD($C_L'$) are shown in figure \ref{fig:clVsT_Re}. A small reduction of the buffet frequency occurs as $\Rey$ is increased beyond the reference value. By contrast, the amplitude increases substantially. Similar minor reduction in frequency and increase in amplitude have been noted at low $\Rey$ for turbulent buffet \citep[see][]{Raghunathan1998,Lee2001}. The spectra also show that the first harmonic of the buffet frequency has significant energy content.

\begin{figure} 
\centering
\includegraphics[trim={0.3cm 1.5cm 0cm 2cm},clip,width=.495\textwidth]{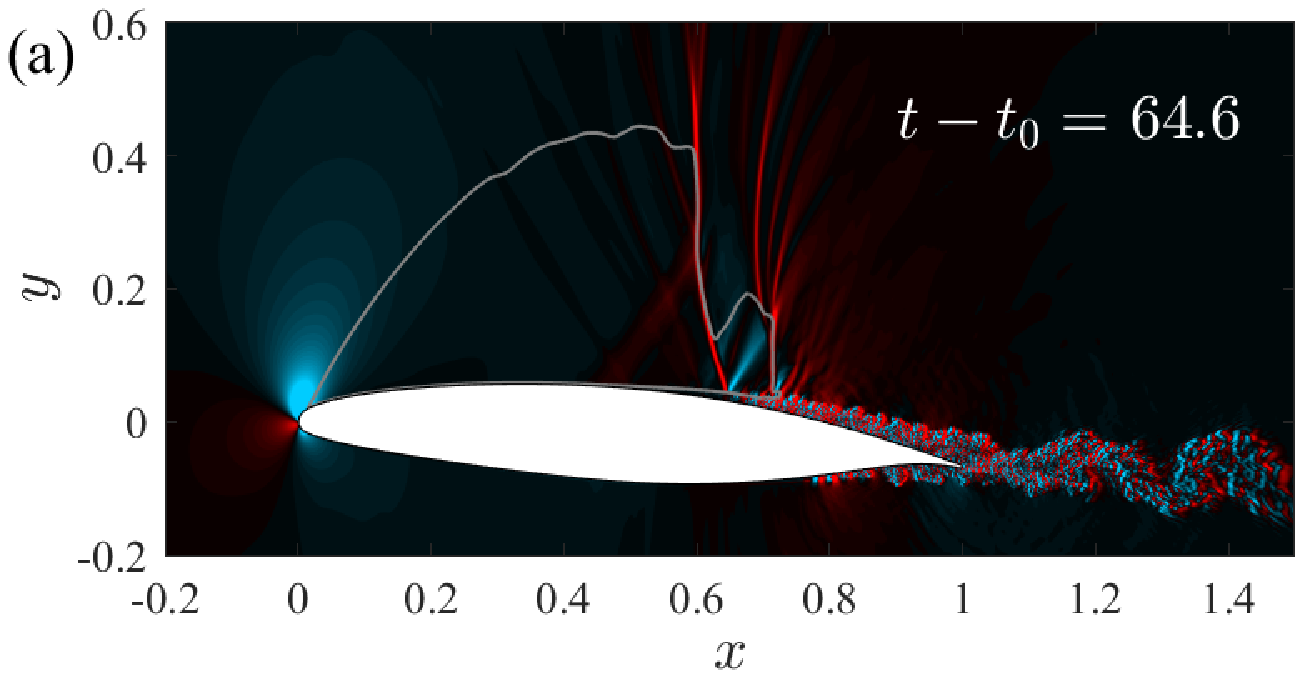}
\includegraphics[trim={0.3cm 1.5cm 0cm 2cm},clip,width=.495\textwidth]{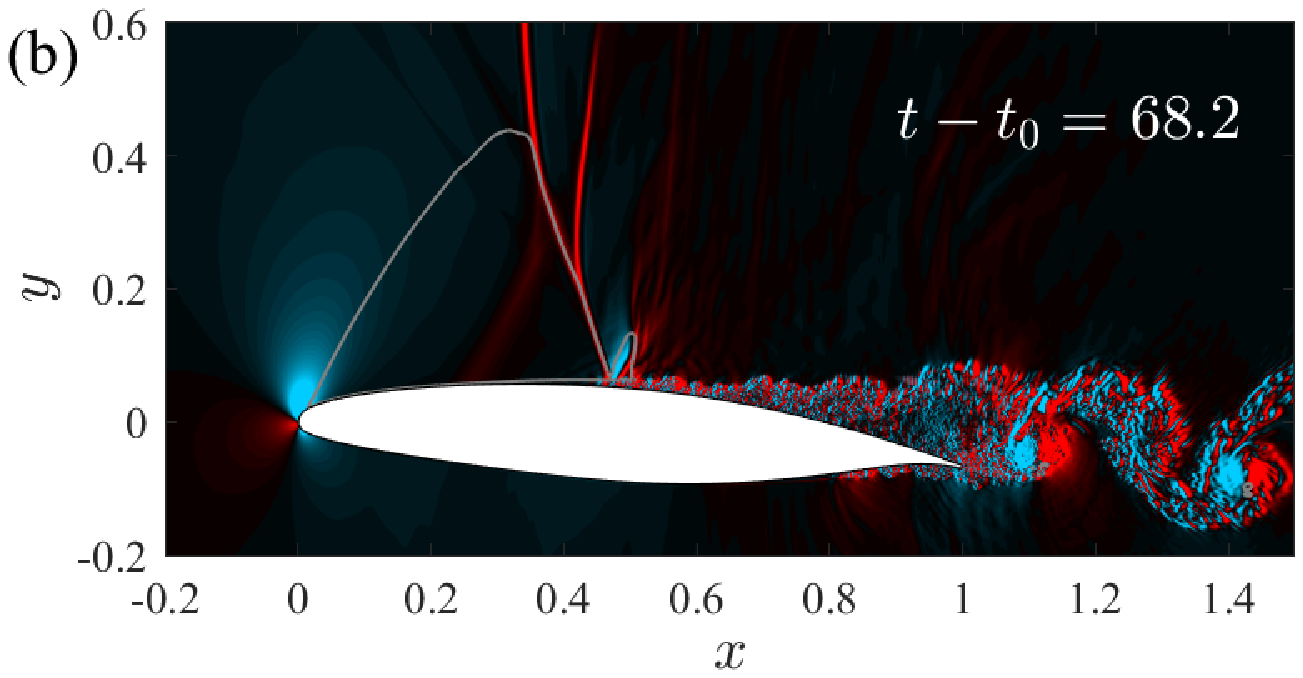}
\includegraphics[trim={0.3cm 1.5cm 0cm 2cm},clip,width=.495\textwidth]{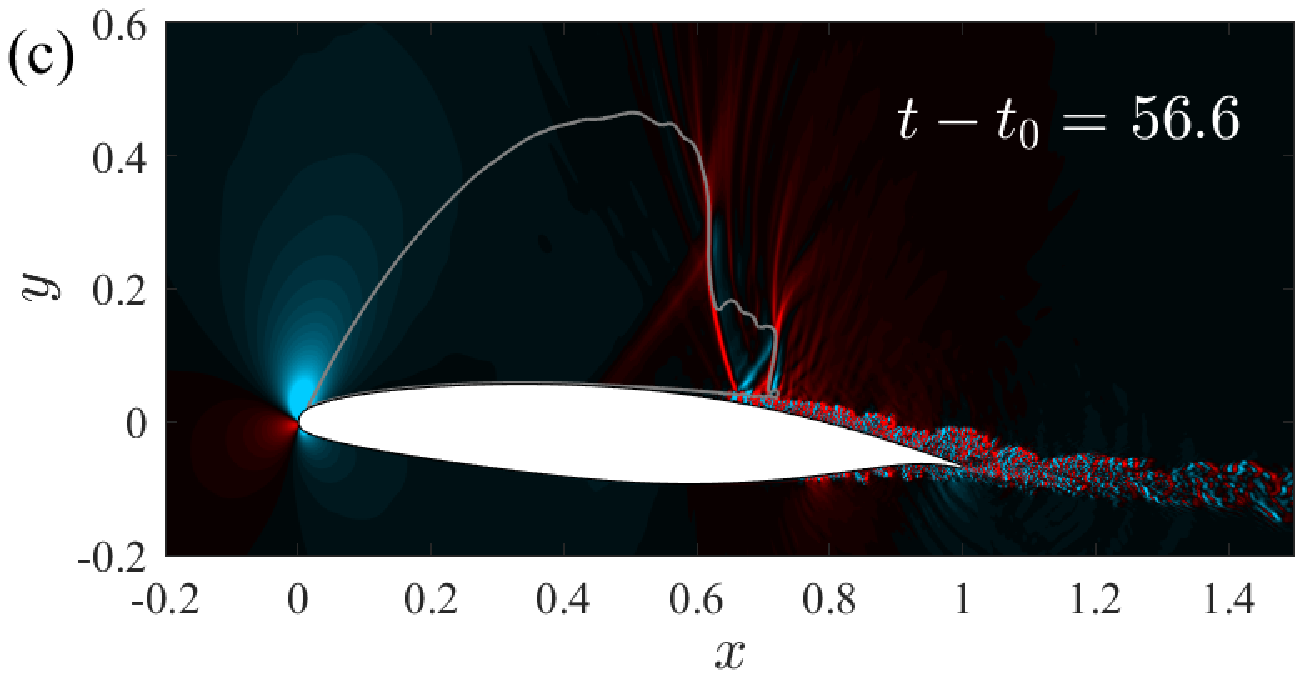}
\includegraphics[trim={0.3cm 1.5cm 0cm 2cm},clip,width=.495\textwidth]{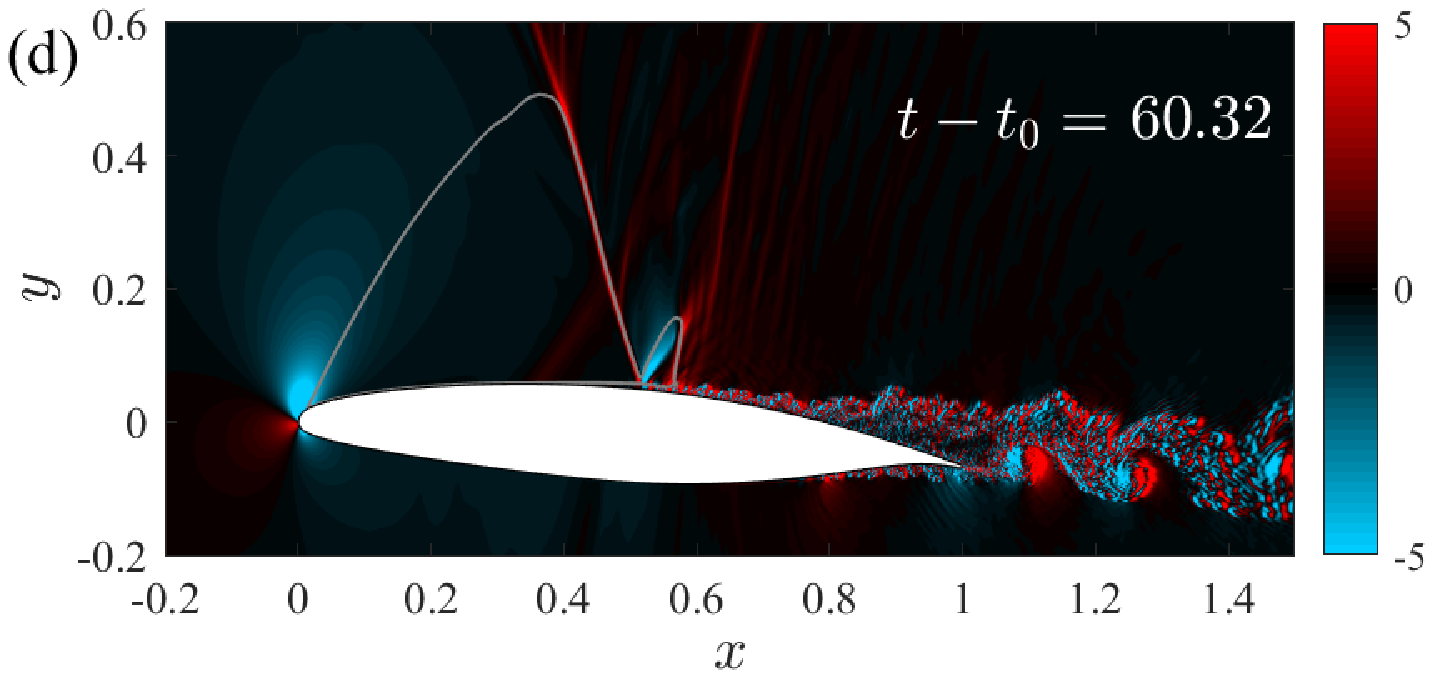}
\caption{Streamwise density gradient contours on the $x-y$ plane shown at high (left) and low (right) lift phases of the buffet cycle for the cases, $\Rey = 1 \times 10^6$ (top) and $1.5\times 10^6$ (bottom).}
    \label{fig:ReDensGrad}
\end{figure}

The streamwise density gradient contours for $\Rey = 1 \times 10^6$ and $1.5\times 10^6$ are shown in figure \ref{fig:ReDensGrad}. It is apparent from the plots that the supersonic to subsonic transition is dominated by a single shock wave structure, which is accompanied by only a small pocket of supersonic region downstream. Similar to the other cases, this dominant shock wave structure has an orientation of a negative slope in the low-lift phase. At all times, we observed shock wave structures to be strongly reduced in number in comparison to the reference case, with the supersonic region not having any strong pressure waves (compare with figure \ref{fig:RefDensGrad}). This suggests that with further increase in $\Rey$, there could be a critical $\Rey$ above which only a single shock wave is present at all times. This is consistent with other studies on buffet since most of these examine a higher $\Rey$ range than that studied here and report only a single shock wave.

\begin{figure} 
\centerline{\includegraphics[trim={0cm 0cm 1cm 0.5cm},clip,width=.495\textwidth]{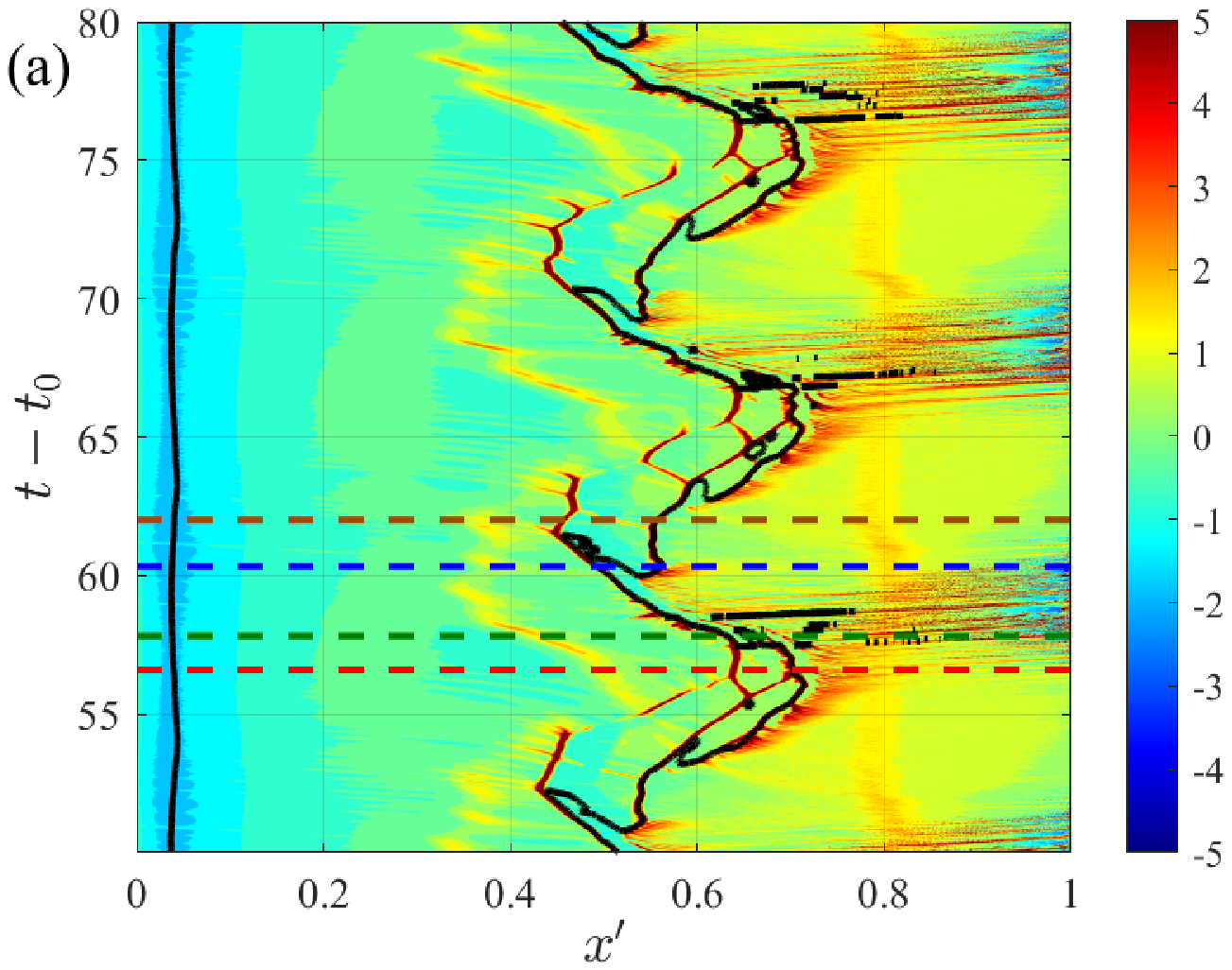}
\includegraphics[width=.495\textwidth]{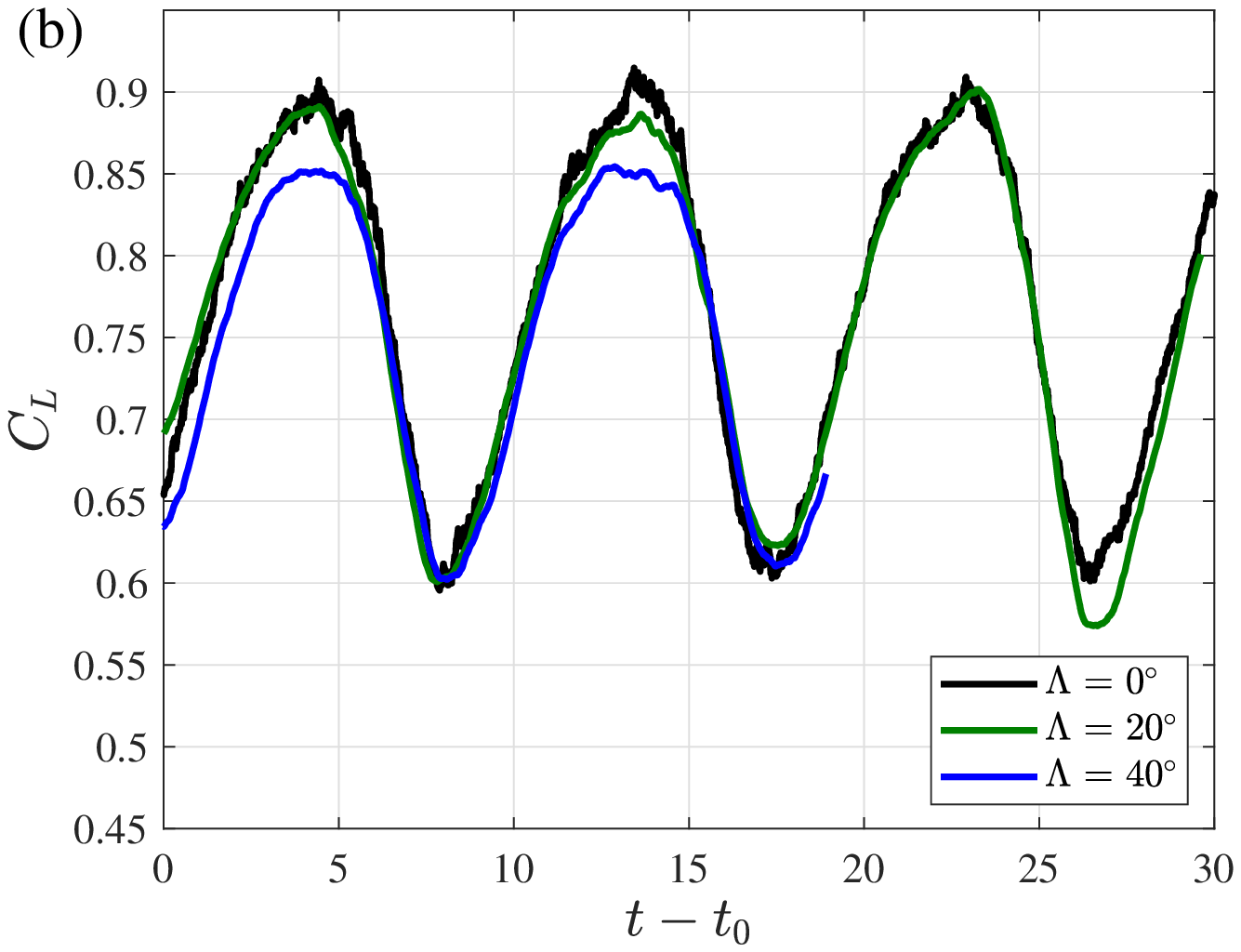}}
\caption{(a) Spatio-temporal variation of streamwise pressure gradient on the suction side of C5 for $\Rey = 1.5\times10^6$. (b) Temporal variation of $C_L$ past initial transients for various sweep angles simulated in the wide domain.}
\label{fig:R1p5e6XTDia}
\end{figure}

The spatio-temporal variation of the streamwise pressure gradient at the highest $\Rey$ simulated are shown in figure \ref{fig:R1p5e6XTDia}\textit{a}. The reduced number of shock wave structures is also apparent here. Also, these structures are seen to become pressure waves within a short distance from where they first appear, in contrast to the large upstream excursion seen for the other cases at the lower $\Rey = 5\times 10^5$ (cf. figures \ref{fig:RefXTDia}, \ref{fig:M0p735XTDia} and \ref{fig:AoA6XTDia}). The time required for pressure waves to propagate from the sonic envelope to the TE was found to be in the range $0.5 \leq t_\mathrm{down} \leq 0.7$. 

%%%%%%%%%%%%%%%%%%%%%%%%%%%%%%%%%%%%%%%%
\subsection{Effect of sweep}
\label{secSweep}
%%%%%%%%%%%%%%%%%%%%%%%%%%%%%%%%%%%%%%%%

The effect of sweep angle on flow features is considered in this section. The temporal variation of the lift coefficient past initial transients is shown in figure \ref{fig:R1p5e6XTDia}\textit{b} for the sweep angles simulated ($\Lambda = 0^\circ$ and $20^\circ$). Note that these were chosen based on the range in which buffet cells are reported in \citet{Iovnovich2015}. Since these simulations were carried out in the numerically expensive wide domain ($L_z = 1$), only a few buffet cycles were simulated. However, it is evident from the limited time simulated for that there is no drastic difference between the three cases, except for a small reduction in the maximum $C_L(t)$ attained. Indeed, as will be shown below, no dominant 3D structures appear in any part of the buffet cycle indicating the absence of buffet cells. With increasing sweep, the maximum lift achieved is slightly reduced, but the minimum lift and the frequency remain approximately the same. Thus, the addition of a strong spanwise velocity component ($U_z \approx 0.36U_\infty$ for $\Lambda = 20^\circ$ and 0.84$U_\infty$ for $\Lambda = 40^\circ$) does not have any significant effect on buffet. The present results differ from those reported in \citet{Iovnovich2015} for infinite swept wings with non-periodic lateral boundary conditions, where a monotonic increase in frequency was observed with increasing sweep angle, including in ranges of $\Lambda$ where no buffet cells occur.

\begin{figure} 
\centering
\includegraphics[trim={1cm 0cm .7cm 0cm},clip,width=.495\textwidth]{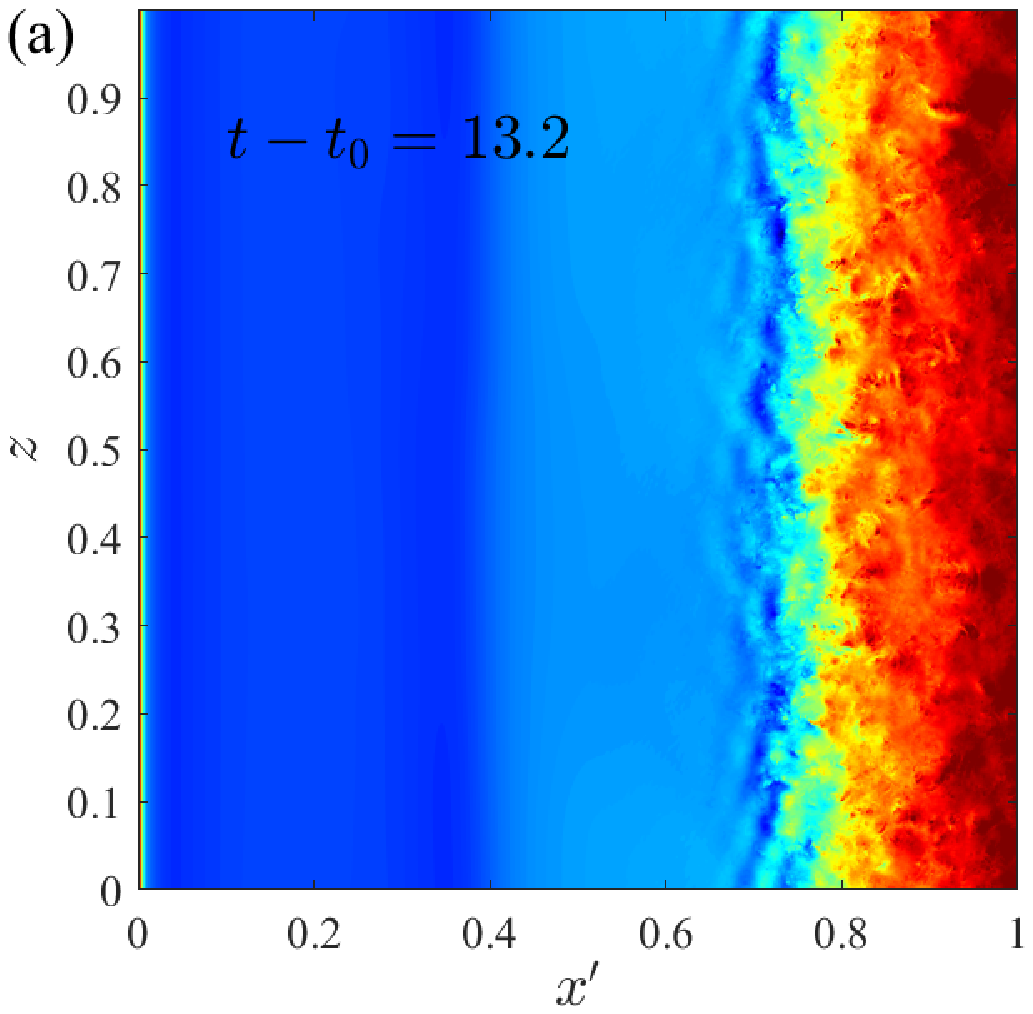}
\includegraphics[trim={1cm 0cm .7cm 0cm},clip,width=.495\textwidth]{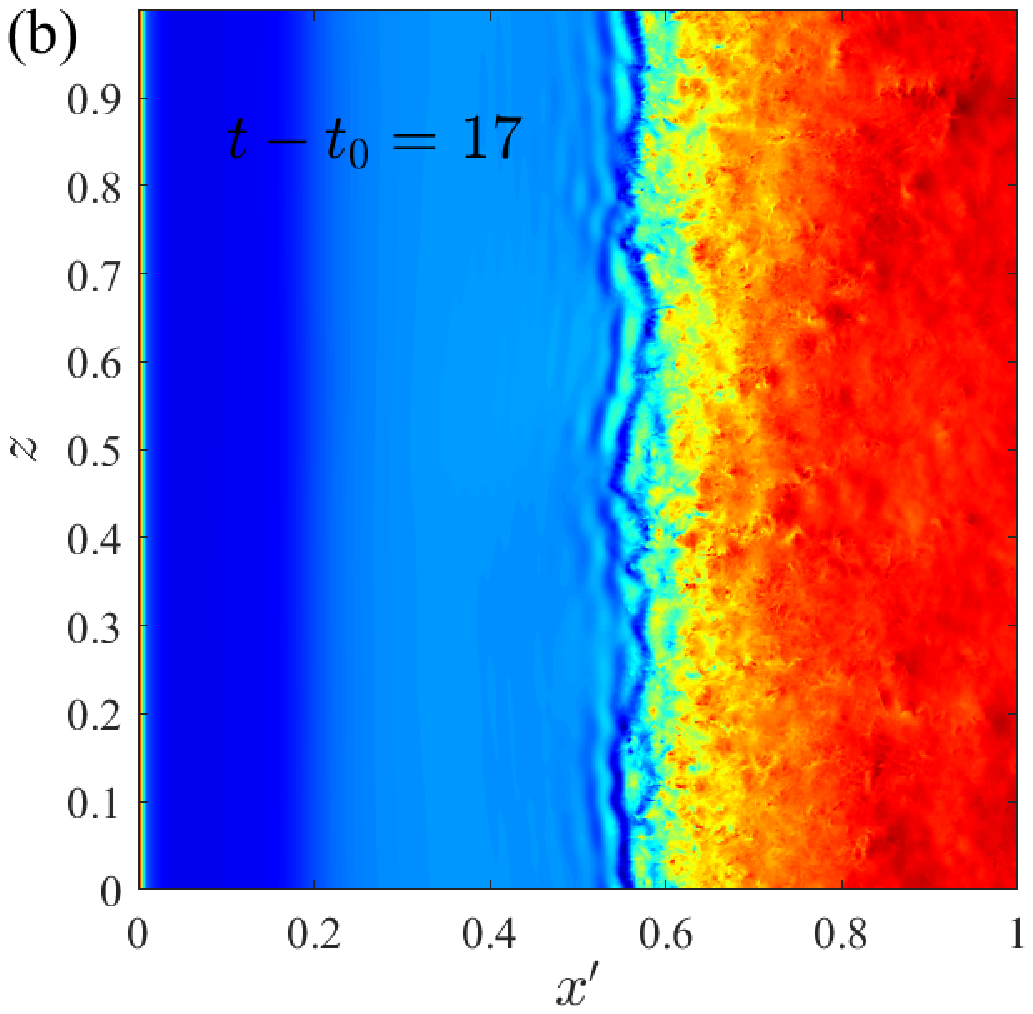}
\includegraphics[trim={1cm 0cm .7cm 0cm},clip,width=.495\textwidth]{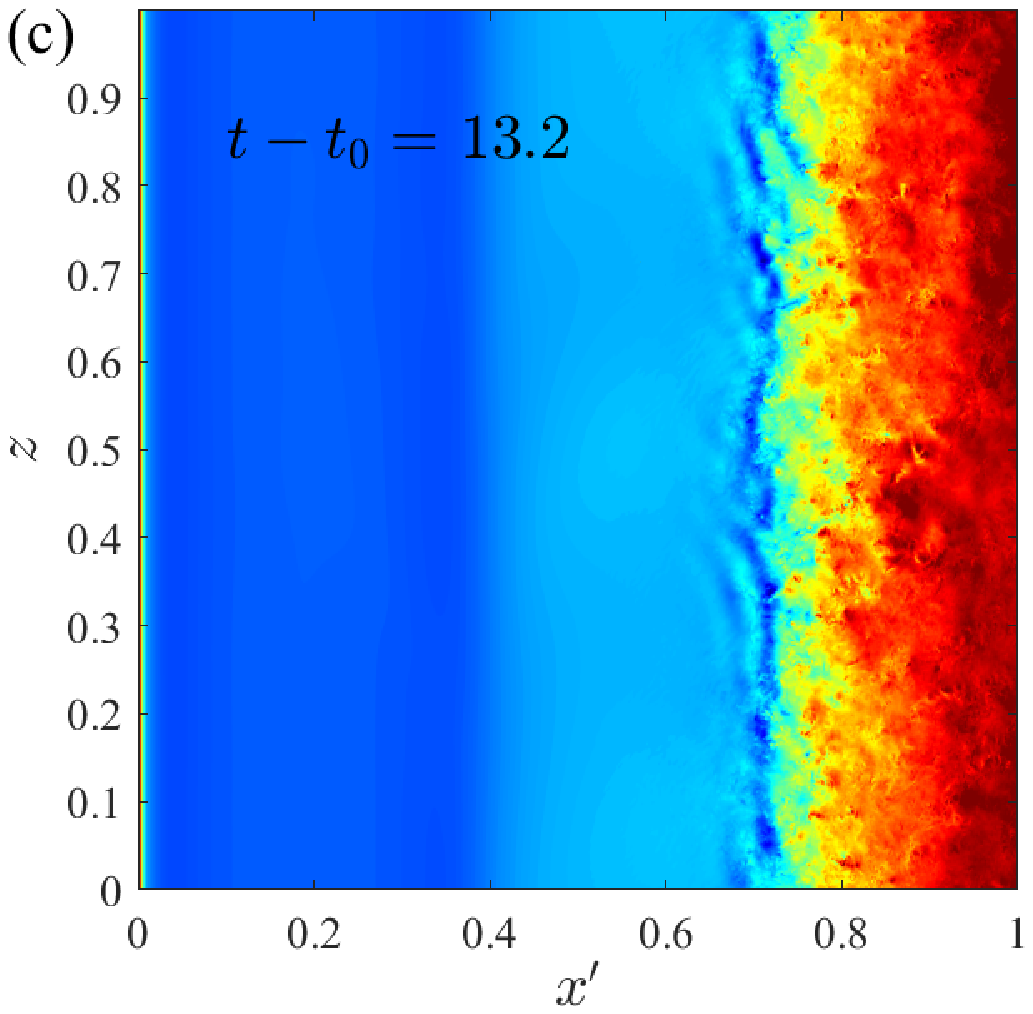}
\includegraphics[trim={1cm 0cm .7cm 0cm},clip,width=.495\textwidth]{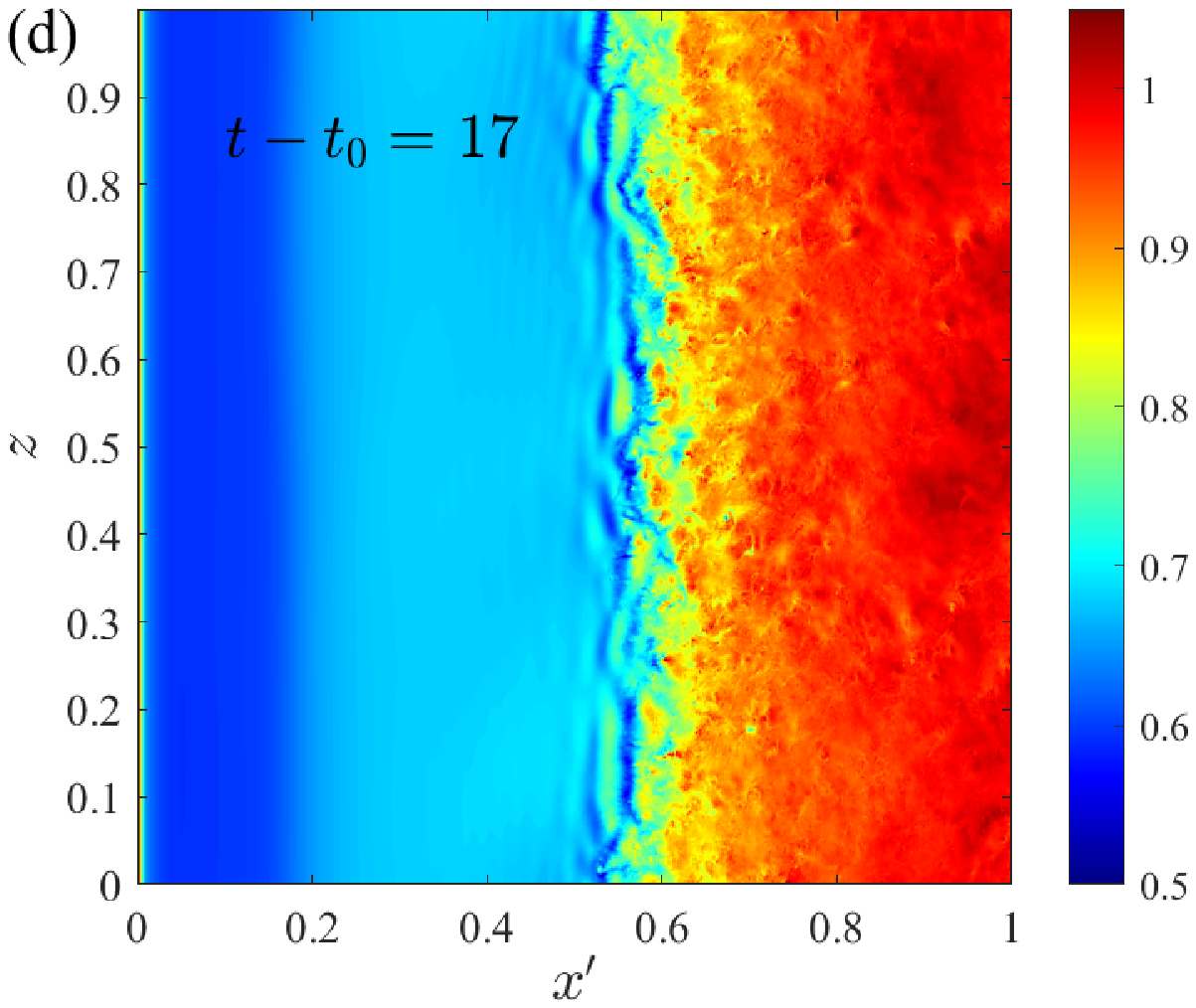}
\caption{Contours of density on the aerofoil surface shown at high- (left) and low-lift (right) phases of the buffet cycle for swept cases, $\Lambda = 20^\circ$ (top) and $\Lambda = 40^\circ$ (bottom).}
    \label{fig:SwWDensAir}
\end{figure}

The spanwise flow features are highlighted for the swept wing cases in figure \ref{fig:SwWDensAir} using contours of density on the aerofoil surface. Weak 3D features seem to appear in the supersonic region of $0.5 \leq x' \leq 0.7$ in the high-lift phase and $0.4 \leq x' \leq 0.6$ in the low-lift phase, although they remain too weak to affect the shock wave structure or the transition features. These low-energy structures were found to convect along the spanwise direction. Thus, we conclude that for the present configuration, buffet is essentially a 2D phenomenon in the range of sweep angles studied. From a visual examination of the flow fields and comparisons of the aerofoil coefficients for different $\Lambda$, we also conclude that the Independence Principle \citep{Selby1983}, which states that the two-dimensional flow features (including the pressure) are independent of the spanwise velocity component (see  \S\ref{secMethod}), remains a good approximation in the parametric range examined.

%%%%%%%%%%%%%%%%%%%%%%%%%%%%%%%%%%%%%%%%
\section{Modal decomposition and reconstructed flow fields}
\label{secSPOD}
%%%%%%%%%%%%%%%%%%%%%%%%%%%%%%%%%%%%%%%%
The presence of multiple shock wave structures, a turbulent boundary layer and vortices leads to a complex flow field that makes it difficult to understand their individual characteristics. To overcome this issue, we present a modal decomposition using SPOD and study the coherent flow features individually. As will be shown, the two dominant coherent structures are the buffet and wake modes associated with the low frequency shock oscillations and high frequency vortex shedding, respectively. A flow field reconstruction based on each of these modes is used to characterise their influence on the dynamics of shock wave boundary layer interactions, which also allows us to examine the different models proposed to explain buffet. For brevity, only the reference case and one case from each parametric variation will be reported below. Among these, we focus mainly on the former and additionally, $\alpha = 6^\circ$, for which the buffet amplitude is maximum. The cases where shock waves appear on both aerofoil surfaces are examined separately in  \S\ref{subSecSPODHighM}.

\subsection{Features of modes}
\label{subSecSPODModeFeatures}

\subsubsection{Temporal features}
\label{subsubsecSPODTemporal}
\begin{figure} 
\centerline{\includegraphics[width=.45\textwidth]{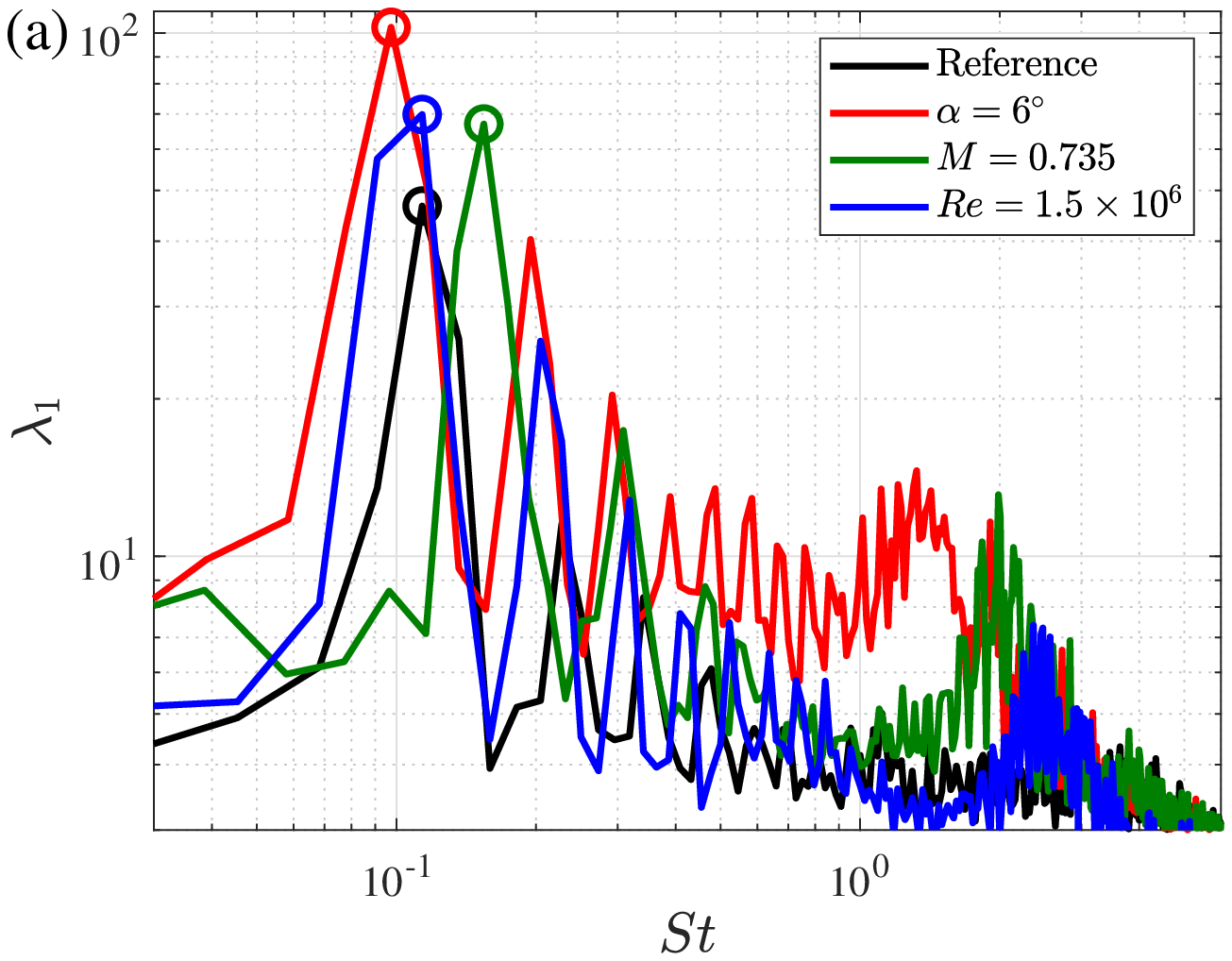}
\includegraphics[width=.5\textwidth]{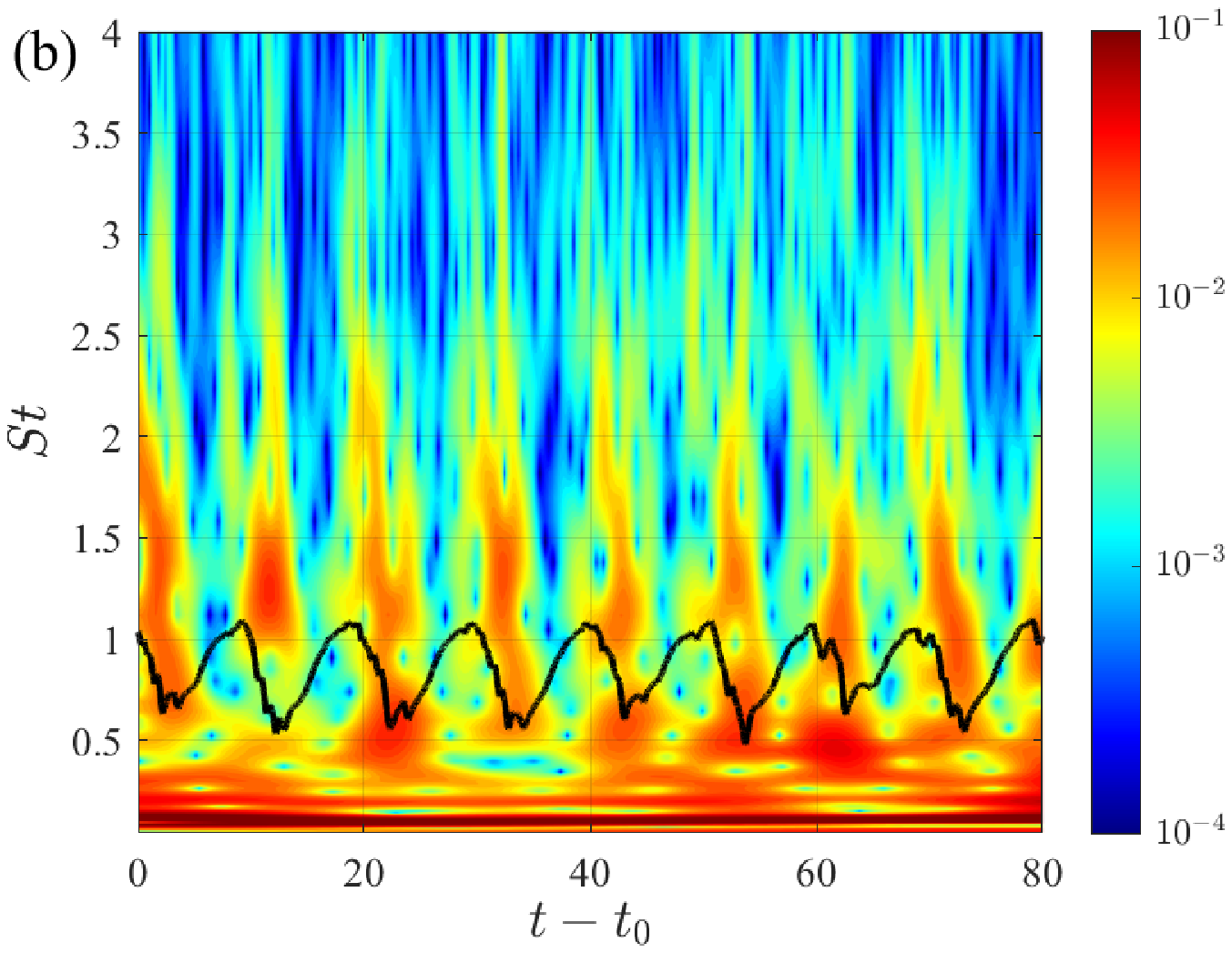}}
\caption{(a) Eigenvalue spectra (logarithmic scale) based on SPOD of the dominant eigenvalue for the reference case and a typical case for each parameter varied. (b) Scalogram based on the lift coefficient ($\log_{10} |W(C_L')|$) with the temporal variation of the same overlaid (black curve) for the case of $\alpha = 6^\circ$.}
\label{fig:RefSPODSpectrum}
\end{figure}

A comparison of the spectra for the dominant eigenvalue, $\lambda_1$, obtained through SPOD for the reference case and one case selected from each parameter variation is shown in Fig \ref{fig:RefSPODSpectrum}\textit{a}. For all cases, we see peaks at a fundamental frequency of $St \approx 0.1$, \textit{i.e.}, the buffet frequency, and additionally, peaks at its harmonics. Significant energy content is also present in a medium frequency range of $1 \leq St \leq 4$ associated with what we will refer to as wake modes. These will be shown below to be associated with vortex shedding  (see figure \ref{fig:RefSPODModes}\textit{c}). A bump in the spectrum similar to that seen here and associated with vortex shedding was reported in the detached eddy simulations of the OAT15A aerofoil in \citet{Grossi2014} and organized eddy simulations of \citet{Szubert2015}. We have also extracted a similar mode at $St \approx 1.87$ for the reference case using dynamic mode decomposition in our previous study \citep{Zauner2020a}. Note that the wake modes were predicted based on a RANS base flow field in \citet{Sartor2015} using resolvent analysis as the most amplified optimal input other than the global mode. Although not frequently reported in experiments, \citet{Szubert2015} have shown that phase-averaging implemented in experiments to capture buffet features can prevent the effects of such modes from being observed. From the figure, it is evident that a change in any parameter ($M/\alpha/\Rey$) leads to an increase in the energy content of the wake modes, indicating a stronger vortex shedding beyond buffet onset. With increasing $\alpha$, the frequency associated with these modes seems to be shifted to lower values. 

To examine for temporal variations in the intensities and frequencies of these modes, a time-frequency analysis using the continuous wavelet transform was also carried out. The scalogram based on the transform of the lift coefficient's fluctuating component, $W(C_L')$, is shown  for the case of $\alpha = 6^\circ$ in figure \ref{fig:RefSPODSpectrum}\textit{b}. Note that the contours are based on the logarithm of the magnitude, \textit{i.e.}, $\log_{10} |W(C_L')|$. The temporal variation of $C_L$ is also overlaid to identify high and low lift phases. A horizontal band in the contour at the buffet frequency ($St \approx 0.1$) confirms that it does not vary with time. However, it is also evident that there are periodic changes of $|W(C_L')|$ in the frequency range $1 \leq St \leq 3$. As seen from figure \ref{fig:RefSPODSpectrum}\textit{a}, this is associated with the wake modes for this case. It is evident from the figure that the intensity of these modes is approximately a maximum in the low-lift phase and vice versa. This is expected based on the vortex shedding behaviour reported in \S\ref{secAoA} (see figure \ref{fig:AoA5DensGrad}). Similar results have been reported for turbulent buffet in \citet{Szubert2015}.

\subsubsection{Spatial features}

\begin{figure} 
\centering
\includegraphics[trim={1.6cm 0cm 2.7cm 0cm},clip,width=.32\textwidth]{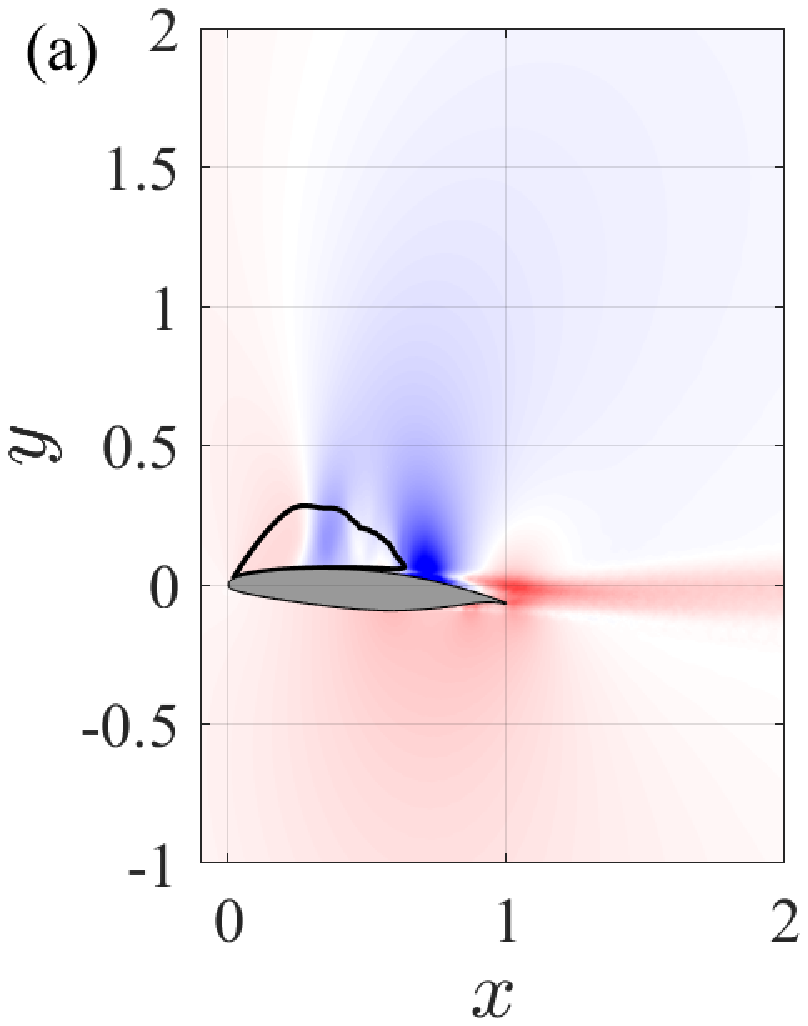}
\includegraphics[trim={1.6cm 0cm 2.7cm 0cm},clip,width=.32\textwidth]{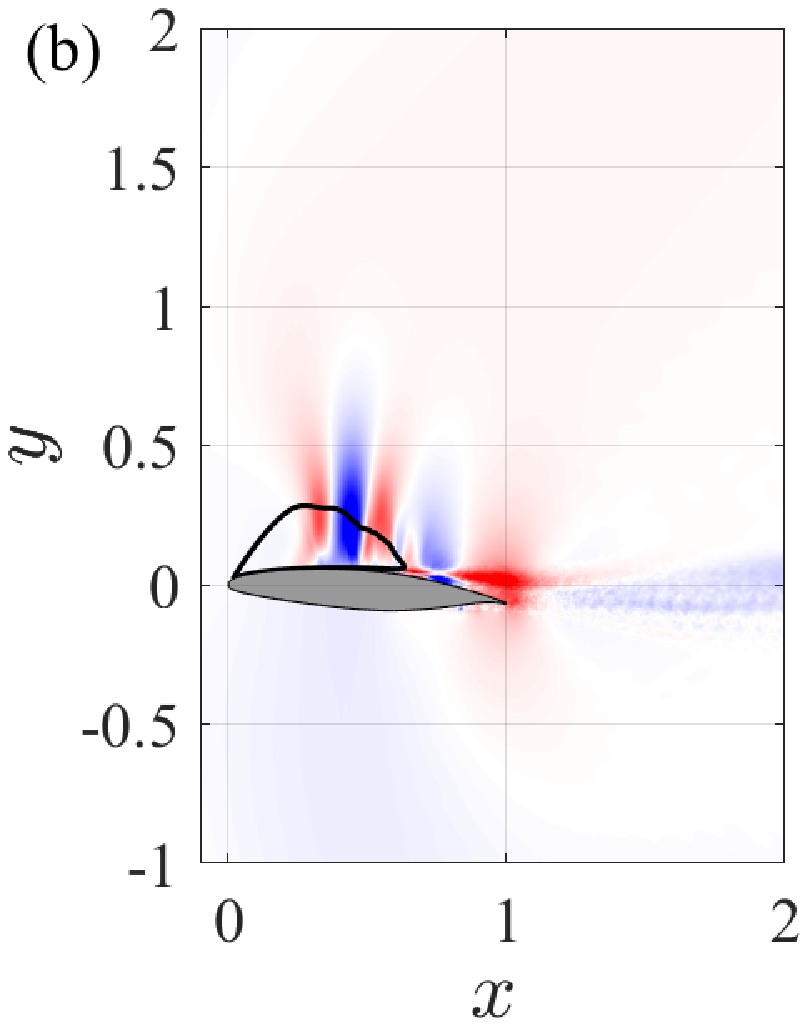}
\includegraphics[trim={1.6cm 0cm 2.7cm 0cm},clip,width=.32\textwidth]{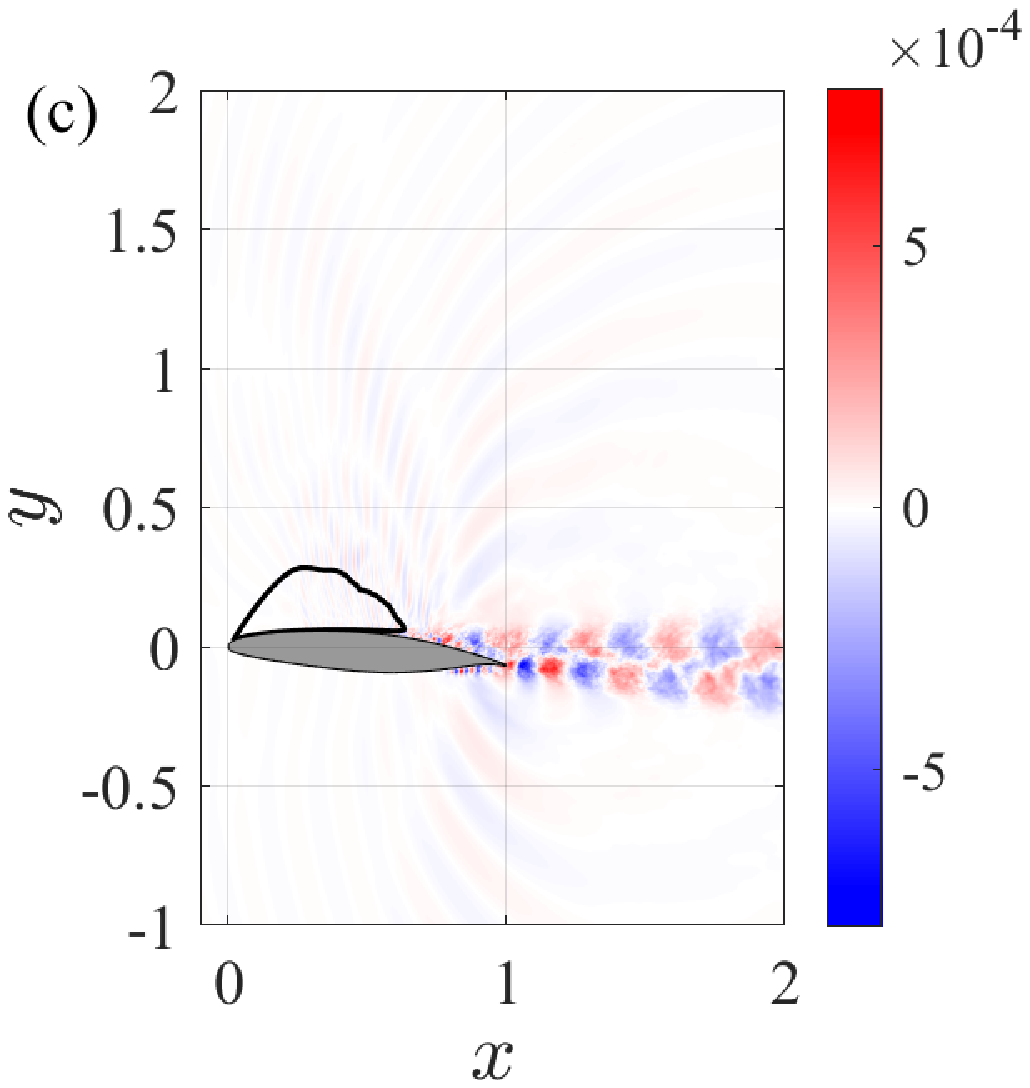}
\caption{Eigenmodes extracted using SPOD for the reference case shown using contour plots of real part of the density field: (a) buffet mode ($St = 0.13$), (b) its first harmonic ($St = 0.27$) and (c) a wake mode ($St = 2.57$).}
    \label{fig:RefSPODModes}
\end{figure}

The spatial structure of the eigenmodes of importance are shown for the reference case in figure \ref{fig:RefSPODModes} using the real part of the density field. From left to right, these are the buffet mode ($St = 0.13$), its first harmonic ($St = 0.27$) and a typical wake mode ($St = 2.57$), where the bump in the spectra attains a local maximum). The sonic line based on the time-averaged local Mach number (solid curve) is shown for reference. For the buffet mode we see that strong density fluctuations are present in the vicinity of the trailing edge and near-wake which are out of phase (\textit{i.e.} opposite sign) with those near the shock foot. Thus, the density increases at the shock foot when it reduces at the TE and in the wake and vice versa. The harmonic mode is of a shorter wavelength relative to the buffet mode. %and might be related to the generation of multiple shock structures.
Interestingly, both these modes show large density variations in the supersonic regions in the fore part of the aerofoil. This should be contrasted with the global mode features reported for the turbulent buffet in \citet[figure 8]{Crouch2009} and \citet[figure 12]{Sartor2015}, where the variations are localised to the vicinity of the shock wave and the shear layer downstream. These turbulent buffet mode features do not change significantly irrespective of where the transition to turbulence occurs (fully turbulent conditions or forced transition at any finite distance upstream of the shock foot), as long as the BL is turbulent at the shock foot \citep[][figure 5]{Garbaruk2021}. Therefore, whether the BL is still laminar at the shock foot seems to be of significance. 

The wake mode has a structure similar to that of the von K\'arma\'n vortex street. In addition to this, waves in the flow field that seem to be generated at the TE and propagating upstream outside the BL on both sides of the aerofoil can be identified. These appear to be the Kutta waves suggested in \citet{Lee1990}. Based on visualisations of the time evolution of the wake mode, the upper limit on the time required for these waves to reach the shock wave starting from the TE was estimated as $t_\mathrm{up} < 1.5$. Thus, for the reference case, Lee's model predicts the buffet time period as $\tau_\mathrm{Lee} = t_\mathrm{up}+t_\mathrm{down} < 2.1$, which is substantially lower than the actual buffet period of $\tau_\mathrm{LES} =8$ observed in the LES (see \S\ref{secRef}). Similar results were observed for other wake modes associated with the spectral bump and for all other cases where buffet was observed, suggesting that Lee's model is invalid. This is further considered in \S\ref{subsecDiscLee}. % 2do: confirm if TBlock=5 gives same result % 

The modification to Lee's model proposed in \citet{Jacquin2009} assumes that $t_\mathrm{up}$ should be replaced with $t^J_\mathrm{up}$, the time required for the waves to travel upstream along the pressure surface and turn around the leading edge to reach the shock foot. Such upward propagating waves are also seen in figure \ref{fig:RefSPODModes}\textit{c} on the pressure side of the aerofoil. However, we observed that they reduce in intensity by more than an order of magnitude upstream of mid chord while they're unidentifiable past the leading edge in the supersonic region, likely because their intensity is reduced to levels similar to that of noise in the SPOD modes. By using the same approximation as \citet{Jacquin2009} that the time spent in the supersonic region is negligible, we observed $t^J_\mathrm{up} < 3$ implying that the predicted time period is $\tau_J = t^J_\mathrm{up}+t_\mathrm{down} < 3.6$ which still remains significantly lower than $\tau_\mathrm{LES}$.

\begin{figure} 
\centering
\includegraphics[trim={1.6cm 0cm 2.7cm 0cm},clip,width=.32\textwidth]{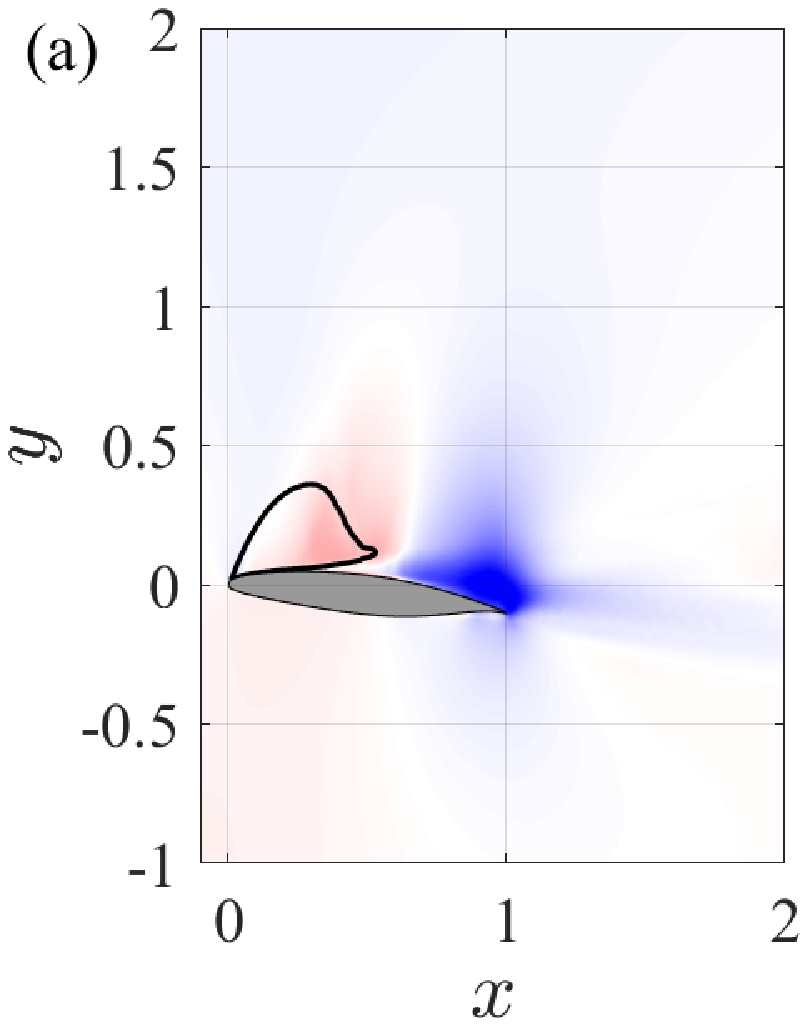}
\includegraphics[trim={1.6cm 0cm 2.7cm 0cm},clip,width=.32\textwidth]{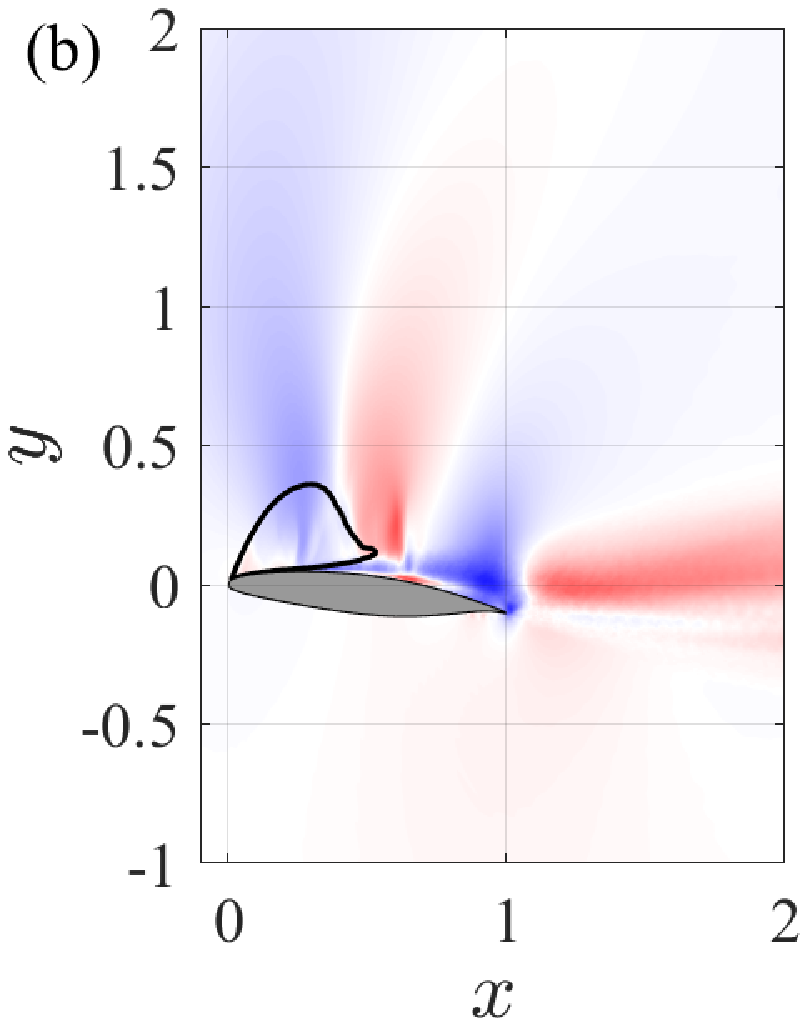}
\includegraphics[trim={1.6cm 0cm 2.7cm 0cm},clip,width=.32\textwidth]{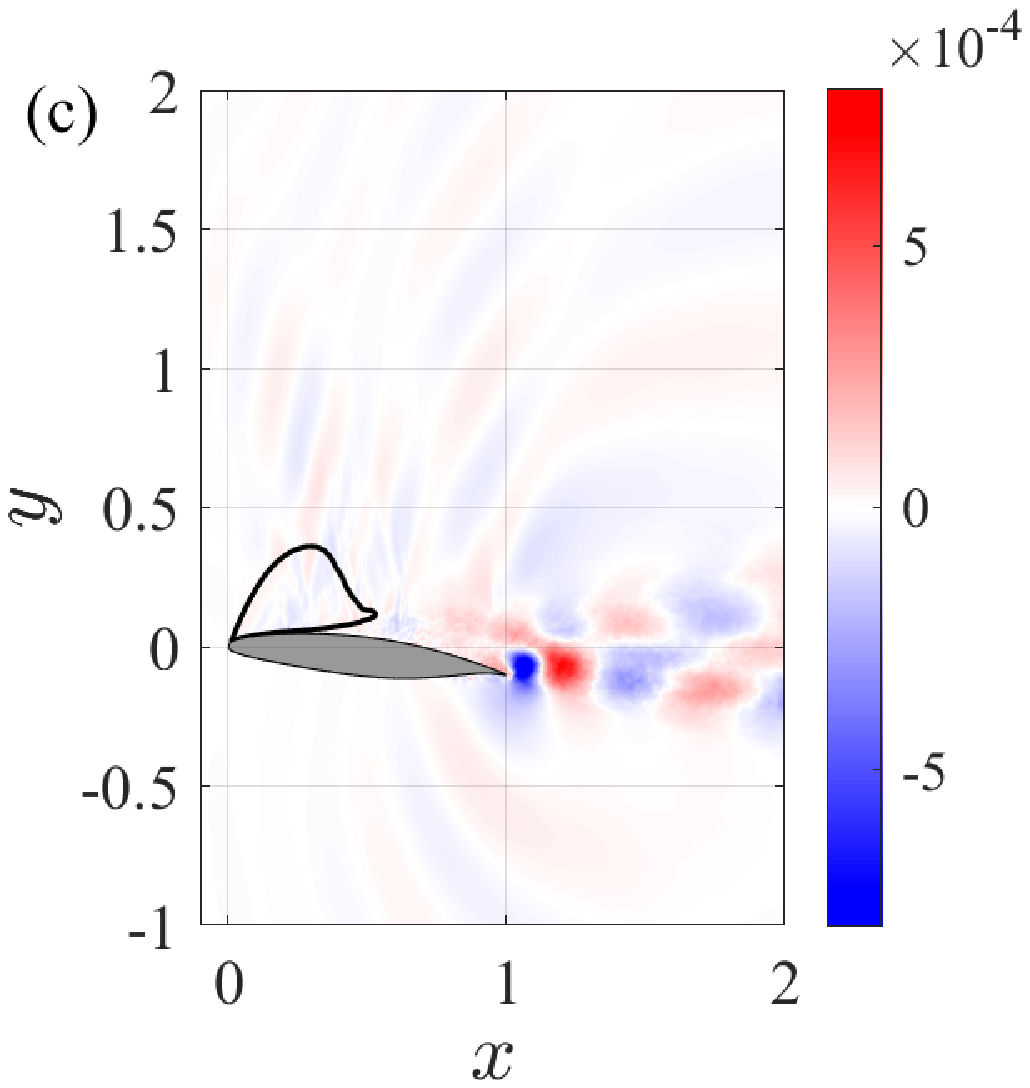}
\caption{Eigenmodes extracted using SPOD for the $\alpha = 6^\circ$ shown using contour plots of real part of the density field: (a) buffet mode ($St = 0.1$), (b) its first harmonic ($St = 0.2$) and (c) a wake mode ($St = 1.3$).}
    \label{fig:SPODModesAoA6}
\end{figure}

Features of the corresponding SPOD modes for $\alpha = 6^\circ$ are shown in figure \ref{fig:SPODModesAoA6}. We see that the wave number associated with each mode is substantially lower with all modes showing a larger spatial structure. A relatively strong separation in the vicinity of the shock foot is apparent, while the wake mode's energy is high at the TE. These results match with the flow features observed in the simulations (cf. figure \ref{fig:RefDensGrad} and \ref{fig:AoA5DensGrad}). Similar to the reference case, upstream propagating waves emanating from the TE can be observed for the wake mode. Based on this, it was estimated that $t_\mathrm{up} < 2$ and $t^J_\mathrm{up} < 3$ implying that $\tau_\mathrm{Lee} < 3.1$ and $\tau_J < 4.1$ compared to $\tau_\mathrm{LES} = 9.9$. 

%%%%%%%%%%%%%%%%%%%%%%%%%%%%%%%%%
\subsection{Modal reconstruction}
%%%%%%%%%%%%%%%%%%%%%%%%%%%%%%%%%

The features of the buffet and wake modes are further scrutinised here by performing a flow field reconstruction, superposing each mode separately with the mean flow field. All quantities described here (lift, velocity, local Mach number, \textit{etc.}) are based on the reconstructed flow field. As noted in  \S\ref{subsubSecSPODImpl}, the temporal variation is described using the phase, $\phi$, with $\phi = 0^\circ$ and $180^\circ$ signifying the phases at which maximum and minimum lift (reconstructed) are attained, respectively.
% Since the maximum shock excursion occurs for the case of $\alpha = 6^\circ$, this will be the focus of the following discussion.

\subsubsection{Buffet mode}
\label{subsubSecFlowRecBuffet}
\begin{figure} 
\centering
\includegraphics[trim={0cm 1.5cm 0cm 2cm},clip,width=.495\textwidth]{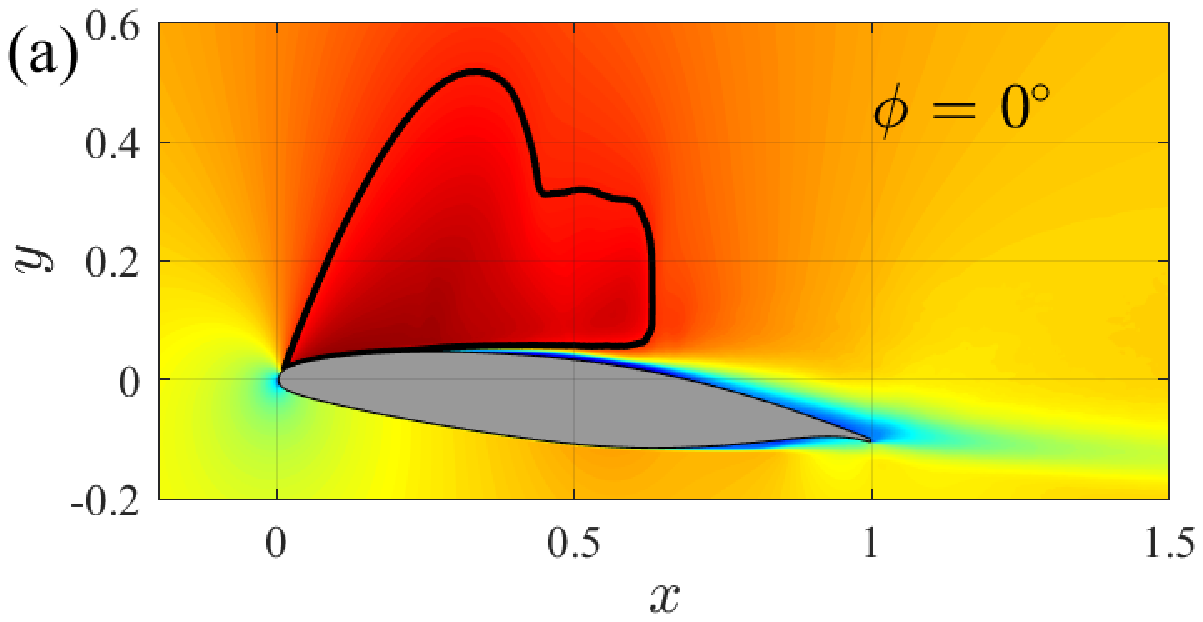}
\includegraphics[trim={0cm 1.5cm 0cm 2cm},clip,width=.495\textwidth]{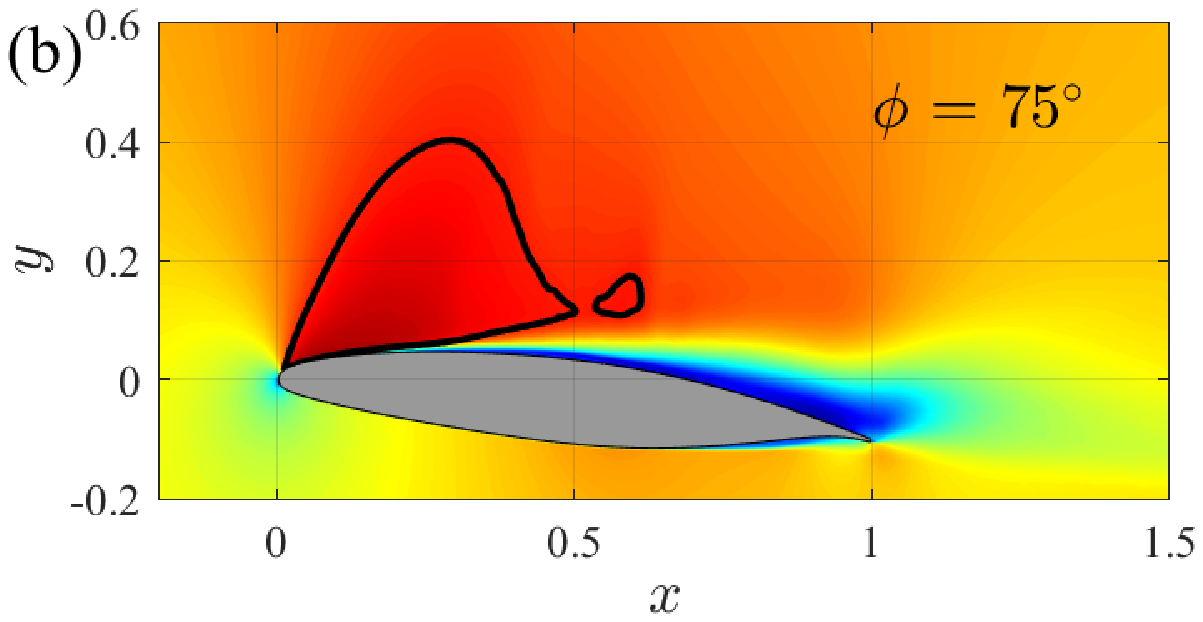}
\includegraphics[trim={0cm 1.5cm 0cm 2cm},clip,width=.495\textwidth]{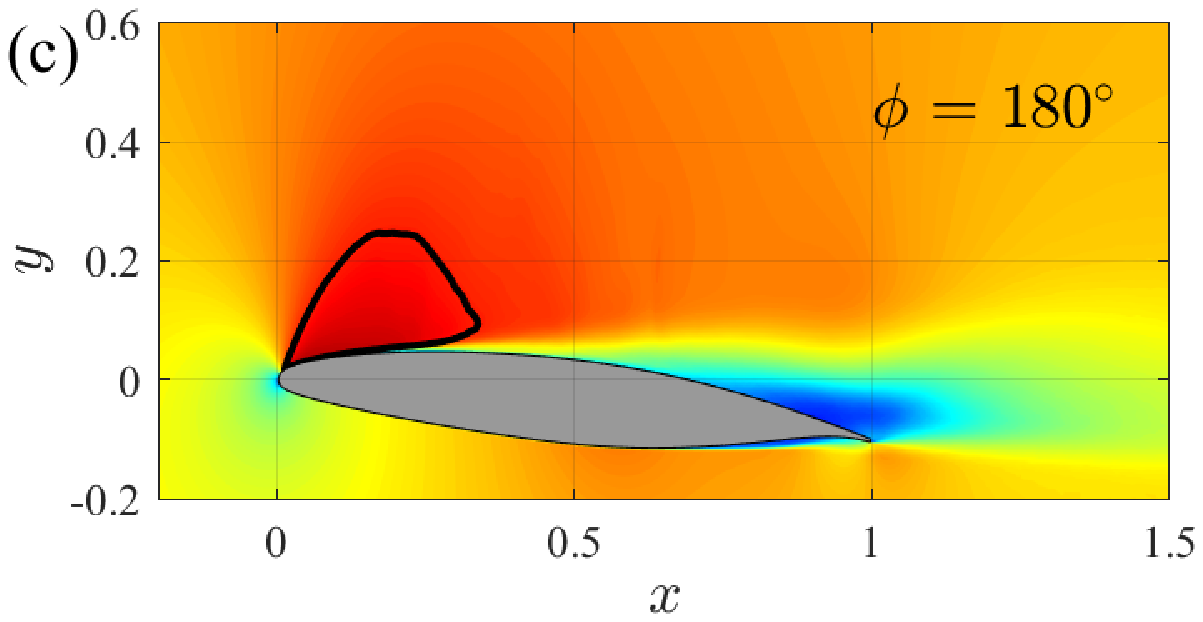}
\includegraphics[trim={0cm 1.5cm 0cm 2cm},clip,width=.495\textwidth]{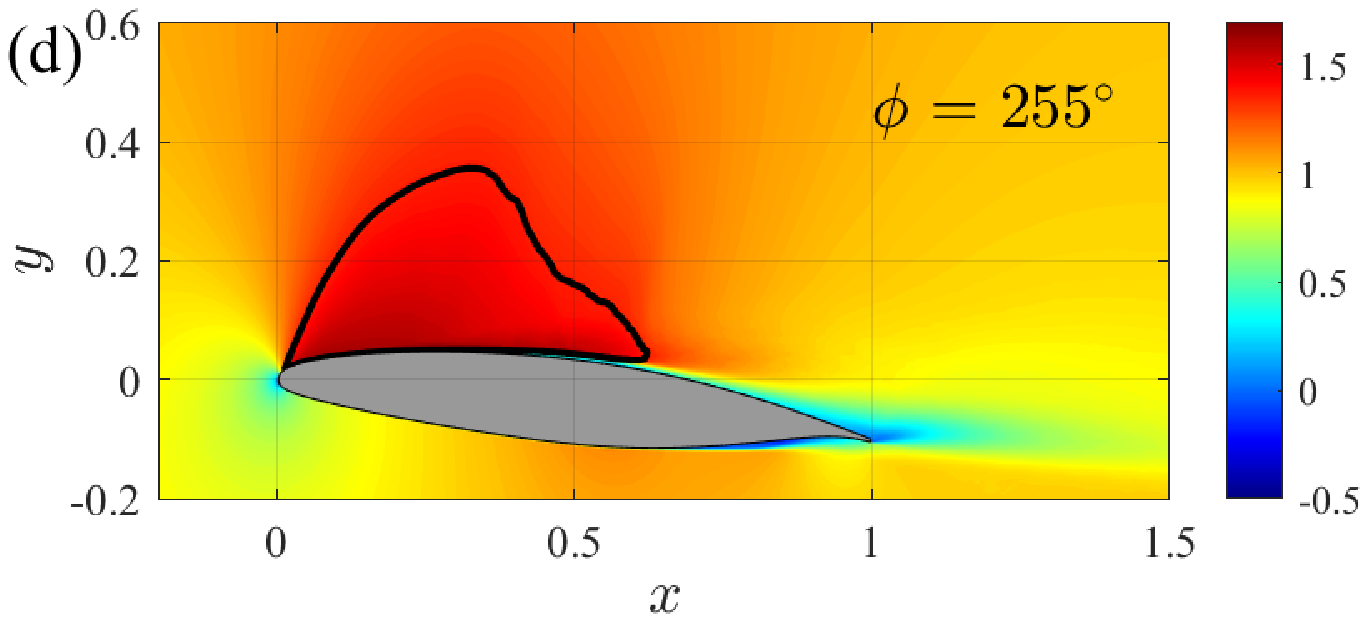}
\caption{Reconstructed flow field based on the buffet mode for $\alpha = 6^\circ$ shown using axial velocity contour at (a) high-lift, (b) low-skin-friction-drag, (c) low-lift, and (d) high-skin-friction-drag phases.}
    \label{fig:SPODReconstr2DAoA6}
\end{figure}

 Contours of axial velocity at different phases are shown in  figure \ref{fig:SPODReconstr2DAoA6} for $\alpha = 6^\circ$. For this case, the phases of $\phi = 73^\circ$ and $253^\circ$ represent the minimum and maximum skin friction drag, respectively. In contrast to the actual flow field in which multiple shock structures are present, the reconstructed flow field contains only a single supersonic region enclosed by the sonic envelope. This is expected, since these multiple shock structures arise at a frequency higher than the buffet frequency (see figure \ref{fig:RefXTDia}). For convenience, the sonic envelope's downstream edge is referred to as a shock wave. The presence of an attached BL downstream of the shock wave when the latter is at its most downstream position (maximum lift phase) can be discerned. A subsequent separation (minimum skin-friction phase) and eventual reattachment as the shock wave moves downstream is also observed. As the shock wave moves upstream, we see that at $\phi = 73^\circ$, the flow is fully separated beyond its foot and up to the TE. As it reaches its most upstream position, the BL is fully attached downstream to the shock wave. We found these features to be common to all cases considered. 

\iffalse 2do: If time permits, add streamlines.\fi

\begin{figure} 
\centering
\includegraphics[width=.495\textwidth]{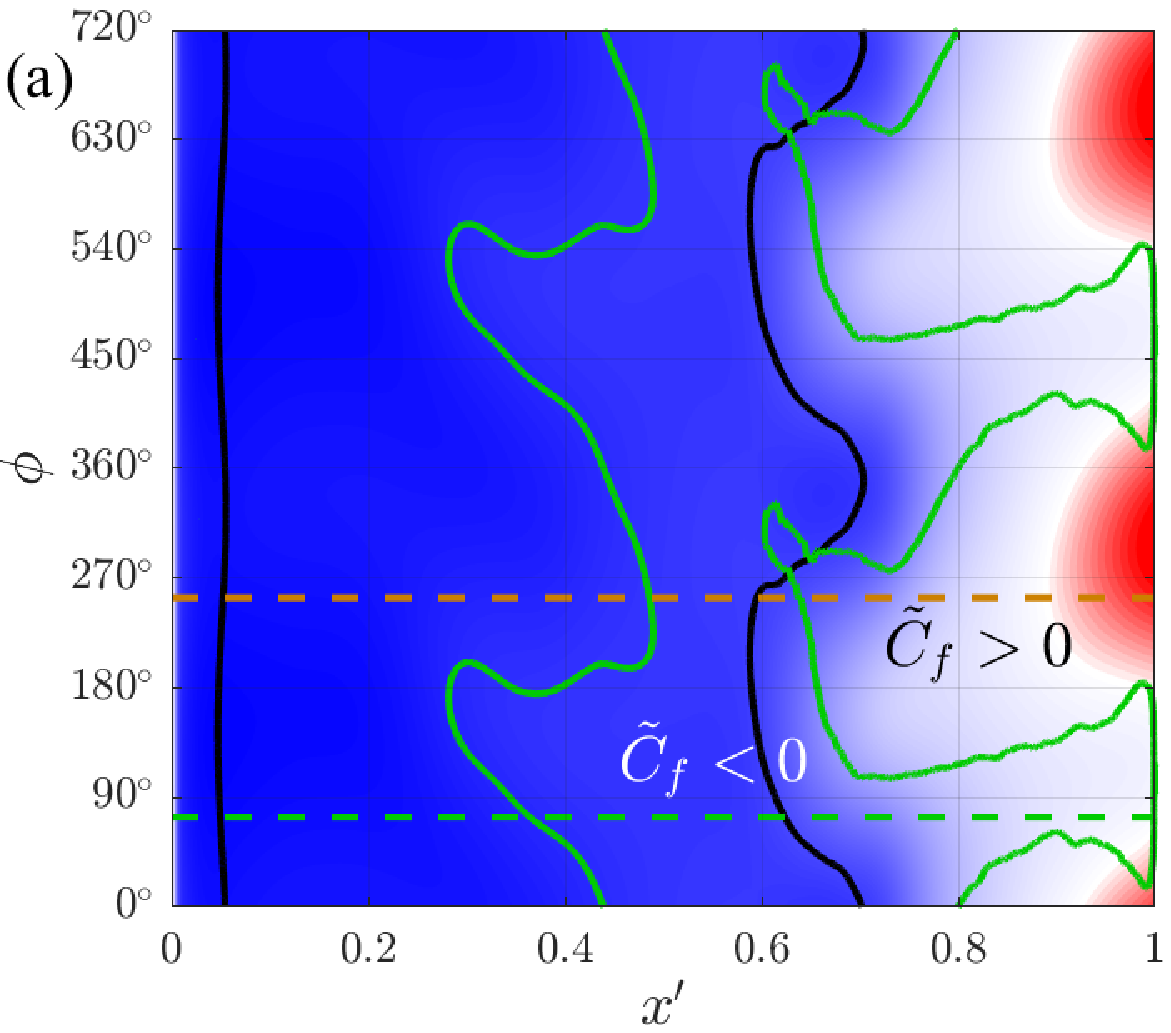}
\includegraphics[width=.495\textwidth]{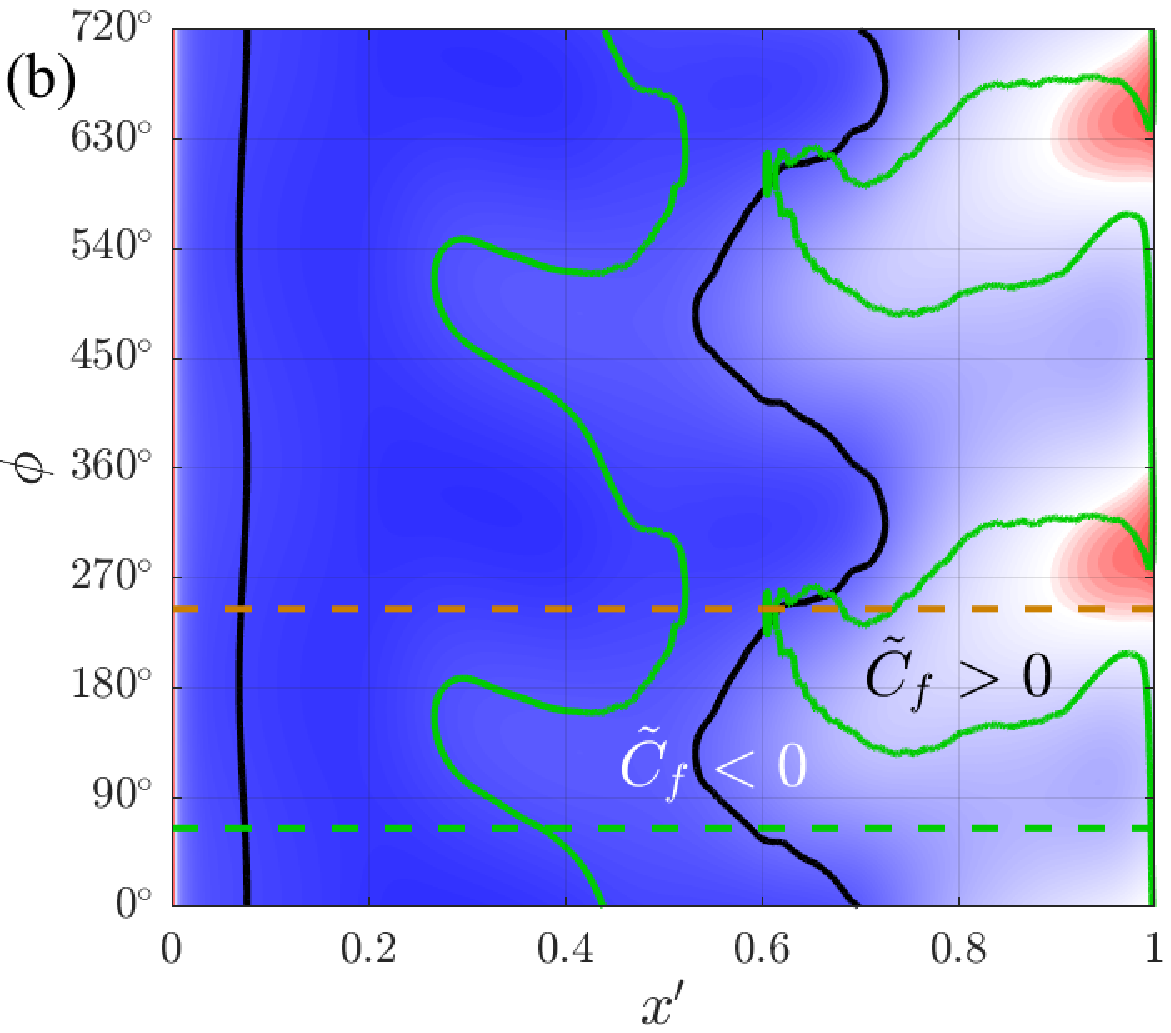}
\includegraphics[width=.495\textwidth]{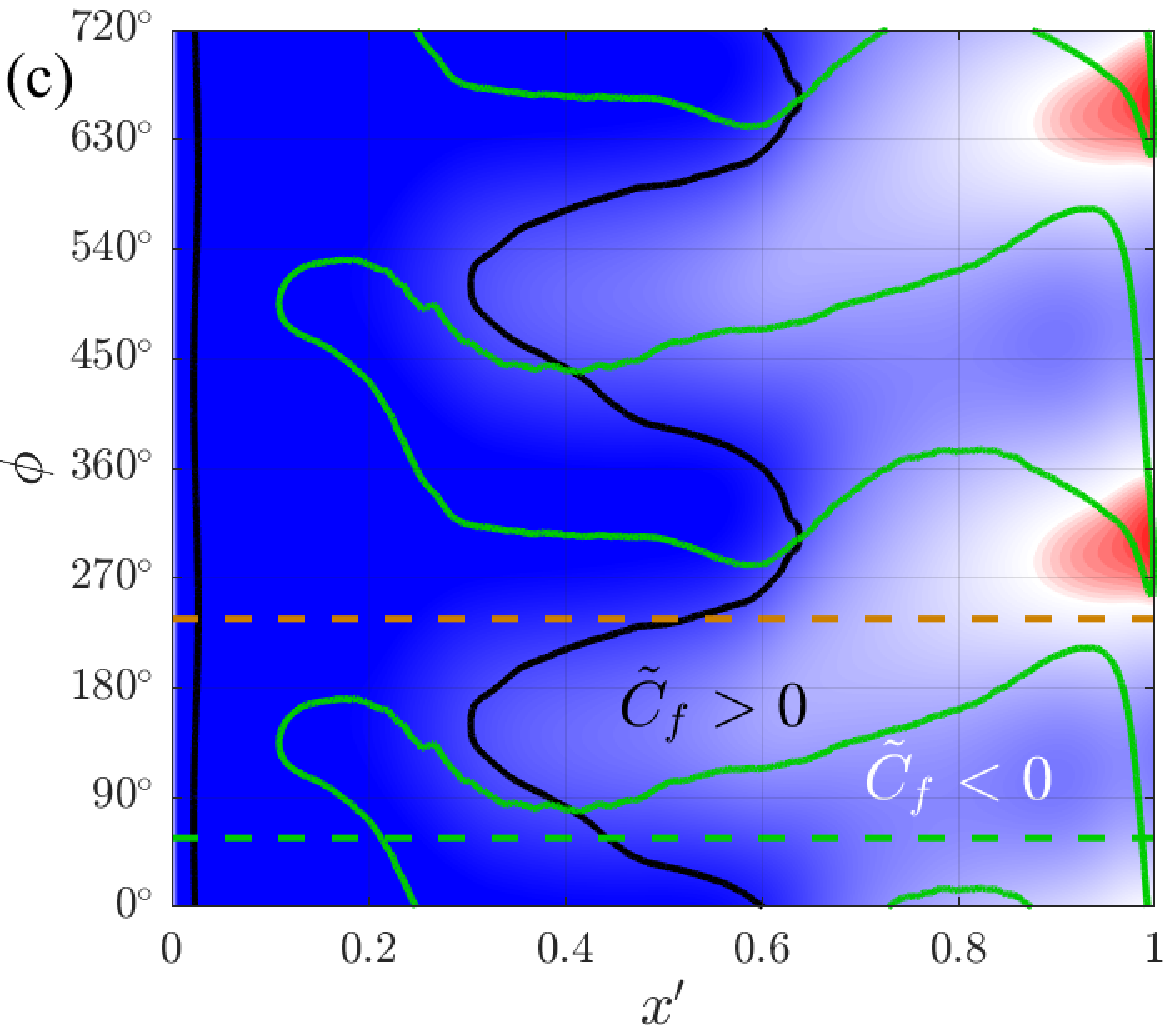}
\includegraphics[width=.495\textwidth]{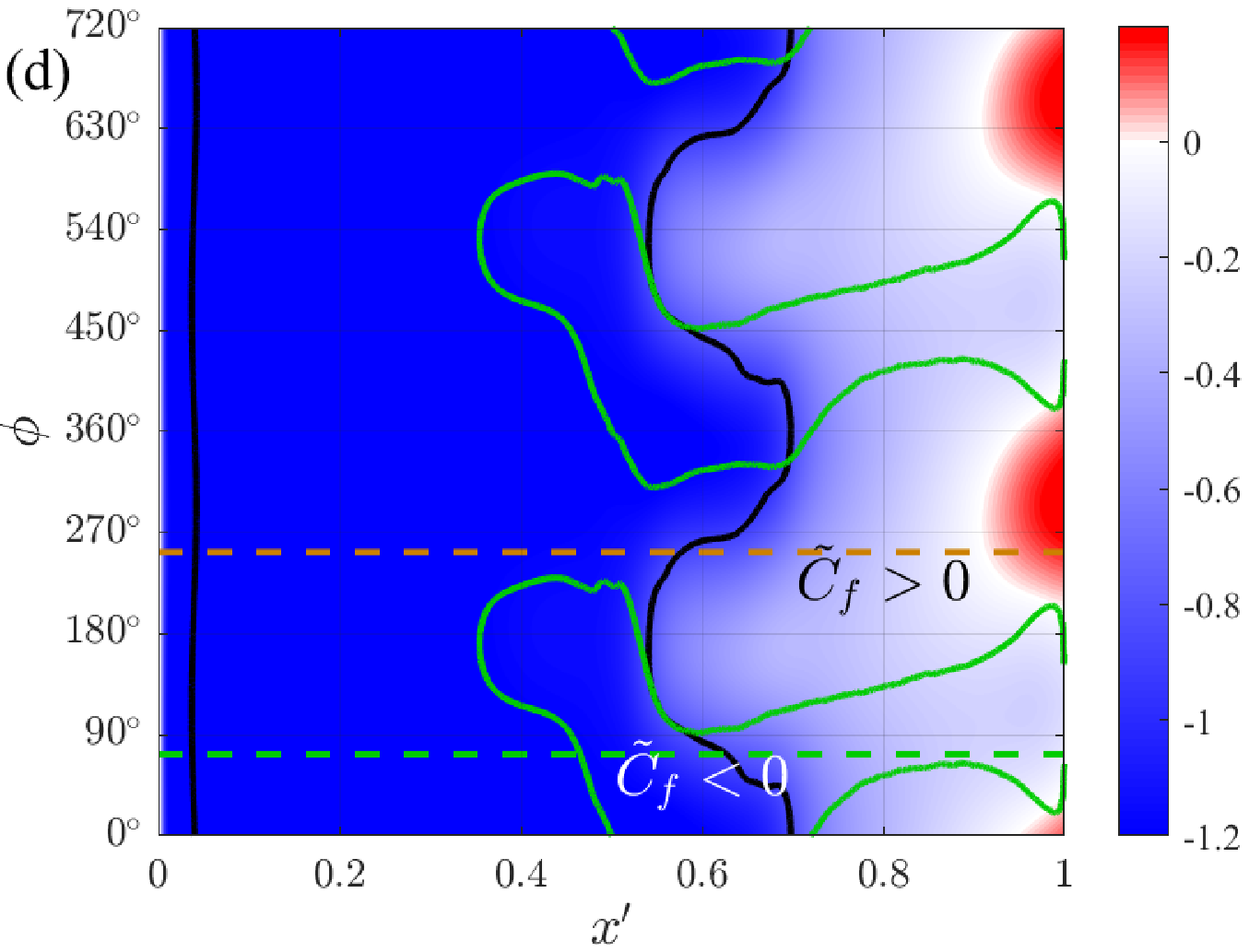}
\caption{Spatio-temporal variation of $\tilde{C}_p$ (reconstructed using buffet mode) on the aerofoil suction surface for (a) reference, (b) $M = 0.735$, (c) $\alpha = 6^\circ$ and (d) $\Rey = 1.5\times10^6$ cases. The isolines $\tilde{M}_\mathrm{loc} = 1$ (solid black curve) and $\tilde{C}_f = 0$ (solid green curve) are also shown.}
    \label{fig:RefSPODReconXT}
\end{figure}

To better understand the dynamical behaviour of buffet, it is useful to examine the variation of flow features on the aerofoil surface at different phases. This spatio-temporal variation is shown using $x'-\phi$ diagrams for the different cases in figure \ref{fig:RefSPODReconXT}. Unlike the preceding sections where the streamwise pressure gradient on the curve C5 was examined to scrutinise shock wave structures, we show here the variation of the pressure coefficient on the aerofoil surface. In addition, we overlay the contour line, $\tilde{C}_f = 0$ (green curve), which delineates flow reversal ($\tilde{C}_f \leq 0$). To include the shock wave position, we also overlay the sonic line based on the local Mach number (black curve). We emphasise that only this quantity is computed based on the curve C5 while the rest are based on the aerofoil surface. As before, the phases at which minimum and maximum $(\tilde{C}_D)_f$ occur are highlighted using dashed horizontal lines. 

The shock wave's most upstream and downstream positions (black curve) occur when the lift is close to its minimum ($\phi = 180^\circ$) and maximum ($\phi = 0^\circ$), although there seems to be a small phase lag between the two. As alluded to in  \S\ref{secRef}, the maximum and minimum $(\tilde{C}_D)_f$ occur when the flow is `least' and `most' separated, as can be inferred from the chordwise extent of separation delineated by the $\tilde{C}_f = 0$ isoline (green curve). Indeed, the BL remains separated up to the TE ($\tilde{C}_f < 0$) at the phase when the minimum $(\tilde{C}_D)_f$ is attained (dashed horizontal green line) for all cases. By contrast, during the phase of maximum $(\tilde{C}_D)_f$ (dashed horizontal brown line), the BL is fully attached from the leading edge to TE for the cases $\Rey = 1.5\times10^6$ and $\alpha = 6^\circ$, while the flow reattaches ($\tilde{C}_f > 0$) downstream to the shock wave (black curve) and remains attached up to the TE for the other cases. Thus, we can conclude that the surface coefficients $\tilde{C}_L$ and $(\tilde{C}_D)_f$ are approximate indicators of the shock wave position and the BL separation, respectively. The reasons for these relations can be explained by observing that a negative $\tilde{C}_p$ exists upstream of the shock wave, and an increase in the chordwise extent of this low pressure region leads to an increased lift. Similarly, as mentioned previously, an attached BL leads to larger skin-friction due to large positive velocity gradients on the aerofoil surface, while a separated BL and flow reversal above the surface contribute to a low $(\tilde{C}_D)_f$. 

\begin{figure} 
\centering
\includegraphics[width=.495\textwidth]{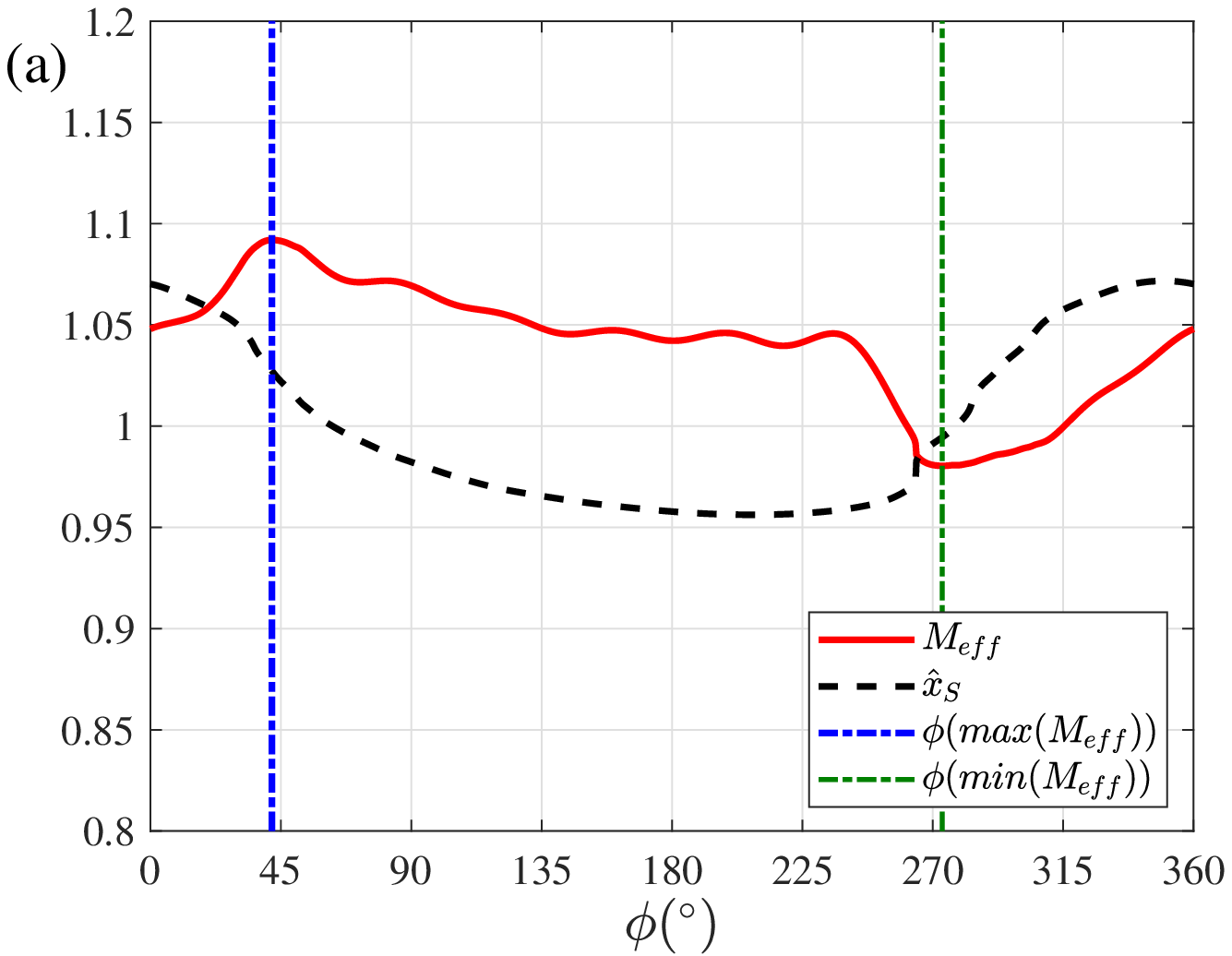}
\includegraphics[width=.495\textwidth]{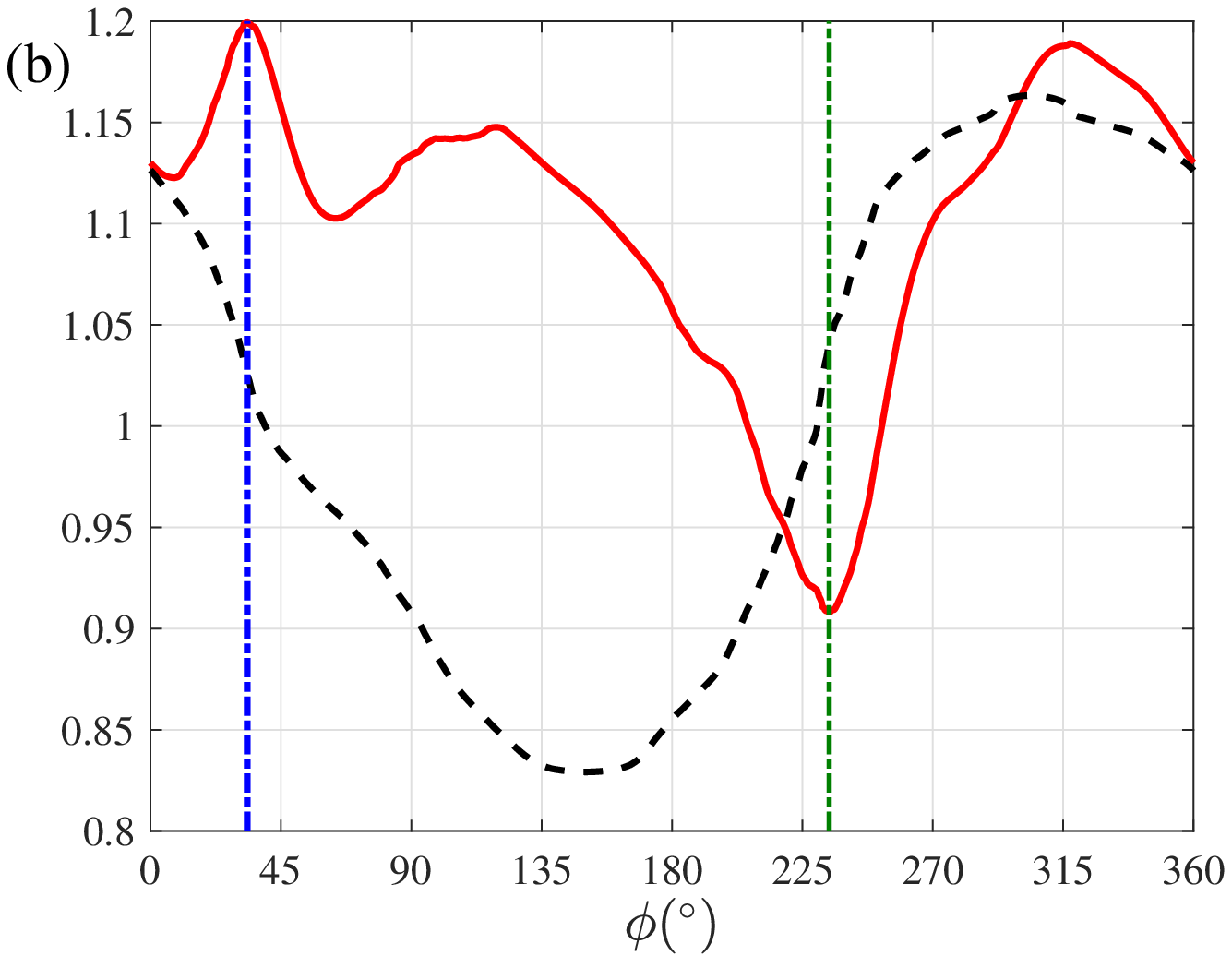}
\caption{Temporal variation of shock position and strength based on modal reconstruction using buffet mode for (a) reference and (b) $\alpha = 6^\circ$ cases.}
    \label{fig:ShockStrengthSPOD}
\end{figure}

The temporal variation of the shock wave strength during the buffet cycle is estimated based on the reconstructed flow field. One estimate of this strength is the ratio of the instantaneous pressure immediately downstream and upstream of the sonic line in the $x-\phi$ diagram, $\tilde{p}(x^+_S,\phi)/\tilde{p}(x^-_S,\phi)$, where $x_S$ represents the shock wave's streamwise position (\textit{i.e.}, $x_S = x(\tilde{M}_\mathrm{loc} = 1)$) and the symbols `$+$' and `$-$' indicate downstream and upstream, respectively. However, this choice was found to be sensitive to the choice of $x^+_S$ and $x^-_S$. This is possibly because the actual flow-field has multiple shock wave structures and the single shock wave seen in the modal reconstruction is only an approximation, implying that choosing points both upstream and downstream of the shock introduces relatively large errors in the estimation of the pressure ratio. Instead, we found the upstream effective Mach number, which is based on the instantaneous upstream velocity in the shock wave frame of reference, to be more robust with regard to the choice of $x^-_S$. This is given by
$$
\tilde{M}_\mathrm{eff}(x^-_S,\phi) = \tilde{M}_\mathrm{loc}(x^-_S,\phi) - \left(\frac{dx_S}{dt}\right)\frac{1}{a(x^-_S)},
$$
 and can then be used to represent the shock strength based on the Rankine-Hugoniot conditions \citep{Gibb1988}. However, note that this approximation neglects the effect of shock wave acceleration. The effective upstream Mach number for the reference and $\alpha = 6^\circ$ cases are shown in figure \ref{fig:ShockStrengthSPOD} for $x_S^- = x_S - 0.1$. For comparison, the variation of shock wave position is also shown, but for clarity, the mean value it takes with phase is altered to unity ($\int_0^{360^\circ} \hat{x}_Sd\phi = 1$) by adding a constant. It is seen from the plots that $\tilde{M}_\mathrm{eff}$ reaches a maximum or minimum when the shock is approximately at the mid-point of its upstream or downstream excursion, respectively. This was observed for all other cases where buffet occurs (not shown for brevity), implying that the shock wave is always strongest/weakest close to the its mean streamwise position during its upstream/downstream motion. A similar result based on Mach number was shown in \citet{Fukushima2018} (their figure 18), while other studies have also noted the same trend \citep{Iovnovich2012, Hartmann2013, Tijdeman1980, Lee2001}. The plots show that in there is a small interval in which $\tilde{M}_\mathrm{eff} < 1$. This is possibly not of physical significance, since there are shock wave structures present throughout the buffet cycle suggesting that it arises due to the approximations in the estimate of shock strength.

From figures \ref{fig:RefSPODReconXT} and \ref{fig:ShockStrengthSPOD}, the following sequence of events can be inferred for all cases:
\begin{enumerate}
    \item When the shock wave is at its most downstream position, the BL is separated upstream of the shock wave, but reattaches downstream, forming a separation bubble.
    \item As the shock moves towards the LE, it strengthens, while the upstream separation point moves in the same direction and the reattachment point moves further downstream (or vanishes), increasing the chordwise extent of the separation region. 
    \item The BL is `most' separated when the shock wave is approximately at the mid-point of its upstream excursion, where its strength is close to its maximum.
    \item Past the mid-point, the BL separation weakens leading to the formation of a separation bubble that reattaches at the shock foot for all cases except $M = 0.735$, where this occurs later. 
    \item Once the shock wave reaches its most upstream position and starts to move downstream, the separation point upstream of the shock wave abruptly moves downstream by almost $\delta x' = 0.1$ for the $M = 0.735$ and reference case, while it completely vanishes for the other cases. 
    \item As the shock wave reaches the midpoint of its downstream excursion it strength is relatively weakened and the flow is `least' separated.
    \item The reattached flow is associated with a pressure rise in the TE region, with this back pressure exceeding the freestream pressure (red contours, $\tilde{C}_p > 0 \Rightarrow \tilde{p} > \tilde{p}_\infty$). The chordwise extent of this region of increased pressure attains a maximum before the shock wave reaches its most downstream position.
\end{enumerate}

The above description indicates that the laminar buffet observed in this study is essentially related to a moving shock wave of temporally varying strength whose motion is accompanied by BL separation and reattachment. Importantly, the following two features are common to all buffet cases studied: a phase lag between the shock wave position and the separation extent and the build up of back pressure. Additionally, it is interesting to note that, for all cases, there are phases in the buffet cycle where the flow separation extends up to the TE, indicating that the aerofoil is intermittently stalled.  However, we emphasise that it is difficult to disentangle cause, correlation and effect. The possible implications of these observations are discussed in \S\ref{secDisc}. 

% Note that the extension of BL separation during the upstream excursion of the shock wave is particularly stark for $M = 0.735$, where a complete separation up to the TE can be observed almost at the phase when the shock wave starts moving upstream. 
% 2do: \citep{Giannelis2018} noted at onset: shock moving upstream = separation moving till TE. once at most upstream, reattachment occurs
%In summary, the upstream motion of the shock wave is initially accompanied by strong separation, but eventually, leads to reattachment as it moves past its mid-point. The downstream motion is characterised initially by an attached BL and later by a pressure rise in the TE region. 
% 2do: have a look at reconstruction only based on first harmonic (no need to report unless something imp comes up)

\subsubsection{Wake mode}
\label{subsecWakeModeSPOD}

\begin{figure} 
\centering
\includegraphics[trim={0cm 1.5cm 0cm 2cm},clip,width=.495\textwidth]{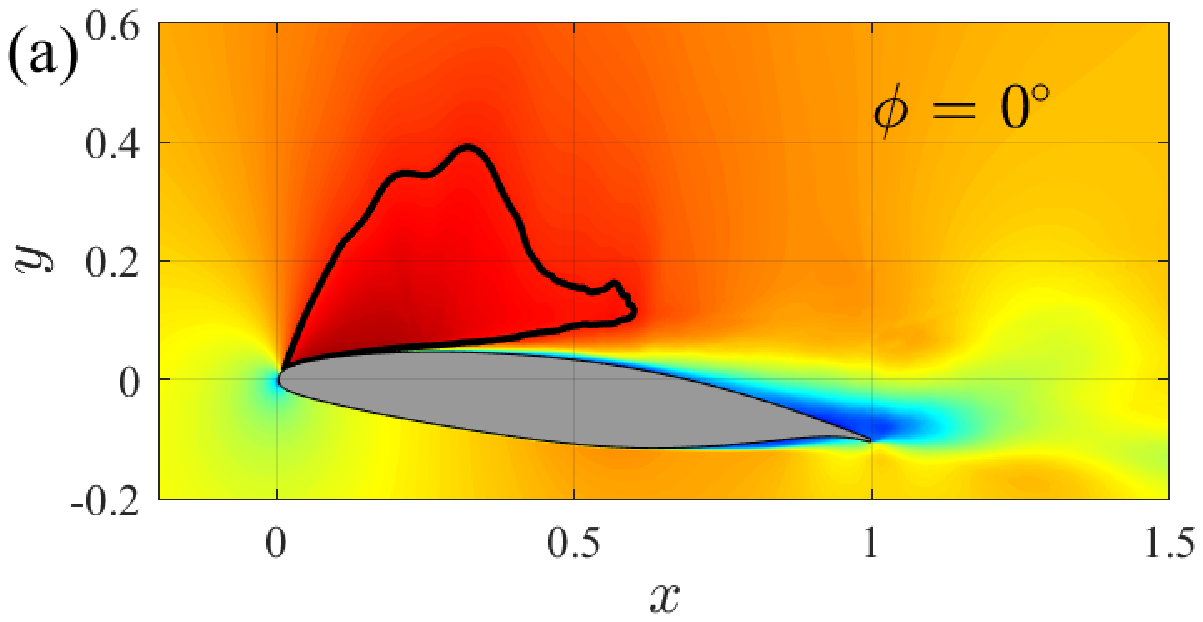}
\includegraphics[trim={0cm 1.5cm 0cm 2cm},clip,width=.495\textwidth]{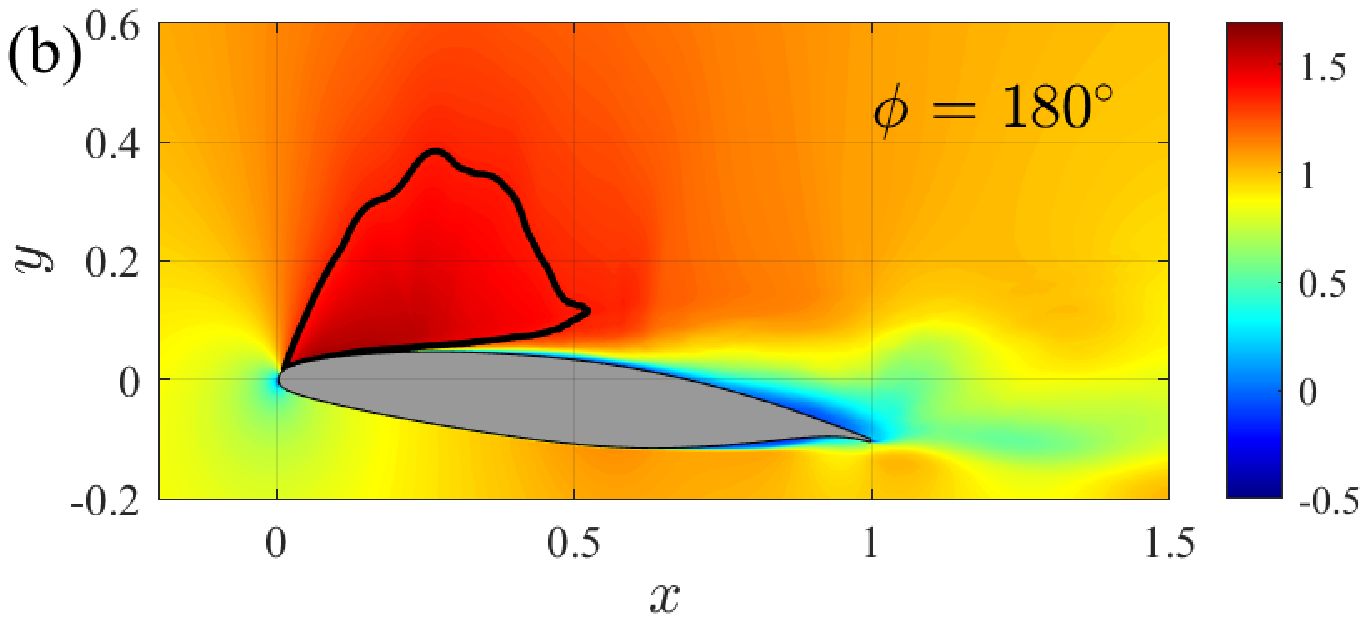}
\caption{Reconstructed flow field based on the wake mode for $\alpha = 6^\circ$ shown using axial velocity contour at (a) high- and (b) low-lift phases.}
    \label{fig:SPODReconstr2DAoA6WakeMode}
\end{figure}

The contours of reconstructed axial velocity field based on the wake mode for the case of $\alpha = 6^\circ$ are shown in figure \ref{fig:SPODReconstr2DAoA6WakeMode} at phases corresponding to the highest ($\phi = 0^\circ$) and lowest ($\phi = 180^\circ$) values of $\tilde{C}_L$. Note that the phase, $\phi$, considered here is based on the time-period of the wake mode and not the buffet mode. Thus, the high-lift phase signifies the phase at which the wake mode interaction with the mean flow induces the maximum lift obtained for the reconstructed flow field and is not related to the buffet mode. In the plots, the sonic line (black curve) indicates that the shock foot extends and retracts at different phases, similar to the shock foot motion observed in \citet{Dandois2018}, implications of which are discussed in \S\ref{subsecDiscLamTurbBuff}.

% %%%%%%%%%%%%%%%%%%%%%%%%%%%%%%%%%%%%%%%%
\subsection{High Mach number features}
\label{subSecSPODHighM}
% %%%%%%%%%%%%%%%%%%%%%%%%%%%%%%%%%%%%%%%%

\begin{figure} 
\centering
\includegraphics[trim={3cm 0cm 3cm 0cm},clip,width=.32\textwidth]{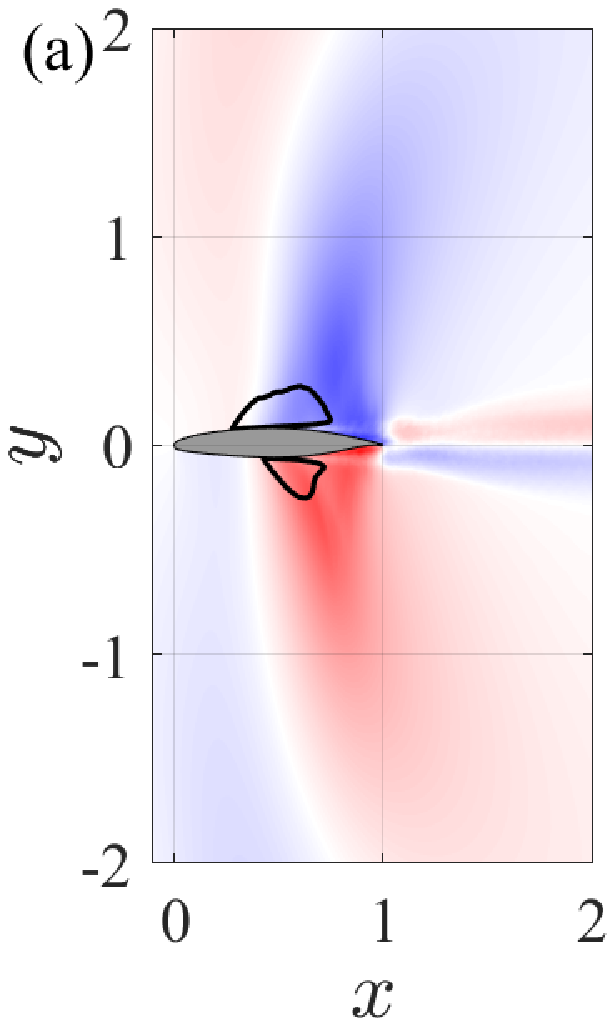}
\includegraphics[trim={3cm 0cm 3cm 0cm},clip,width=.32\textwidth]{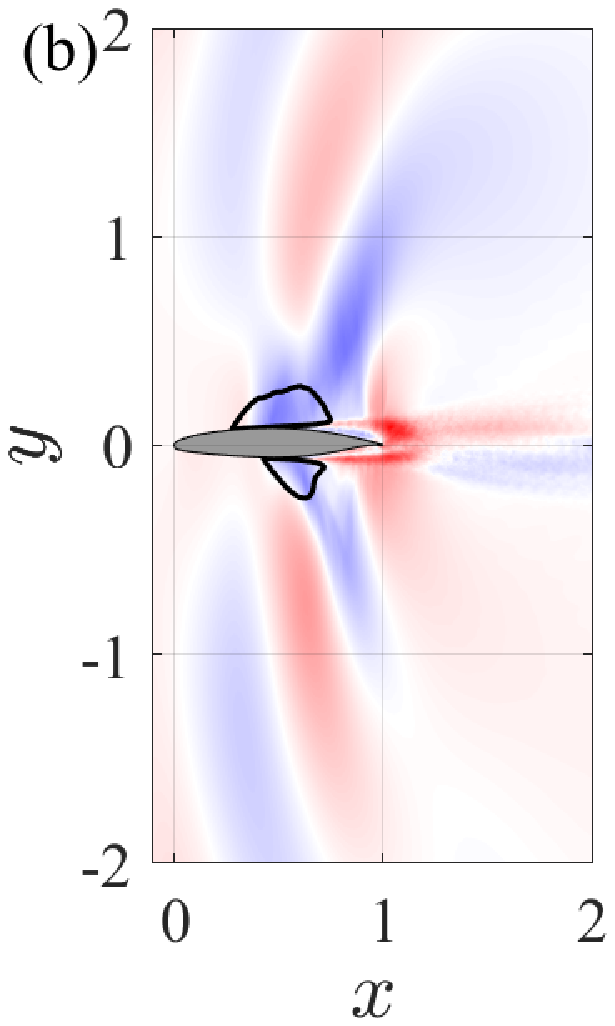}
\includegraphics[trim={3cm 0cm 3cm 0cm},clip,width=.32\textwidth]{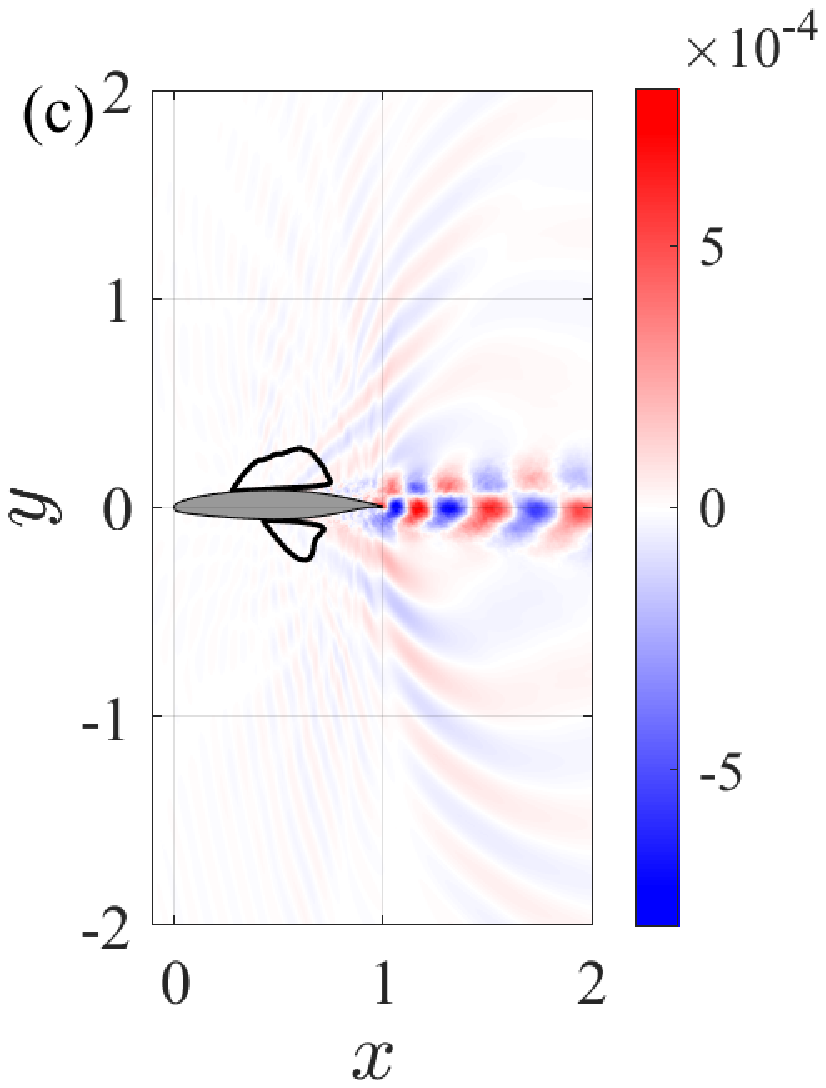}
\caption{Eigenmodes extracted using SPOD for the A0M8 case ($\alpha= 0^\circ$, $M=0.8$) are shown using contour plots of the real part of the density field: (a) buffet mode ($St = 0.16$), (b) its first harmonic ($St = 0.31$) and (c) a wake mode ($St = 1.59$).}
    \label{fig:0DegSPODModes}
\end{figure}

The spatial structure of the buffet mode, its first harmonic and a typical wake mode for the A0M8 case ($\alpha = 0^\circ$ and $M = 0.8$) are shown in figure \ref{fig:0DegSPODModes} using the real part of the density field. It is evident that the buffet mode is approximately antisymmetric about the $y-$axis, which indicates that the flow behaviours on the suction and pressure sides are out of phase with each other. By contrast, the first harmonic exhibits symmetric features. 

\begin{figure} 
\centerline{\includegraphics[width=.495\textwidth]{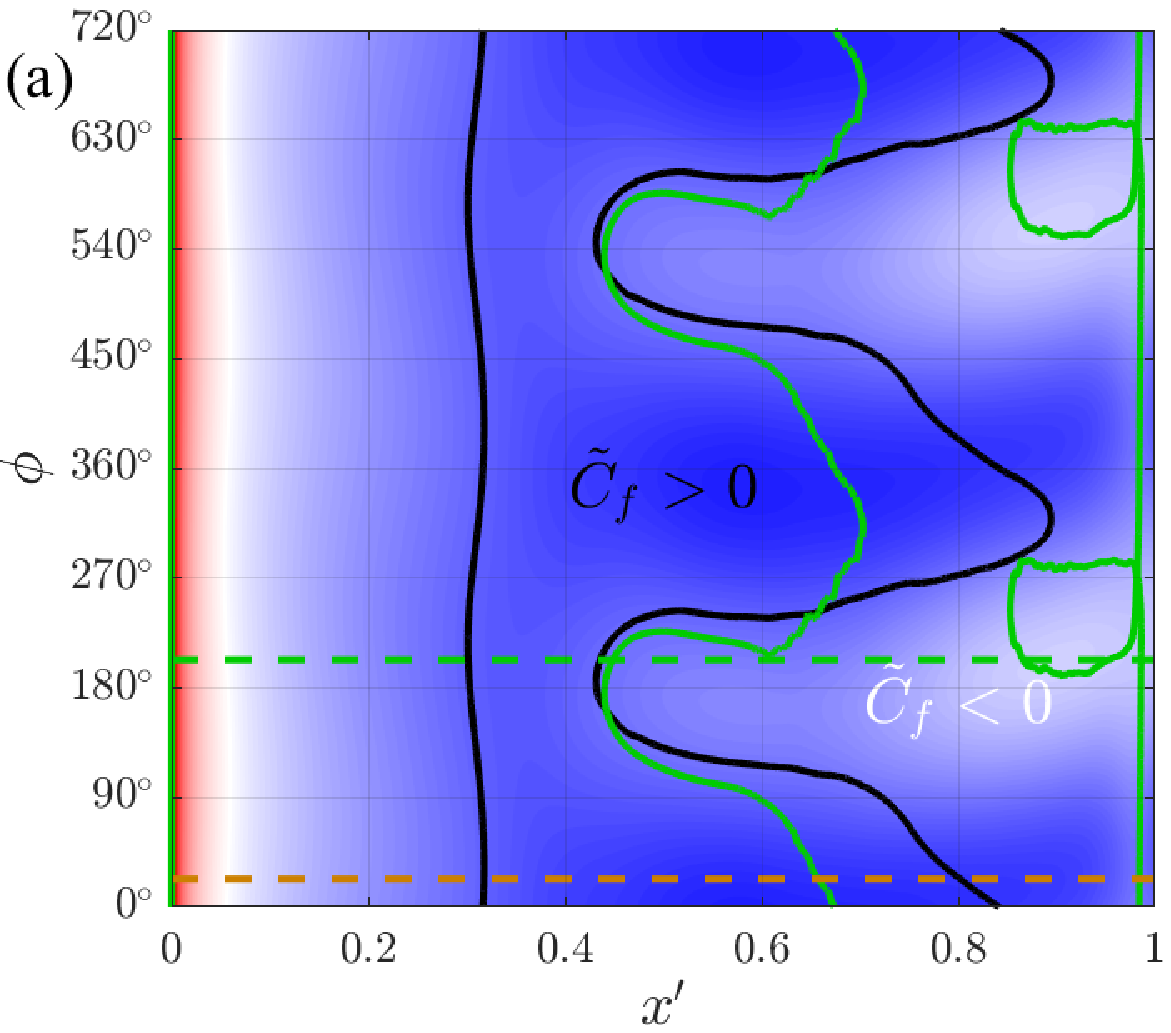}
\includegraphics[width=.495\textwidth]{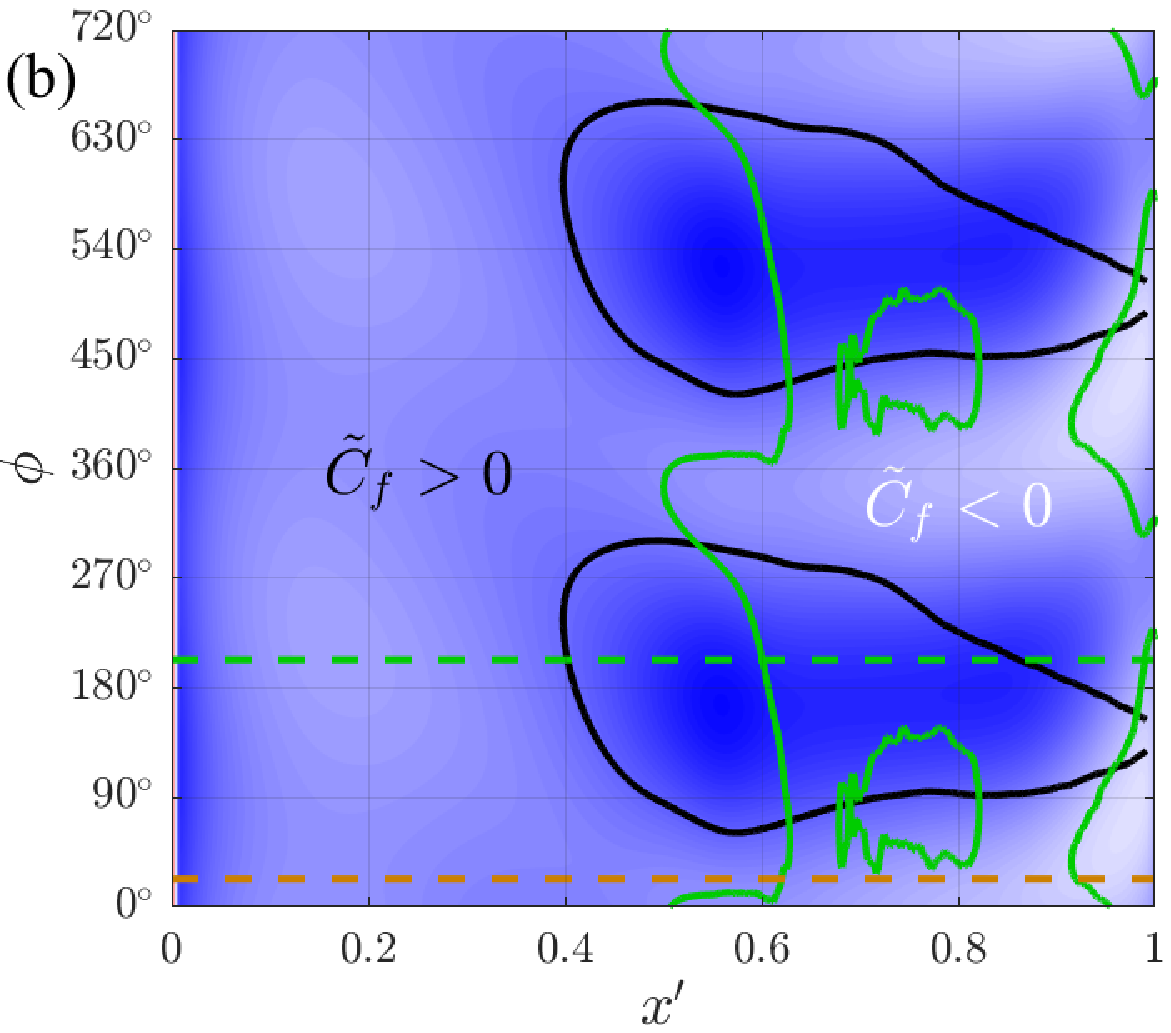}}
\centerline{\includegraphics[width=.495\textwidth]{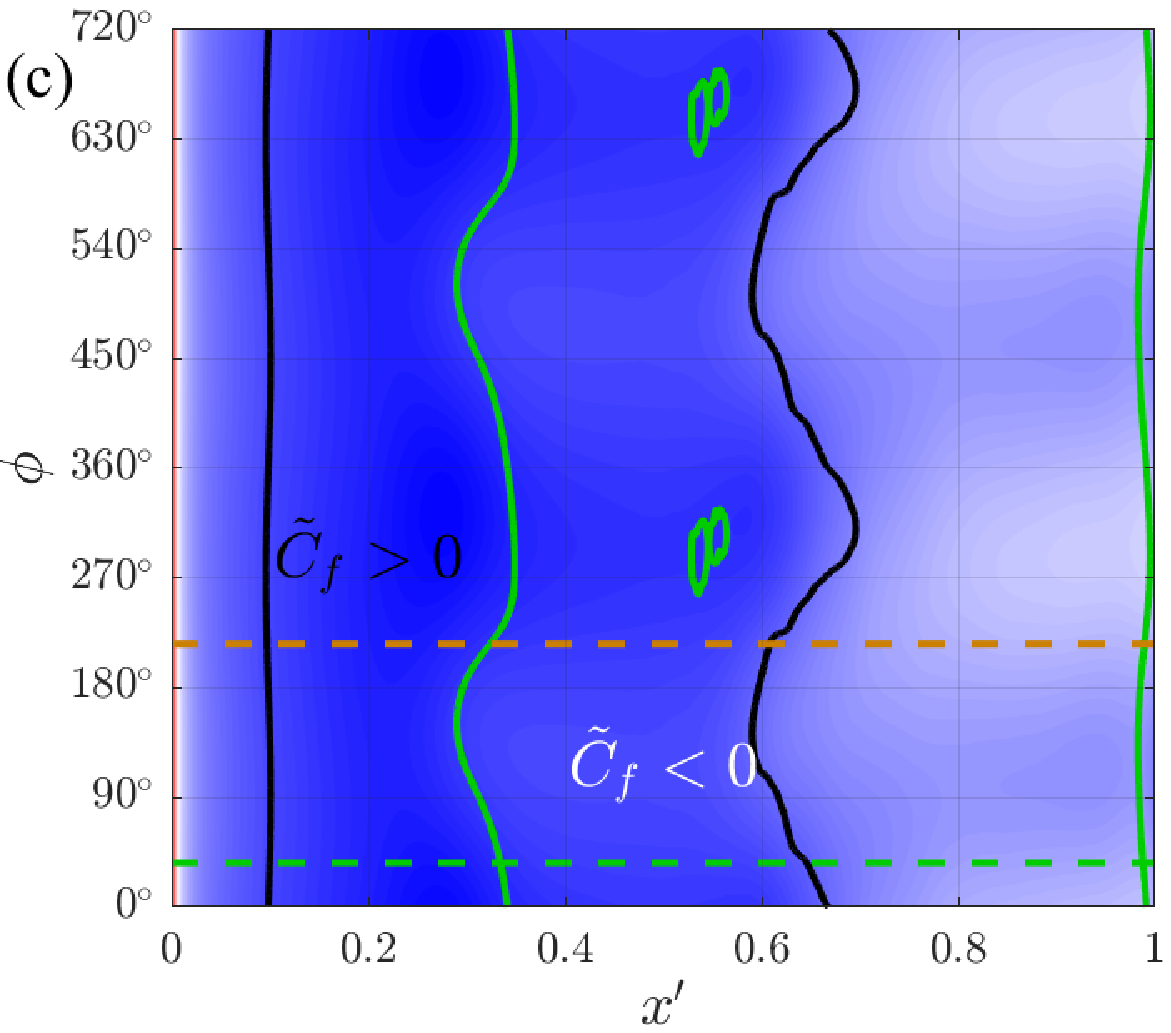}
\includegraphics[width=.495\textwidth]{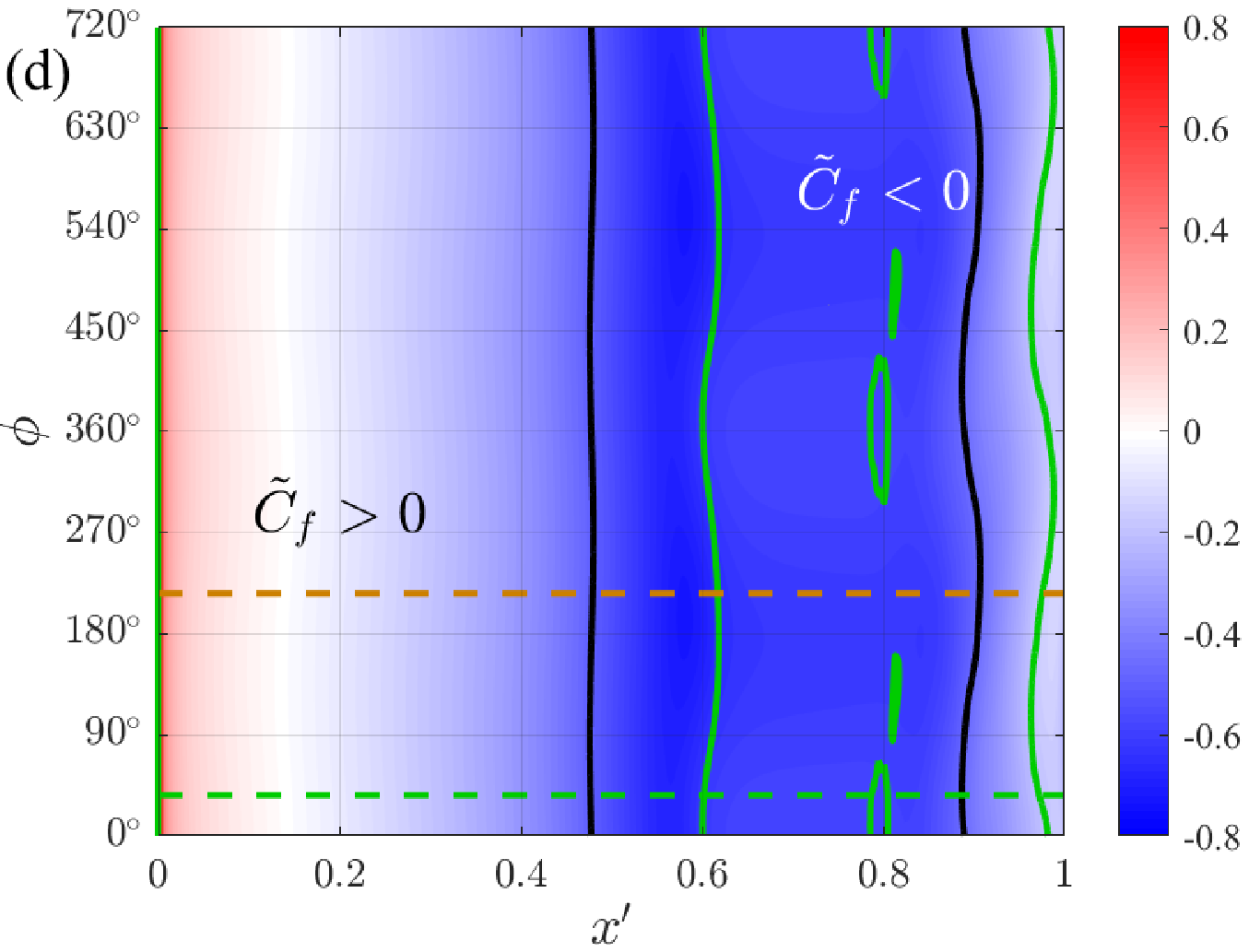}}
\caption{Same as figure \ref{fig:RefSPODReconXT}, but for the cases of $\alpha = 0^\circ$ (top) and $\alpha = 4^\circ$ (bottom) at $M = 0.8$ on suction (left) and pressure (right) surfaces.}
    \label{fig:SPODM8ShockPosn}
\end{figure}

Reconstructions based on the buffet mode at $M = 0.8$ for the cases of $\alpha = 0^\circ$ (A0M8 case) and $\alpha = 4^\circ$ ($M=0.8$ case) are shown in figure \ref{fig:SPODM8ShockPosn}. It is evident from the figure that the motion of shock wave (black curve) on the suction side is of similar amplitude as that observed for other $M$ (cf. figure \ref{fig:RefSPODReconXT}). However, the flow is also seen to be separated downstream of some chordwise position for almost all times, especially at $\alpha = 4^\circ$. Importantly, this implies that strong buffet can occur even when the BL is permanently separated. Note also that for the $M = 0.8$ and $\alpha = 4^\circ$ case, the amplitudes of lift fluctuations are relatively low as compared to lower $M$ (see figure \ref{fig:clVsT_HighM}\textit{b}), although the amplitude of shock wave motion remains approximately the same. This implies that the determination of buffet occurrence merely based on temporal variations in $C_L$ or $C_p$ might not be always accurate due to the low variation in these coefficients when the BL is permanently separated. 

As noted before, a phase difference between the shock wave positions on the suction and pressure sides is present for both cases. For $\alpha = 4^\circ$, although the amplitude of the streamwise excursion of the shock wave on the pressure side is relatively weak in comparison to that on the suction side, the former was found to lag the latter by $\approx 69^\circ$. For the $\alpha = 0^\circ$ case, in which the shock wave vanishes for part of the cycle, this phase difference was computed using the most downstream shock position. This was computed as approximately $179^\circ$, which is similar to that seen for biconvex aerofoils \citep{McDevitt1976}. A phase difference between the shock wave positions and the extent of flow separation is also seen in figure \ref{fig:SPODM8ShockPosn}, albeit, significantly reduced. An increase in back pressure close to when the shock wave reaches its most downstream position can also be discerned. Thus, these two features seem to be common to all buffet cases simulated here.

%%%%%%%%%%%%%%%%%%%%%%%%%%%%%%%%%%%%%%%%
\section{Discussion}
\label{secDisc}
%%%%%%%%%%%%%%%%%%%%%%%%%%%%%%%%%%%%%%%%

%%%%%%%%%%%%%%%%%%%%%%%%%%%%%%%%%%%%%%%%
\subsection{The relation between laminar and turbulent buffet}
\label{subsecDiscLamTurbBuff}
%%%%%%%%%%%%%%%%%%%%%%%%%%%%%%%%%%%%%%%%
An important conclusion on the relation of laminar and turbulent buffet can be made based on the present results. Laminar buffet, as identified in \citet{Dandois2018}, is characterised by the BL remaining laminar up to the shock foot during the entire buffet cycle. Based on their simulations of the OALT25 at a specific flow setting, the authors concluded that unlike the turbulent buffet, which occurs as a ``global instability of the flow with intermittent boundary-layer separation and reattachment between the shock and the trailing edge", the laminar buffet occurs due to a ``separation bubble breathing phenomenon associated with a vortex shedding mechanism". In the present study, the previous sections clearly show that we have a laminar buffet, with the BL laminar until the vicinity of the shock foot for all cases studied (\textit{e.g.}, figure \ref{fig:RefDensGrad}). However, the results of the SPOD reconstruction of the buffet mode indicate that the laminar buffet observed here is driven by boundary-layer separation and reattachment (see figure \ref{fig:RefSPODReconXT}), a characteristic feature of turbulent buffet. 

The modal reconstruction based on the wake mode shows that this mode can cause a localised motion of the shock foot (figure \ref{fig:SPODReconstr2DAoA6WakeMode}), while the vortices shed are characteristic of bubble breathing observed in shock wave boundary layer interactions. Thus, we predict that the self-sustained oscillations reported in \citet{Dandois2018} are not associated with the buffet mode, but are due to the wake mode. These two modes can both coexist for the same flow settings, as shown here for the V2C and it is possible that for the OALT25, the wake mode alone is the predominant unstable mode under certain conditions, while the buffet mode is incipient. This is also suggested by the experimental results for the same aerofoil reported in \citet{Brion2020}, where an increase in $\alpha$ or $M$ at $\Rey = 3\times 10^6$ leads to the emergence of a minor low frequency peak at $St \approx 0.05$, indicating incipient laminar buffet. Based on the above considerations, we conclude that laminar buffet is essentially the same as turbulent buffet with regard to the physical mechanisms that sustain the low-frequency oscillation. This allows for using the present results to make general comments on the buffet phenomenon.

%However, based on the present results, we should also note that there may be some interesting differences between the two, including the occurrence of multiple shock wave structures and the relatively larger area of influence of the buffet mode \citep[compare figures \ref{fig:RefSPODModes} and \ref{fig:SPODModesAoA6} with figure 4][p. 10]{Crouch2019}. 

%%%%%%%%%%%%%%%%%%%%%%%%%%%%%%%%%%%%%%%%
\subsection{The relation between Type I and II buffet}
\label{subsecDiscTypeINII}
%%%%%%%%%%%%%%%%%%%%%%%%%%%%%%%%%%%%%%%%
In the previous sections, we have shown that buffet also occurs at zero incidence ($M = 0.8$ and $Re = 5\times 10^5$) with shock waves present on both aerofoil surfaces exhibiting oscillatory motion such that they are out of phase with each other. These aspects are the defining characteristics of Type I buffet, albeit for a supercritical aerofoil instead of a biconvex/symmetrical aerofoil as is typically reported. However, the flow features seen for this case, including the presence of multiple shock wave structures and the temporal variation of $C_L$, also resemble those observed for Type II buffet discussed for $0.7 \leq M \leq 0.775$ at $\alpha = 4^\circ$. This indicates that Type I and Type II buffet potentially have the same underlying mechanisms. Similar inferences can also be made from \citet{McDevitt1985} for the symmetric NACA 0012 aerofoil. The results of that study indicate that with increasing $M$, the onset incidence angle for Type II buffet reduces  (see their figure 34, p. 74) suggesting that this buffet type could occur at zero incidence if $M$ is sufficiently large \citep[see also][figure 10]{Crouch2007}. Thus, while it is still useful to differentiate buffet into these two types to highlight the presence or absence of shock waves on the pressure side, the two appear to be governed by the same mechanisms irrespective of the aerofoil used. This implies that any model for buffet should be capable of explaining either buffet type.

 Building on earlier studies \citep{TijdemanReport, McDevitt1985}, \citet{Gibb1988} proposed a model for Type I buffet based on an instability arising from the shock-wave boundary layer interactions which couples with the wake deflection, \textit{i.e.}, if a perturbation causes the shock wave on one side, say the upper surface, to move upstream, it can cause it to strengthen leading the BL to separate at its foot which would cause the wake to deflect upward. This would cause the flow on the lower surface to accelerate, causing the shock wave on that surface to move downstream further deflecting the wake upward which in turn promotes the upstream excursion of the the upper shock wave. This would continue until the upper shock wave moves sufficiently far upstream that the local Mach number it encounters is small, causing it to weaken, which in turn leads to BL reattachment and subsequent downstream motion. This model requires that shock waves are present on both aerofoil surfaces and play an active role in sustaining buffet. However, this model fails for Type II buffet, where there is a negligible temporal variation in the coherent flow features on the pressure side (\textit{e.g.}, see pressure side, figure \ref{fig:SPODReconstr2DAoA6}). Similarly, models proposed exclusively for Type II buffet must be reinterpreted to explain Type I buffet.  
 
 %%%%%%%%%%%%%%%%%%%%%%%%%%%%%%%%%%%%%%%%
\subsection{The wake mode and the model of Lee}
\label{subsecDiscLee}
%%%%%%%%%%%%%%%%%%%%%%%%%%%%%%%%%%%%%%%%
As shown in \S\ref{subSecSPODModeFeatures}, the time periods of buffet predicted by the models proposed in \citet{Lee1990} and \citet{Jacquin2009} are substantially lower than those observed in the simulations. While this quantitative difference implies that the models are inaccurate, there are physical reasons to suspect that the proposed physical mechanism itself is invalid. As indicated by the results of SPOD, the waves that are observed in these simulations are due to the wake modes. These modes are essentially related to vortex shedding implying that they are primarily dependent on the position of the BL separation. While the buffet mode strongly influences BL separation, the results from the variation of $M$ indicate that the buffet mode is not influenced by the intensity of vortex shedding. Firstly, wave propagation can be discerned for cases where no buffet is present, as seen for the case $M =0.65$ (figure \ref{fig:M65N69DensGrad}\textit{a}). Also, as $M$ is increased from $M = 0.7$ to 0.85, we observe a monotonic increase in vortex shedding intensity, but the energy of the buffet mode initially increases when $0.7 \leq M \leq 0.735$ and substantially reduces when $0.735 \leq M \leq 0.85$. Additionally, the frequencies associated with the buffet and wake modes differ by an order of magnitude (here, $St \approx 0.1$ and $1 \leq St \leq 4$, respectively), with only the former being predicted in a global linear stability analysis \citep{Crouch2007} while a resolvent analysis is required to predict the latter \citep{Sartor2015}. Furthermore, unsteady RANS simulations that capture essential buffet features, usually do not capture vortex shedding which is caused by convective instabilities. For example, it was shown in \citet{Grossi2014} using delayed DES that vortex shedding accompanies buffet for the same conditions where URANS simulations do not show any. Similarly, a modal decomposition of the URANS flow field in \cite{Poplingher2019} did not show any vortical structures of significant energy. Thus, with the observations that the dominant waves accompanying buffet are related only to the wake modes, but that such modes do not directly influence buffet, we find no convincing evidence for buffet models based on feedback loops involving wave propagation. Note that the validity of such models has also been questioned in \citet{Paladini2019a} based on other physical arguments. The authors, using a localised selective frequency damping approach, showed that the subsonic region above the BL (where the upstream propagating waves travel) and the pressure side do not play a role in the instability and suggested that the feedback must occur within the separated BL.   
 
%%%%%%%%%%%%%%%%%%%%%%%%%%%%%%%%%%%%%%%%
\subsection{Phase lags in buffet}
\label{subsecBuffetMech}
%%%%%%%%%%%%%%%%%%%%%%%%%%%%%%%%%%%%%%%%

The phase lag between BL separation and shock wave location that is persistent for all cases where buffet occurs (see figure \ref{fig:RefSPODReconXT}) seems to be a characteristic feature of buffet, although it does not appear to have been highlighted previously. For example, it is noted in \citet{Jacquin2009} that ``the boundary layer is separated when the shock is in the upstream location [...] and attached when it moves downstream". However, indications that this phase lag exists can be inferred from some previous studies. Re-examining the results from \citet{Xiao2006a} (see their figure 11) for buffet on the BGK supercritical aerofoil, we can infer that the chordwise extent of separation is minimum at $\phi\approx90^\circ$ and maximum at $\phi \approx 270^\circ$, which further corroborates the importance of this phase difference. Similarly, examining figure 8 in \citet{Grossi2014} (OAT15A aerofoil), we see that the streamwise extent of flow separation is a minimum when the shock wave is approximately at the mid-point of its downstream motion. Similar inferences can be made by considering figures 17f and 17g in \citet{Fukushima2018}. As with any harmonic oscillator system, phase lags between competing forces can act as the drivers of self-sustained oscillations. Thus, we suggest that this lag could be an essential physical feature for the development and sustenance of buffet, although the exact role it plays remains unclear, as discussed in the following section, and requires scrutiny.

%%%%%%%%%%%%%%%%%%%%%%%%%%%%%%%%%%%%%%%%
\subsection{The role of shock waves in buffet}
\label{subsecDiscShockwaveInBuffet}
%%%%%%%%%%%%%%%%%%%%%%%%%%%%%%%%%%%%%%%%
The results from the flow reconstruction show two common features that are conserved for all cases: the phase lag discussed above and the build up of back pressure close to when the shock wave starts to move upstream. These results indicate that there are three possibilities regarding the role of the shock wave in buffet. The first is that the shock wave plays only a passive role and buffet can arise as a global instability that causes large-scale flow oscillations even in its absence. There are several results from different studies that suggest this. Similar oscillations at $St \approx 0.2$ have been reported for periodic bubble bursting on aerofoils at incidence angles close to stall in the \textit{incompressible} regime which occur simply due to viscous-inviscid interactions and can be accompanied by vortex shedding at a higher $St \approx 2.2$ \citep{Sandham2008}. Additionally, it can be inferred from the two-dimensional simulations performed in \citet{Bouhadji2003} of a flow over a NACA 0012 aerofoil at zero incidence ($\Rey = 10^4$) that Type I buffet occurs in the absence of shock waves for a certain range of freestream Mach numbers (see their figures 4d and 6c). Also, \citet{Paladini2019a} have concluded using different approaches that the detached BL is the ``active key of buffet instability'', while the ``shock is a slave zone but behaves as a stiffness on the instability phenomenon''. Furthermore, the periodic build up of pressure near the TE seen here for all cases and the fact that such changes in back pressure can lead to periodic shock motion \citep{Bruce2008} also supports this possibility, suggesting that the shock wave passively responds to back pressure changes. 

Alternatively, the shock wave might be crucial only in the part of the buffet cycle where it moves upstream and leads to a strong shock-induced separation, while playing a passive role as it moves downstream, merely responding to a pressure recovery of the reattached boundary layer. This is supported by the strengthening of the shock wave as it moves upstream and the phase lag between the shock wave position and the separation extent that is present for all cases. Furthermore, \citet{Tijdeman1980} have noted that there are situations where the shock wave vanishes during its downstream excursion (``Type B buffet"), suggesting that it might not be crucial in this part of the cycle. Note that similar features are also observed here for some cases (see figure \ref{fig:0AoAXTDia}).

The third possibility is that of a coupled interaction between shock wave and BL separation, with the shock wave's presence being important throughout the buffet cycle. As noted previously, the coupling between shock wave position, strength and flow separation have been physically explained using the wedge, curvature, displacement and dynamic effects \citep{Tijdeman1980,Iovnovich2012}, which suggest that the interaction between the shock wave and the BL is present throughout the buffet cycle. Then, the periodic build up of pressure near the TE could simply be a consequence of these interactions: for example, due to shock wave motion causing BL reattachment and pressure recovery. Given these possibilities, an aim for future studies could be to find the minimal physical model that can distinguish between cause and effect while predicting the key features seen in the present investigation.

%%%%%%%%%%%%%%%%%%%%%%%%%%%%%%%%%%%%%%%%
\section{Conclusions}
\label{secConc}
%%%%%%%%%%%%%%%%%%%%%%%%%%%%%%%%%%%%%%%%

In this study, Large-Eddy Simulations of transonic buffet under free transition conditions have been performed for infinite wing configurations based on Dassault Aviation's supercritical laminar V2C profile. Parameters were individually varied from a baseline reference value of $M = 0.7$, $\alpha = 4^\circ$, $Re = 5\times 10^5$ and $\Lambda = 0^\circ$, with the ranges reported being $0.5 \leq M \leq 0.9$, $3^\circ \leq \alpha \leq 6^\circ$, $2\times 10^5 \leq Re \leq 1.5\times 10^6$ and $0^\circ \leq \Lambda \leq 20^\circ$. For all cases, the flow remained laminar from the leading edge to approximately the shock foot, while multiple shock wave structures were present in the flow field. With increasing Mach number, buffet onset and offset were observed at $M = 0.7$ and $M = 0.9$, respectively, while the frequency of buffet increases approximately monotonically in between. For $M \geq 0.8$, shock waves were observed on both surfaces of the aerofoil, while the boundary layer remained permanently separated, although large amplitude excursions of the shock waves could still be observed at $M = 0.8$. This implies that strong buffet can persist even when the aerofoil is fully stalled. With increasing incidence angle or Reynolds number, buffet amplitude increases, but the buffet frequency does not change significantly in the ranges studied. The number of shock wave structures reduce with increasing $\Rey$, implying that further increases in $\Rey$ would lead to a single shock wave. Sweep has negligible effect on buffet features and three-dimensional buffet cells are absent for the flow conditions studied. Indeed, the results satisfy the Independence Principle approximately. 

Based on the results at high $M$, a separate case of $M = 0.8$ at zero incidence was simulated. Buffet was observed with shock waves present on both aerofoil surfaces which oscillate with a 180 degree phase difference between each other implying the occurrence of a Type I buffet \citep{Giannelis2017}, but on a supercritical aerofoil. This case resembles others where the shock wave appears only on the suction side (Type II), with attributes like the frequency and the occurrence of multiple shock wave structures being common to both. Thus, these two buffet types are suggested to be governed by the same physical mechanisms. From this, it is proposed that models like that of \citet{Gibb1988} that are exclusively proposed to explain one type and not the other could be invalid.

Spectral orthogonal decomposition (SPOD) was used to extract coherent features of the flow field. In addition to a mode at the lower buffet frequency ($St  \approx 0.1$), a bump was observed in the spectrum in the range $1 \leq St \leq 5$, associated with vortex shedding in the wake. A modal reconstruction of the flow field based only on a single SPOD mode was implemented to examine the individual influence of each mode on the flow field. Reconstruction based on the low frequency buffet mode showed two features that are common to all cases. The first is a phase lag between the shock wave position and the streamwise extent of flow separation. The second is the the periodic build up of pressure near the trailing edge when the shock wave is close to its downstream most position. These aspects require further scrutiny, but it is evident from the results that the laminar buffet simulated here is related to intermittent flow separation and reattachment and large amplitude shock wave motion, implying that it is similar to turbulent buffet. By contrast, a reconstruction based only on the wake mode implemented here resembles the bubble-breathing phenomenon reported in \citet{Dandois2016} for the OALT25 aerofoil, suggesting that bubble-breathing is not related to laminar buffet as proposed in that study.

The waves associated with the model proposed in \citet{Lee1990} were identified and the buffet time period was predicted based on estimates of the time required for these to move from the shock foot to the trailing edge and back. The predicted time period was found to be substantially lower as compared to that observed in the LES for all cases studied indicating that the model is inaccurate. While such quantitative comparisons have been provided in previous studies, physical reasons to suspect the model's validity is provided here by showing that these waves are primarily associated with vortex shedding (wake modes) which do not directly influence the buffet mode. The modified version of this model proposed in \citet[figure 15, p. 14]{Hartmann2013} is also shown to make erroneous assumptions: for all cases, it was seen that the vortices reaching the trailing edge are most intense when the shock wave is most upstream, which is opposite to what is assumed.

The main advantages of the present study are the high-fidelity approach, the wide range of parameters used and the use of modal decomposition and reconstruction to isolate individual effects of coherent structures on the global flow field. These allowed for identifying generic features of buffet and assessing various models and mechanisms proposed to explain it. Based on these, it is proposed that further understanding of the physical mechanisms underlying buffet can be achieved by scrutinising the causal relationship between shock waves and flow separation. \\

\textbf{Acknowledgements}: We would like to acknowledge the computational time on ARCHER and ARCHER2 (UK supercomputing facility) provided by the UK Turbulence Consortium (UKTC) through the EPSRC grant EP/R029326/1. We also acknowledge the use of the IRIDIS High Performance Computing Facility, and associated support services at the University of Southampton, in the completion of this study. The V2C airofoil geometry was kindly provided by Dassault Aviation. This study was funded by the Engineering and Physical Sciences Research Council (EPSRC) grant, ``Extending the buffet envelope: step change in data quantity and quality of analysis” (EP/R037167/1).

\bibliographystyle{jfm}
\bibliography{jfm}

\end{document}